\documentclass[aps,prf,onecolumn,superscriptaddress,longbibliography]{revtex4-1}
\usepackage{graphicx}
\usepackage{dcolumn}
\usepackage{bm}
\usepackage{amsmath}

\begin{document}

\title{The role of grain dynamics in determining the onset of sediment transport}

\author{Abram H. Clark}
\affiliation{Department of Mechanical Engineering and Materials Science, Yale University, New Haven, Connecticut 06520, USA}
\author{Mark D. Shattuck}
\affiliation{Benjamin Levich Institute and Physics Department, The City College of the City University of New York, New York, New York 10031, USA}
\author{Nicholas T. Ouellette}
\affiliation{Department of Civil and Environmental Engineering, Stanford University, Stanford, California 94305, USA}
\author{Corey S. O'Hern}
\affiliation{Department of Mechanical Engineering and Materials Science, Yale University, New Haven, Connecticut 06520, USA}
\affiliation{Department of Physics, Yale University, New Haven, Connecticut 06520, USA}
\affiliation{Department of Applied Physics, Yale University, New Haven, Connecticut 06520, USA}

\begin{abstract}

Sediment transport occurs when the nondimensional fluid shear stress $\Theta$ at the bed surface exceeds a minimum value $\Theta_c$. A large collection of data, known as the Shields curve, shows that $\Theta_c$ is primarily a function of the shear Reynolds number ${\rm{Re}}_*$. It is commonly assumed that $\Theta>\Theta_c({\rm{Re}}_*)$ occurs when the ${\rm Re}_*$-dependent fluid forces are too large to maintain static equilibrium for a typical surface grain. A complimentary approach, which remains relatively unexplored, is to identify $\Theta_c({\rm{Re}}_*)$ as the applied shear stress at which grains cannot stop moving. With respect to grain dynamics, ${\rm{Re}}_*$ can be viewed as the viscous time scale for a grain to equilibrate to the fluid flow divided by the typical time for the fluid force to accelerate a grain over the characteristic bed roughness. We performed simulations of granular beds sheared by a model fluid, varying only these two time scales. We find that the critical Shields number $\Theta_c({\rm Re}_*)$ obtained from the model mimics the Shields curve and is insensitive to the grain properties, the model fluid flow, and the form of the drag law. Quantitative discrepancies between the model results and the Shields curve are consistent with previous calculations of lift forces at varying ${\rm Re}_*$. Grains at low ${\rm Re}_*$ find more stable configurations than those at high ${\rm{Re}}_*$ due to differences in the grain reorganization dynamics. Thus, instead of focusing on mechanical equilibrium of a typical grain at the bed surface, $\Theta_c({\rm{Re}}_*)$ may be better described by the stress at which mobile grains cannot find a stable configuration and stop moving.

\end{abstract}

\maketitle

\section{Introduction}
\label{sec:intro}

A fluid that flows over a granular bed exerts a shear stress on the grains and, if the flow is sufficiently strong, will entrain grains in the flow. This process is responsible for shaping much of the natural world. Understanding and controlling the erosion of sediments by flowing water are significant for a range of ecological and agricultural problems~\citep{knisel1980,Walling1983,renard1997}. Thus, the nature of the onset and cessation of grain motion in the presence of a fluid shear flow has been the subject of extensive research dating back many decades (for example, see recent reviews by \citet{Dey2014} and \citet{buffington1997}), but it is still not fully understood. This problem involves nontrivial coupling between several physical processes that are each difficult to characterize. Predicting the dynamics of granular materials is challenging, even for very simple cases like frictionless disks~\citep{xu2006}. In the natural world, the geological processes that produce the granular materials in question yield grains with varying size, shape, roughness, and other material properties~\citep{barrett1980}. The mechanics of the flow that impart stress to the bed are also nontrivial, given both the wide range of channel geometries in natural streams and rivers~\citep{schumm1960,parker1979} and possibly turbulent conditions. Additionally, the fluid inside the bed is also moving, as the bed can be viewed as a porous material, and is governed by Darcy flow~\citep{nield2013}, although with a complicated boundary condition linking it to the turbulent flow at the bed surface.

\begin{figure}

\centering \includegraphics[width=0.5\columnwidth]{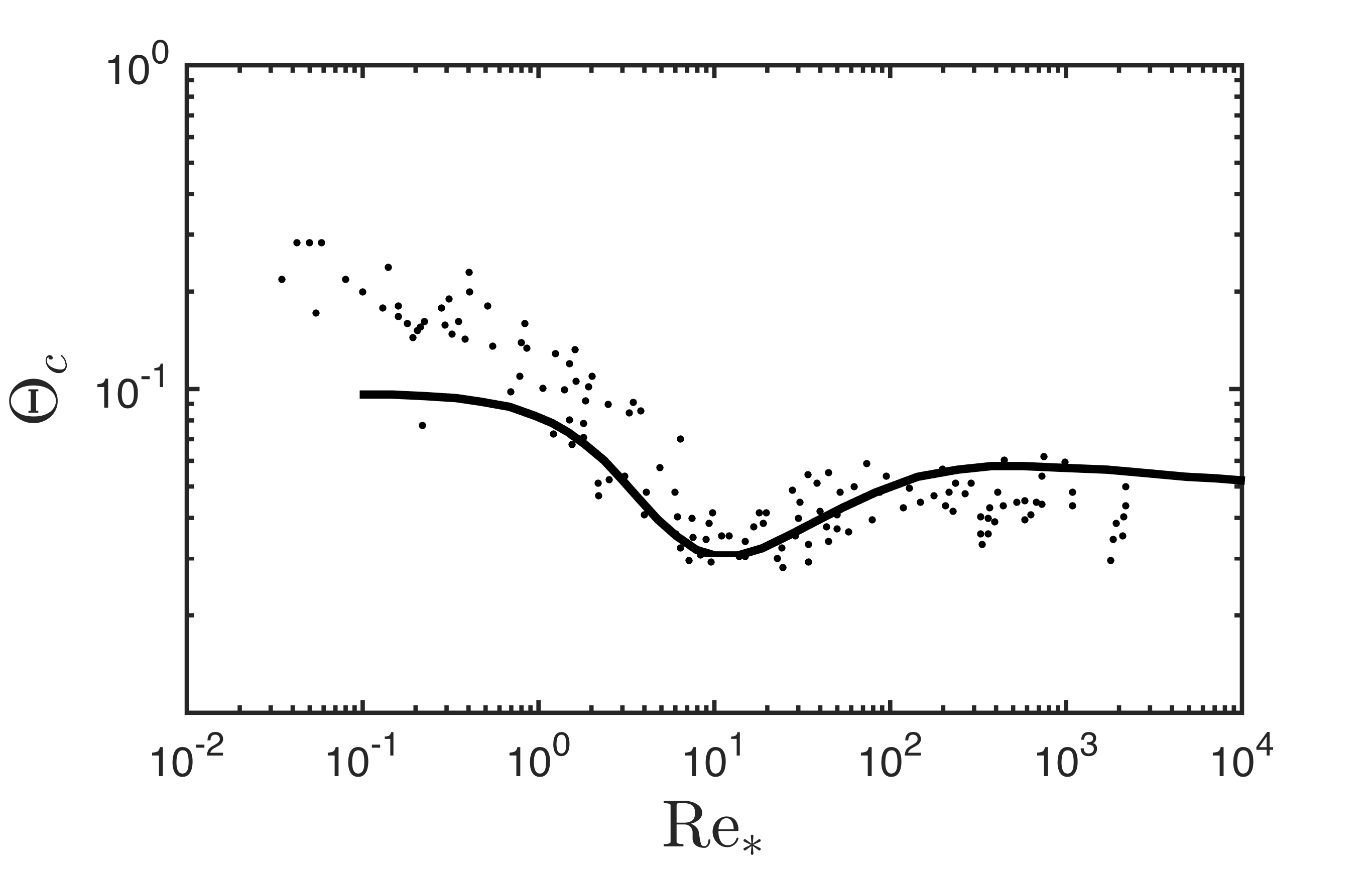}

\caption{A collection of experimental and field data from \citet{Dey2014} showing the variation of the minimum Shields number for grain motion $\Theta_c$ with the shear Reynolds number ${\rm Re}_*$.  The solid black line represents the theoretical curve for monodisperse sediment derived by~\citet{wiberg1987}; see text for discussion.}
\label{fig:Shields-curve}
\end{figure}

Despite the apparent complexity of this problem, there is evidence that the boundary in parameter space between mobile and static beds can be described relatively simply. In particular, a collection of data dating back over a century suggests that the onset of grain motion can be captured by only two nondimensional parameters~\citep{gilbert1914, casey1935, shields1936, USWES1936, white1940, vanoni1946, meyer1948, neill1967, grass1970, white1970, karahan1975, mantz1977, yalin1979, Dey2014, buffington1997}. First, the Shields number $\Theta=\frac{\tau}{\Delta\rho g D}$ compares the horizontal shear stress $\tau$ exerted on the bed surface by the fluid to the downward gravitational stress $\Delta \rho g D$, where $\Delta \rho=\rho_g-\rho_f$, $\rho_g$ and $\rho_f$ are the mass densities of the grains and fluid, $g$ is the gravitational acceleration, and $D$ is the typical grain diameter. The minimum Shields number $\Theta_c$ required for grain motion is typically plotted as a function of the shear Reynolds number ${\rm Re}_*=u_*D/\nu$, where $u_*^2=\tau/\rho_f$ and $\nu$ is the kinematic viscosity of the fluid. Figure~\ref{fig:Shields-curve} shows data for $\Theta=\Theta_c$ versus ${\rm Re}_*$ taken from~\citet{Dey2014}, who compiled data from a range of sources~\citep{gilbert1914,casey1935,shields1936,USWES1936,white1940,vanoni1946,meyer1948, neill1967,grass1970,white1970,karahan1975,mantz1977,yalin1979}. These data, often referred to as the Shields curve, were collected over a wide range of different flows, spanning the range from laminar to fully turbulent, and for many channel geometries and grain properties. Although the data are scattered, they cluster around a master curve. However, why the Shields curve takes this particular form and why that form is so robust against variation of other parameters remain open questions.

\subsection{Prior descriptions of the Shields curve}
\label{sec:prior-descriptions}
There have been a number of approaches aimed at explaining the shape of the Shields curve (see \citet{Dey2014} for a comprehensive treatment). To date, the most successful descriptions are hydraulic or empirical scaling formulas~\citep{shields1936,miller1977,mantz1977,yalin1972,yalin1979,vanrijn1993}; see~\citet{Paphitis2001} for a review of these models. This approach was pioneered by~\citet{shields1936}, who originally found that $\Theta_c$ varies with ${\rm Re}_*$ and noted that ${\rm Re}_*$ controls the ratio of the boundary roughness (set by the grain size $D$) to the size of the viscous sublayer. The Shields curve can then be broken into regions where grains are, as compared to the viscous sublayer, completely submerged (${\rm Re}_*<2$), near the top ($2<{\rm Re}_*<10$), partially protruding ($10<{\rm Re}_*<1000$), and fully protruding (${\rm Re}_*>1000$). These regions appear to coincide with distinct regimes of the Shields curve. Others have tried a similar approach, where $\Theta_c$ is plotted as a function of different nondimensional parameters. For instance, the Yalin number $\Xi = {\rm Re}_*/\sqrt{\Theta_c}$~\citep{yalin1972} and dimensionless grain diameter $D_*= D(g'/\nu^2)^{1/3}$~\citep{vanrijn1993}, where $g' = \frac{\rho_g - \rho_f}{\rho_g} g$ is the buoyancy-reduced gravitational acceleration, have been used to eliminate the shear stress dependence on the horizontal axis. These approaches have the advantages of being based in grain scale fluid mechanics and being highly predictive, but they still require an empirical fit to the data shown in Fig.~\ref{fig:Shields-curve}.

Attempts at a more theoretical derivation~\citep{Iwagaki1956,wiberg1987,Ling1995,Dey1999,dey2000} of the shape of the Shields curve are typically based on static force and torque balance of a typical grain that resides on the surface of the bed (see \cite{Dey2014} for a thorough review). Such approaches have been successful at capturing certain features of the Shields curve, but not the quantitative shape over the full range of ${\rm Re}_*$. Thus, the variation of $\Theta_c$ with ${\rm Re}_*$ is assumed to be a purely fluid-driven effect. The Shields curve denotes the maximum stress at which a grain on the bed surface in a typical geometric configuration remains in static equilibrium given the ${\rm Re}_*$-dependent contributions from lift, drag, and turbulence. However, since these calculations consider a single grain in a particular local environment, they are highly sensitive to the details of that environment, which is often called a ``pocket'' with pocket angles $\psi$ that specify the orientation of grain-grain contacts~\citep{kirchner1990,andrews1994,buffington1992,lamb2008}. 

A notable example of this approach by~\citet{wiberg1987} considered a quasi-two-dimensional case where a grain sits in a pocket and motion is initiated when the downstream forces exceed the resistive forces. That is, grain motion occurs when the ratio of the downstream force applied by the fluid on the grain $F_d$ to the vertical forces $F_{g'} - F_l$ (the buoyancy reduced gravitational force minus fluid-induced lift force) are equal to $\cot \psi$ (see Fig.~\ref{fig:cartoon}). Thus, initiation of motion occurs when 
\begin{equation}
\Theta_c \propto \left(\frac{F_d}{F_{g'}}\right)_c = \cot \psi \frac{1}{1+\cot \psi (F_l / F_d)_c},
\label{eqn:Wiberg-scaling}
\end{equation}
where the subscript $c$ denotes the critical condition to initiate motion. The only inputs to the calculation are the pocket angle $\psi$ (see Fig.~\ref{fig:cartoon}) and the form of the fluid flow at varying ${\rm Re}_*$, which modulates the ratio $(F_l/F_d)_c$ and causes variation in $\Theta_c$. \citet{wiberg1987} assumed a logarithmic fluid profile for ${\rm Re}_*>100$ and a form proposed by \citet{Reichardt1951} for ${\rm Re}_*<100$ that has been shown to agree well with experiments~\citep{Schlichting2003}. They assumed a drag force $F_d = C_d \frac{1}{2} \rho_f \bar{u}^2 A_x$ and lift force $F_l=C_l \frac{1}{2} \rho_f (\bar{u}_t^2-\bar{u}_b^2) A_x$, where $C_d$ is the ${\rm Re}_p$-dependent drag coefficient, $\bar{u}$ is the height-dependent velocity profile, subscripts $t$ and $b$ denote the respective values at the top and bottom of the grain, $A_x$ is the cross-sectional (frontal) area of a grain, and $C_l$ is a constant lift coefficient. They then self-consistently solved Eq.~\eqref{eqn:Wiberg-scaling} by calculating $F_d$ and $F_l$ from the appropriate fluid profiles at different values of ${\rm Re}_*$. Small grains (${\rm Re}_* \ll 1$) are buried deep in the viscous sublayer, and lift forces are small. As ${\rm Re}_*$ increases, grains begin to protrude out of the viscous sublayer, and lift forces increase rapidly compared to drag forces. This regime corresponds to the global minimum in the Shields curve at ${\rm Re}_*\approx 10$. Large grains (${\rm Re}_* \gg 10$) protrude far into the overlying logarithmic flow profile. In this case, lift forces persist, but their effect becomes less pronounced relative to the drag force. In Fig.~\ref{fig:Shields-curve}, we show a solution to this equation taken from~\citet{wiberg1987}, where the bed roughness $k_s$ is equal to the grain diameter $D$ and $\psi = 30^\circ$. This curve captures the global minimum in the Shields curve at ${\rm Re}_*\approx 10$, corresponding to conditions where the grain size is roughly equal to the size of the viscous sublayer. The overall magnitude of this curve is proportional to $\cot \psi$, which is essentially a fit parameter and is very sensitive to the local grain geometry. This curve agrees well with data for ${\rm Re}_* > 1$, but it underestimates $\Theta_c$ at low ${\rm Re}_*$. A smaller value of $\psi$ would better capture the data at low ${\rm Re}_*$, but there is no clear physical reason to choose a different $\psi$ for small ${\rm Re}_*$.

\subsection{The role of grain dynamics}
\label{sec:dimensional-analysis}

However, a theoretical description for $\Theta_c({\rm Re}_*)$ that includes ${\rm Re}_*$-dependent grain dynamics may be able to explain how beds could be stronger at low ${\rm Re}_*$. For example, grain motion can be temporary, as individual grains that are unstable at a particular shear force can find more stable locations. Depending on the preparation history of the bed, grains may move initially when a shear flow is applied, but the bed may grow stronger as grains search and find more stable configurations~\citep{kirchner1990,charru2004,lobkovsky2008,hong2015}. That is, mobilized grains can often find a more stable pocket than the original one and stop moving, and this effect is usually neglected in prior theoretical considerations. How does ${\rm Re}_*$ affect the dynamics of grains as they search for stability? How do the results of models that include the grains' search for collective stability compare to the Shields curve? This manuscript will address these two important questions.

\begin{figure}
\centering 
\includegraphics[width=0.5\columnwidth]{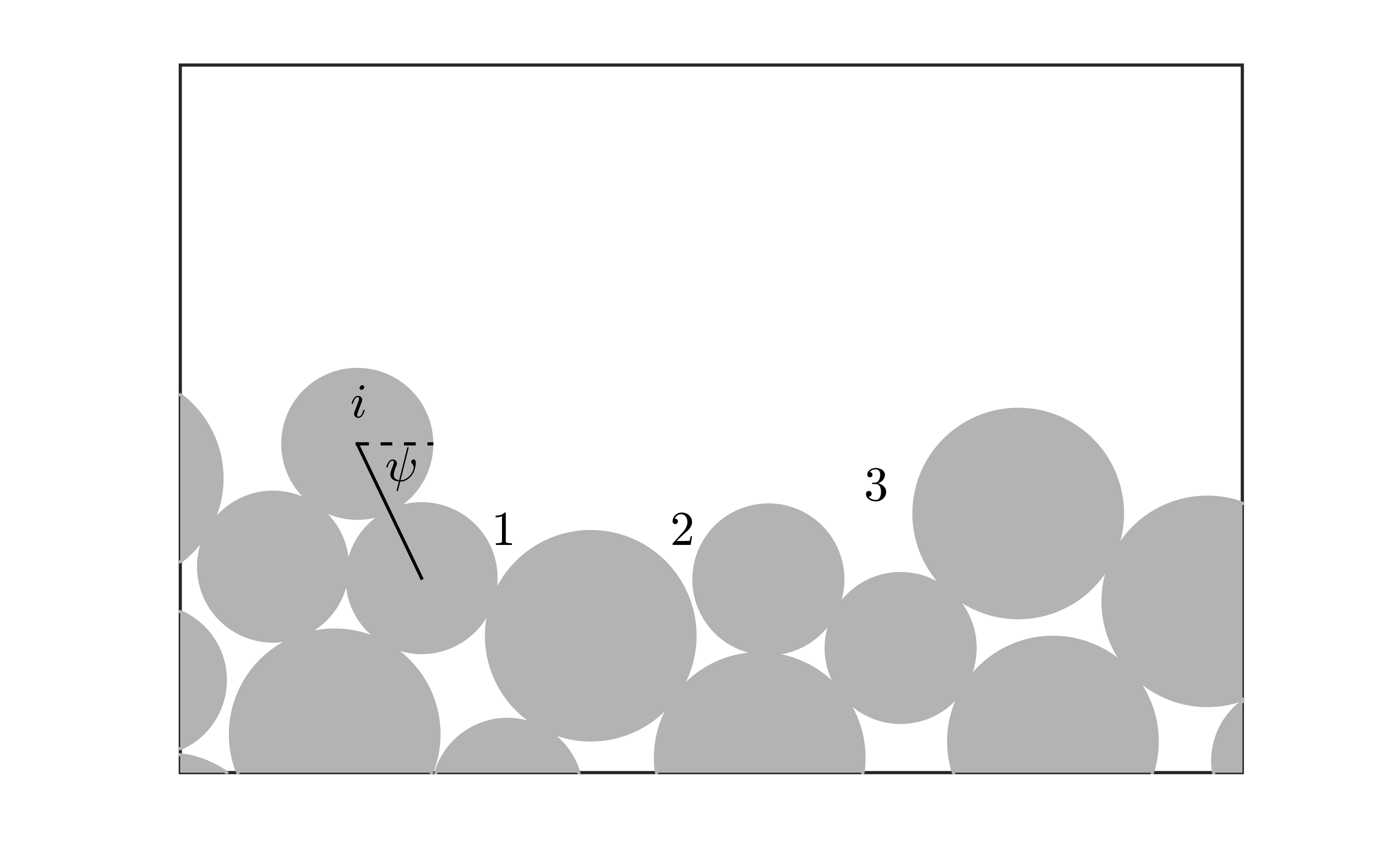}
\caption{Grain $i$ will move when the ratio of horizontal (rightward) to vertical (downward) forces is equal to $\cot \psi$. Possible landing sites (pockets) are marked 1, 2, and 3. As we discuss in the text, whether the grain will stop after motion is initiated depends both on the geometrical details of the pockets as well as the dynamics of the grain motion, which vary with ${\rm Re}_*$.}
\label{fig:cartoon}
\end{figure}

To illustrate the effects of ${\rm Re}_*$-dependent grain dynamics, we first consider a simple example of a grain sitting on top of a 2D bed, shown in Fig.~\ref{fig:cartoon}. Grain $i$ will become unstable when the ratio of the horizontal fluid force (rightward) and vertical forces (gravity minus lift forces) is $\cot \psi$. After this grain becomes unstable (i.e., not in force and torque balance) at the surface of the bed, the subsequent dynamics depend strongly on ${\rm Re}_*$. The mobilized grain can then land and remain in one of the other pockets on the bed surface, labeled 1-3, depending on geometry, grain properties, and the amount of momentum it has acquired. That is, for the grain to land in one of the pockets, the grain must be stable in that pocket at the given $\Theta$ and it must be moving sufficiently slowly to stop in the pocket. 

If we approximate the motion of a mobile grain as that of a sphere in a uniform fluid flow, then the drag force is given by $F_D = \frac{1}{2} C_d \rho_f  A_x V^2$, where $A_x =\frac{\pi}{4} D^2$ is the cross sectional (frontal) area of the sphere, the drag coefficient $C_d \approx \frac{24}{\rm{Re}_p} + 0.4$, the particle Reynolds number ${\rm Re}_p = V D/\nu$, and $V$ is the slip velocity between the fluid and the sphere. The first term, $24/{\rm Re}_p$, in $C_d$ captures Stokes drag (linear in $V$) and the second term, 0.4, captures inertial drag (quadratic in $V$). The dynamics can thus be written as
\begin{equation}
\left(\rho_g\frac{\pi}{6}D^3\right)\frac{dV}{dt}=-(3\pi\rho_f\nu D)V-\left(\frac{\pi}{20}\rho_f D^2\right)V^2,
\label{eqn:grain-dynamics}
\end{equation}
where $\rho_g\frac{\pi}{6}D^3$ is the grain mass. The solution to this equation is
\begin{equation}
V(t)=V_0 \frac{\exp(-t / \tau_\nu)}{1+\frac{{\rm Re}_p^0}{60} [1-\exp(-t / \tau_\nu)]},
\label{eqn:V(t)-solution}
\end{equation}
where $\tau_\nu \propto \frac{\rho_g D^2}{\rho_f \nu}$ is the viscous equilibration time scale, ${\rm Re}_p^0=V_0 D/\nu$ is the initial particle Reynolds number, and $V_0$ is the slip velocity at $t=0$. If ${\rm Re}_p^0 \ll 1$, then $V(t)=V_0 \exp(-t / \tau_\nu)$, and the viscous time scale specifies the dynamics. If ${\rm Re}_p^0 \gg 1$, then $\tau_\nu$ becomes large, and a Taylor expansion of the exponential terms in Eq.~\eqref{eqn:V(t)-solution} yields $V(t) = V_0 \left(1+ \frac{t}{\tau_I} \right)^{-1}$, where $\tau_I \propto \frac{\rho_g D}{\rho_f V_0}$. This solution is also obtained by integrating Eq.~\eqref{eqn:grain-dynamics} with the Stokes drag term set to zero. We note that even for ${\rm Re}_p^0 \gg 1$, $\tau_\nu$ still dominates the final portion of the dynamics.

Assuming a constant acceleration $\Theta g'$, the dimensionless shear stress $\Theta$ takes a characteristic time $\tau_\Theta \propto \sqrt{\frac{D}{\Theta g'}}$ to accelerate a grain through a distance $D$, which is the typical spacing between successive collisions with the bed. Here, $g'$ is the buoyancy-reduced gravitational acceleration, so $\Theta g'$ is the typical acceleration that a grain first experiences after it becomes unstable. We note that $\tau_\Theta$ can also be written as $\tau_\Theta \propto \sqrt{\frac{\rho_g}{\rho_f}}\frac{D}{u_*}$. The form we have chosen emphasizes its connection to horizontal acceleration of grains along the bed surface. However, the acceleration of a mobile grain will eventually be cut off by equilibrating to the fluid flow. Thus, the bed collision time scale $\tau_\Theta$ should be compared to the fluid equilibration time scales $\tau_\nu$, which is associated with the viscous component of the drag force, and $\tau_I$, which is associated with the inertial component of the drag force. It can be shown that ${\rm Re}_*$ compares the viscous equilibration time scale to the bed collision time scale,
\begin{equation}
{\rm Re}_* =\sqrt{\frac{\rho_f}{\rho_g}}\frac{\tau_\nu}{\tau_\Theta}.
\label{eqn:Re*}
\end{equation} 

For ${\rm Re}_* \ll 1$, a weakly mobilized grain quickly equilibrates to the fluid flow. Thus, it is not significantly accelerated between interactions with the bed, it acquires very little momentum, and its dynamics are viscous-dominated. In this case, the effect of grains bouncing over geometrically stable pockets should be negligible. If the grain finds a geometrically stable location, it will stop. When ${\rm Re}_* \gg 1$, a mobilized grain is accelerated between successive interactions with the bed, acquiring momentum $p\sim \Theta m g' \tau_\Theta$, and its dynamics are acceleration-dominated. For ${\rm Re}_*\gg 1$, the inertial time scale is dominant, meaning that $\tau_I/\tau_\Theta$ is the relevant ratio, instead of $\tau_\nu/\tau_\Theta$. This ratio is a constant, $\tau_I / \tau_\Theta = \sqrt{\rho_g / \rho_f}$. Physically, this means that further increasing ${\rm Re}_*$ does not cause the grains to be accelerated for longer times. However, a typical ratio $\rho_g / \rho_f \approx 3$ for rocks, minerals, and soils still yields $\tau_I>\tau_\Theta$. This means that, for ${\rm Re}_*\gg 1$, weakly mobilized grains are significantly accelerated between interactions with the bed~\cite{roseberry2012,schmeeckle2014}, which is not true at ${\rm Re}_* \ll 1$. This framework provides a mechanism whereby grains stop more easily at low ${\rm Re}_*$ than at high ${\rm Re}_*$, given the same value of $\Theta$. Note that this argument implies that $\tau_I$ plays a secondary role, and that much of the relevant physics can be captured by a viscous drag law, neglecting the inertial component. We return to this point in our argument below. We also note that our dimensional analysis focuses on a comparison of a viscous damping time $\tau_\nu$ to an acceleration time scale $\tau_\Theta$. This is similar in spirit to a Stokes number~\cite{joseph2001,yang2006,Schmeeckle2001}, which plays a crucial role in the degree of energy loss in fluid-mediated grain-grain collisions. To compare to the data in Fig.~\ref{fig:Shields-curve}, we chose to formulate our results in terms of ${\rm Re}_*$.

To test this interpretation, we here present the results of discrete-element method (DEM) simulations. While many DEM-based approaches include as much physical realism as possible~\citep{schmeeckle2003,carneiro2011,duran2012,capecelatro2013,nabi2013}, we here simplify the problem to isolate the role of ${\rm Re}_*$ as the ratio of the two time scales controlling grain dynamics. Thus, we are neglecting many physical effects such as ${\rm Re}_*$-dependent lift forces~\cite{wiberg1987}, cohesive forces~\cite{mehta1989}, turbulent fluctuations~\cite{schmeeckle2014,diplas2008}, coherent structures~\cite{robinson1991,adrian2007,hardy2009,vowinckel2016}, added mass forces~\cite{drake2001}, and Basset forces~\cite{nino1998}. In the present study, noncohesive grains are driven by a model fluid shear flow, which does not vary with ${\rm Re}_*$, but is coupled to the velocity of the grains through a drag law, which sets ${\rm Re}_*$. In previous work~\cite{clark2015hydro}, we investigated how $\Theta_c$ varied with a particle Reynolds number using a purely linear drag law in 2D with frictionless, purely elastic grain-grain interactions. Interestingly, even this simple model captured certain features of the Shields curve, namely plateaus at low and high particle Reynolds number with a decrease in between. In this work, we use the improved dimensional analysis presented above to explicitly connect the form of the drag law to ${\rm Re}_*$, in order to more directly compare the results of the DEM simulations to the Shields curve. Additionally, we vary the spatial dimension (2D to 3D), inelastic grain-grain interactions, friction, irregular grain shape, and the form of the drag law (linear to quadratic). We also include ${\rm Re}_*$-independent lift forces and vary their magnitude. By varying these parameters, our goal is to understand the minimal subset of parameters that control the shape of $\Theta_c({\rm Re}_*)$ from the perspective of ${\rm Re}_*$-dependent grain dynamics.

\section{Methods}
\label{sec:methods}

\subsection{Equations of motion and grain-grain interactions}

\begin{figure*}
\raggedright \hspace{10 mm} (a) \hspace{50 mm} (b) \hspace{50 mm} (c) \\
\centering
\includegraphics[trim=0mm 0mm 0mm 5mm, clip, width=0.3\textwidth]{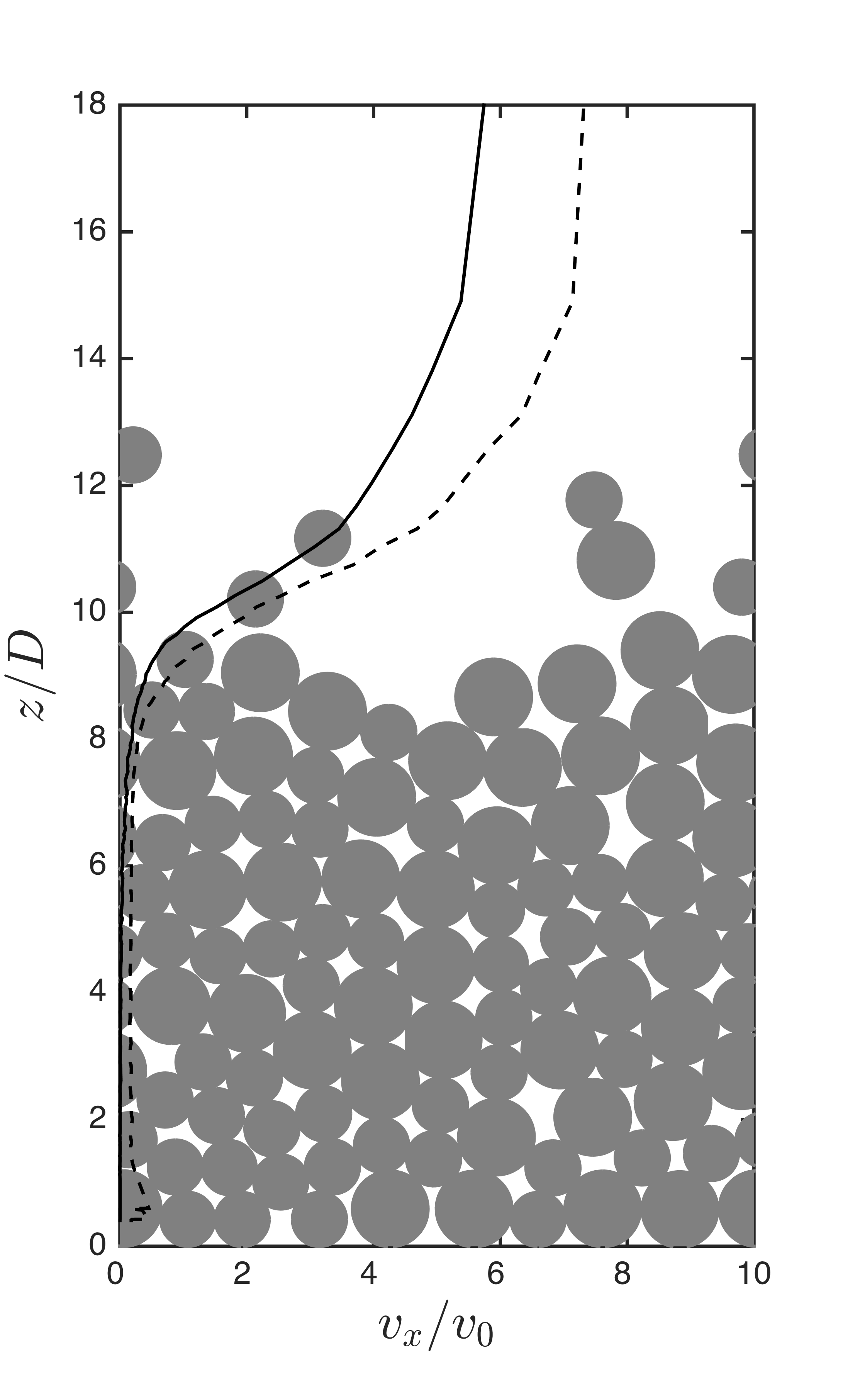}
\includegraphics[trim=0mm 0mm 0mm 5mm, clip, width=0.3\textwidth]{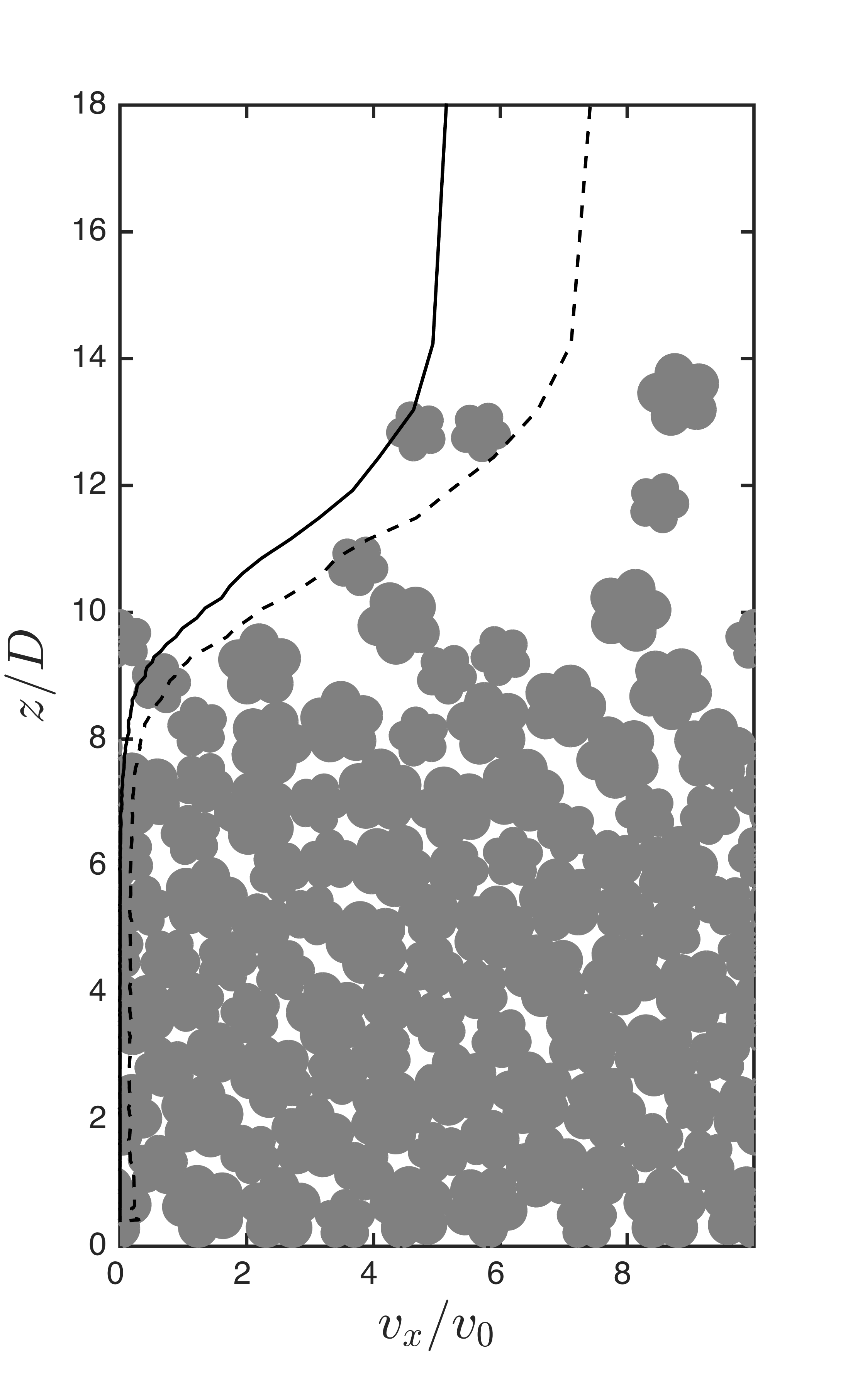}
\includegraphics[trim=0mm 0mm 0mm 5mm, clip, width=0.3\textwidth]{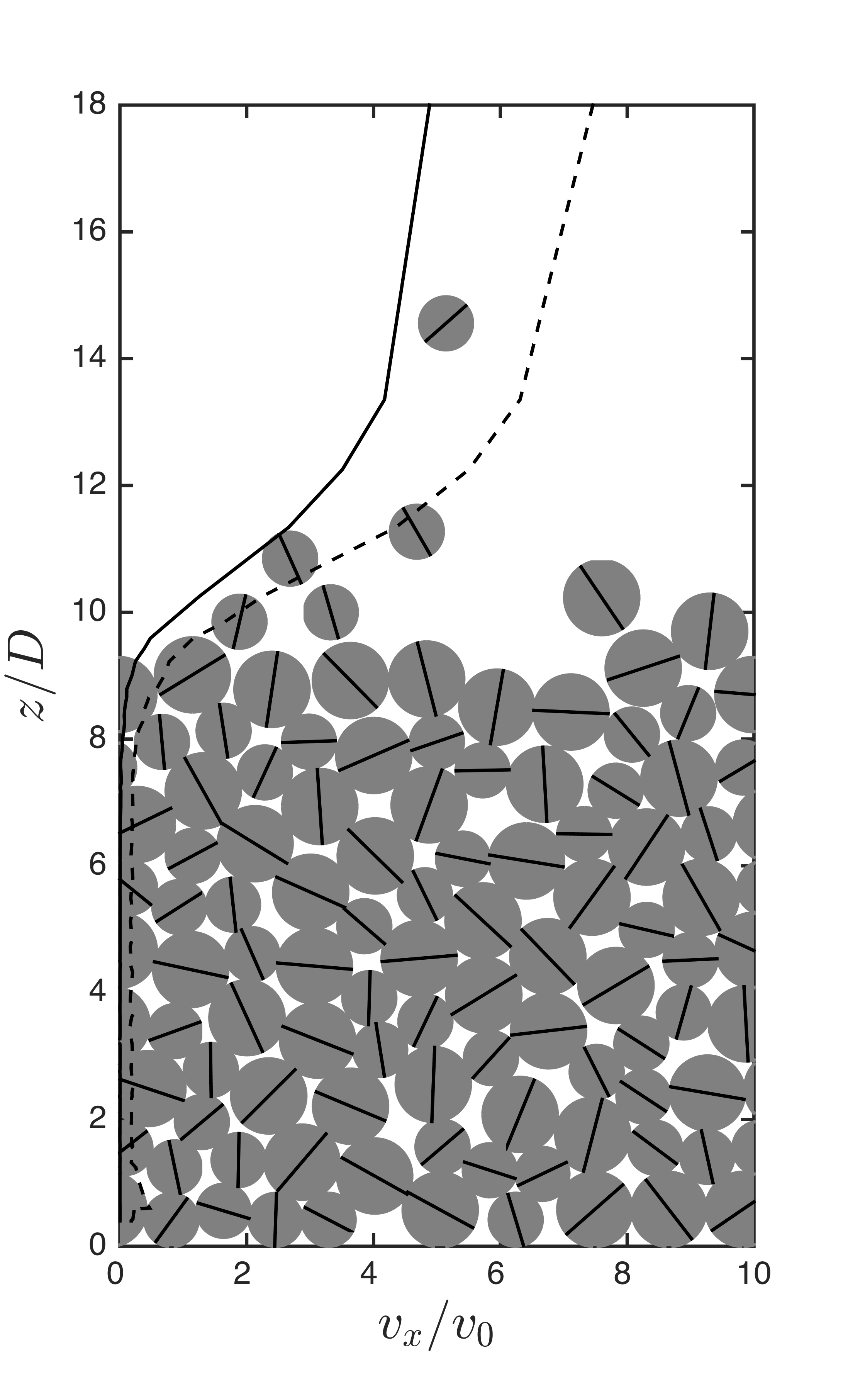}
\caption{Panels (a-c) show snapshots of simulations using (a) frictionless disks; (b) grain clusters from Fig.~\ref{fig:bumpy-packing} with $\mu_{\rm eff}=0.6$; and (c) disks with Cundall-Strack friction~\citep{cundall79}, with $\mu=0.6$. All three simulations shown here are at $\Theta =0.25$, ${\rm Re}_* \approx 1.6$, and restitution coefficient of $e_n = 0.8$. The vertical axis gives the height $z/D$ above the lower boundary, where $D$ is the mean grain diameter, and the horizontal axis gives the horizontal velocity $v_x/v_0$ of grains and the fluid, where $v_0$ is the characteristic fluid velocity at the bed surface. Solid and dashed lines show the time-averaged horizontal component of grain velocity $v_x^g$ and fluid velocity $v_x^f$, respectively, during a short simulation.} 
\label{fig:bumpy-flow}
\end{figure*}

We study systems composed of $N/2$ large and $N/2$ small grains with diameter ratio $1.4$ in 2D~\citep{Perera1999,Speedy1999} and $1.2$ in 3D~\citep{Zhang2014}. These size ratios are chosen to maintain structural disorder in the bed. In our analysis, we use the average diameter to evaluate dimensionless quantities such as ${\rm Re}_*$. Our domain has periodic boundaries in the stream-wise direction, as well as in the cross-stream direction in 3D. We use no upper confining boundary and a rigid lower boundary with infinite friction so that the horizontal velocities of all grains touching it are fixed to zero. We integrate Newton's equations of motion for each grain, including rotational and translational degrees of freedom, using a sixth-order Gear predictor-corrector integration scheme for the case of Cundall-Strack friction~\citep{cundall79} for disks in 2D and a modified velocity Verlet integration scheme for all other systems. The total force on each grain is given by the vector sum of contact forces from other grains, a gravitational force, and a drag force from a fluid that moves horizontally, so that
\begin{equation}
m_i\frac{d\vec{v}_i}{dt} =\sum_j \vec{F}^c_{ij} - m_ig'\hat{z} + \vec{F}_f.
\label{eqn:force-law-rot}
\end{equation}
The total torque on each grain is only due to tangential contact forces, so that
\begin{equation}
I_i\frac{d\vec{\omega}_i}{dt} = \sum_j \vec{s}_{ij}\times\vec{F}^c_{ij}.
\label{eqn:torque}
\end{equation}
Here, the sum over $j$ only includes grains contacting grain $i$, $\vec{s}_{ij}$ is the vector connecting the center of grain $i$ to the point of contact between grains $i$ and $j$, $m_i$ is the mass of the grain ($m_i \propto D_i^2$ in 2D and $m_i \propto D_i^3$ in 3D), $I_i$ is the moment of inertia of the grain, $D_i$ is the diameter of the grain, $\vec{v}_i$ is the velocity of the grain, $m_ig'$ is the buoyancy-corrected grain weight, $\hat{z}$ is the upward normal vector, and $\vec{F}_f$ is the drag force from a model fluid flow, which we discuss below. For the frictionless, elastic case~\citep{clark2015hydro}, $\vec{F}^c_{ij}= \vec{F}^r_{ij}$, where $\vec{F}^r_{ij}=K\left(1-\frac{r_{ij}}{D_{ij}}\right)\theta\left(1-\frac{r_{ij}}{D_{ij}}\right)\hat{r}_{ij}$ is the pairwise (linear repulsive spring) force on grain $i$ from grain $j$, where $K$ is the grain stiffness, $r_{ij}$ is the separation between the centers of the grains, $D_{ij}=(D_i + D_j)/2$, $\hat{r}_{ij}$ is the unit vector connecting their centers, and $\theta$ is the Heaviside step function. We set the nondimensional stiffness $\frac{K}{m g'}>3\times 10^3$ to be sufficiently large that our results become independent of $K$. In this study, we modify the contact force $\vec{F}^c_{ij}$ to include dissipative grain-grain interactions and tangential forces. The dissipative force is given by $\vec{F}^d_{ij}=\gamma_v \frac{m_i m_j}{m_i+m_j} (\vec{v}_i-\vec{v}_j)\cdot \hat{r}_{ij}$, where the dissipation rate $\gamma_v = \frac{-2 \log e_n}{\tau_c}$, $\tau_c=\frac{\pi \sqrt{m}}{2K}$ is the grain-grain collision time, $m$ is the mean grain mass, and $e_n$ is the coefficient of normal restitution~\citep{schafer96}. This form for the normal dissipation is often used to model energy losses that arise from contact mechanics, such as viscoelasticity, internal heating, or internal vibrational modes of grains. Here, it is likely that the fluid in the intergrain gap dominates the energy loss during a collision, and the effective $e_n$ depends on the relative impact velocity of the grains and the viscosity of the fluid via the Stokes number ${\rm St}=\frac{\rho_g v_{ij} D}{\rho_f \nu}$~\cite{yang2006,joseph2001,Schmeeckle2001,nino1998}, where $v_{ij}$ is a relative velocity between colliding grains. We hold $e_n$ fixed for each individual simulation, independent of the relative grain velocity at contact or local fluid behavior. We will include velocity and viscosity dependence of the coefficient of restitution in future studies. Tangential forces in granular beds arise via two mechanisms: nonspherical grain shape and microscopic friction. We approximate these two mechanisms using a grain-asperity model~\citep{Buchholtz1994,papanikolaou2013}, shown in Fig.~\ref{fig:bumpy-flow}(b), and the Cundall-Strack model for friction~\citep{cundall79}, shown in~\ref{fig:bumpy-flow}(c). Further details are provided in Appendix~\ref{app:friction}. 

\subsection{Details of the fluid drag}
\label{sec:fluid-drag}

\begin{figure*}
\raggedright \hspace{0 mm} (a) \hspace{38 mm} (b) \hspace{38 mm} (c) \hspace{38 mm} (d) \\
\centering
\includegraphics[trim=0mm 0mm 0mm 0mm, clip, width=0.24\textwidth]{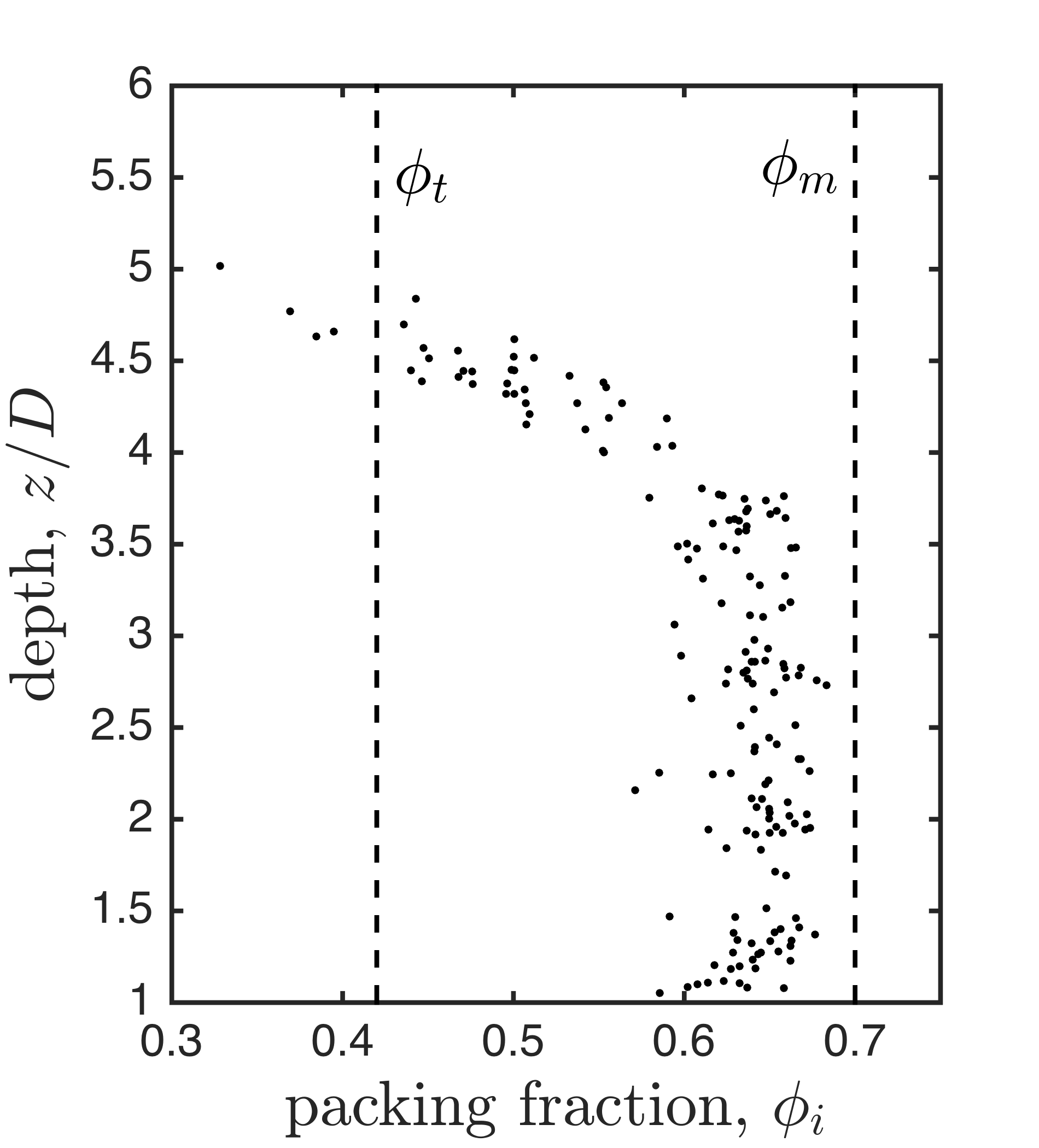}
\includegraphics[trim=0mm 0mm 0mm 0mm, clip, width=0.24\textwidth]{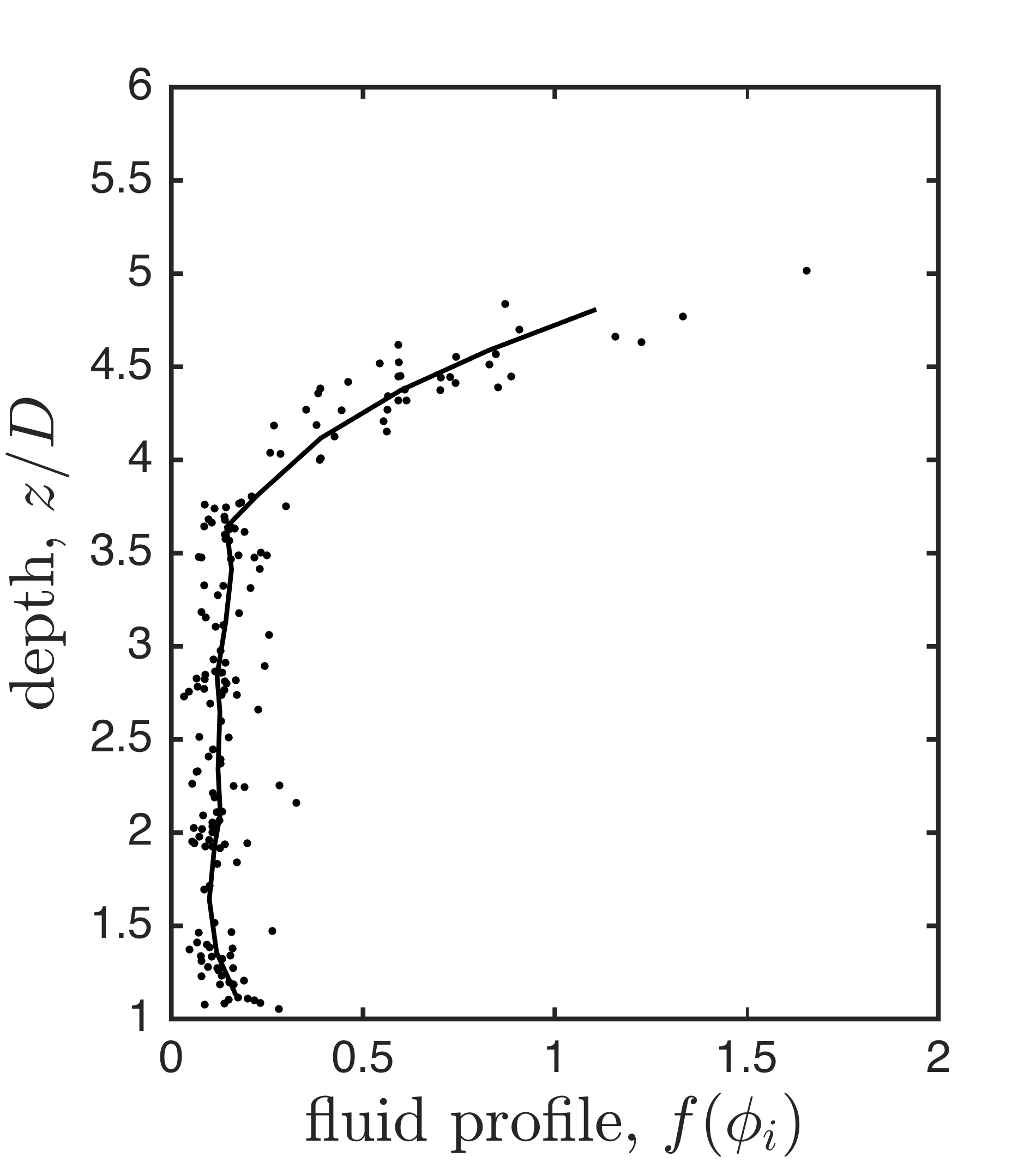}
\includegraphics[trim=0mm 0mm 0mm 0mm, clip, width=0.24\textwidth]{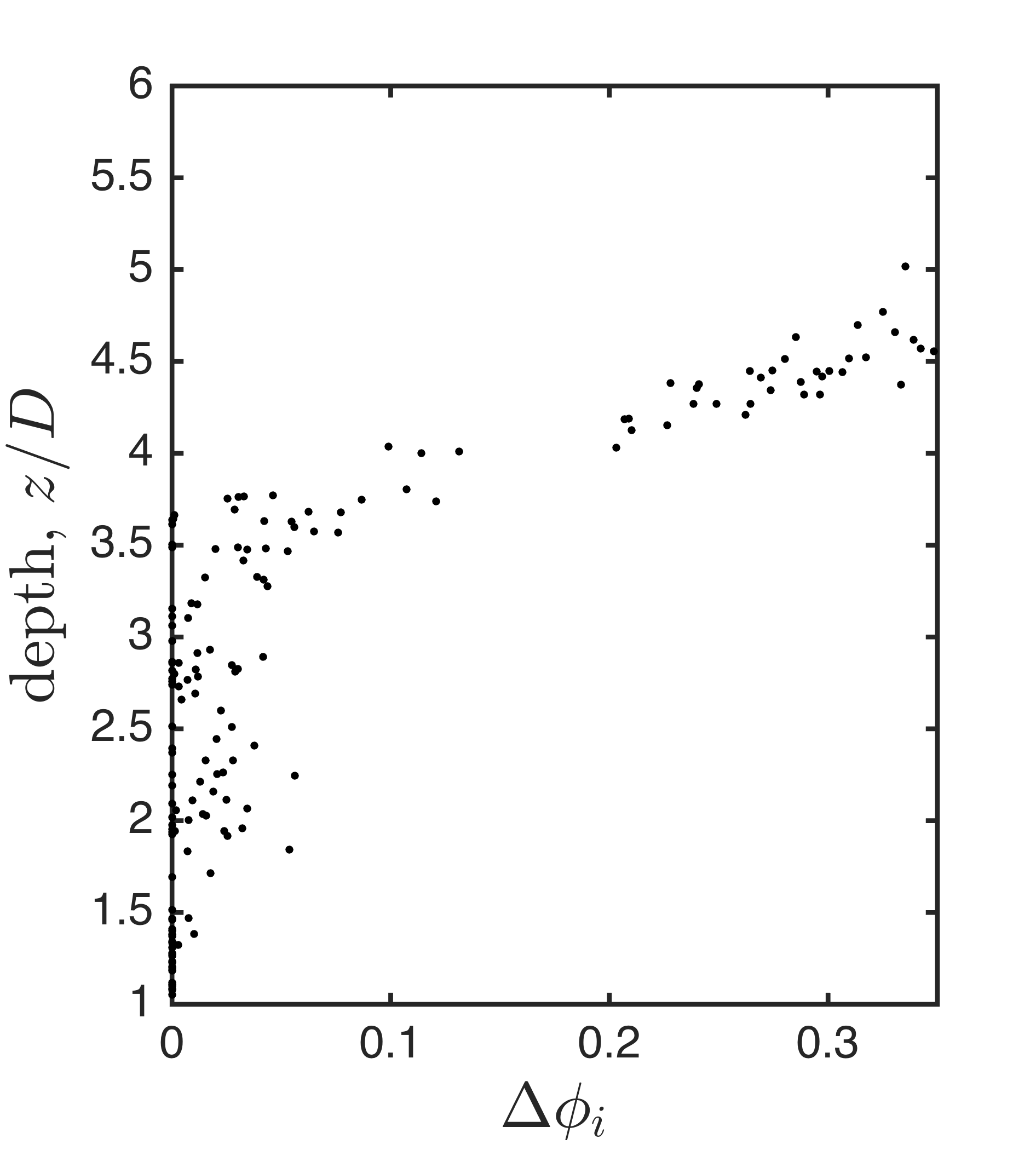}
\includegraphics[trim=0mm 0mm 0mm 0mm, clip, width=0.24\textwidth]{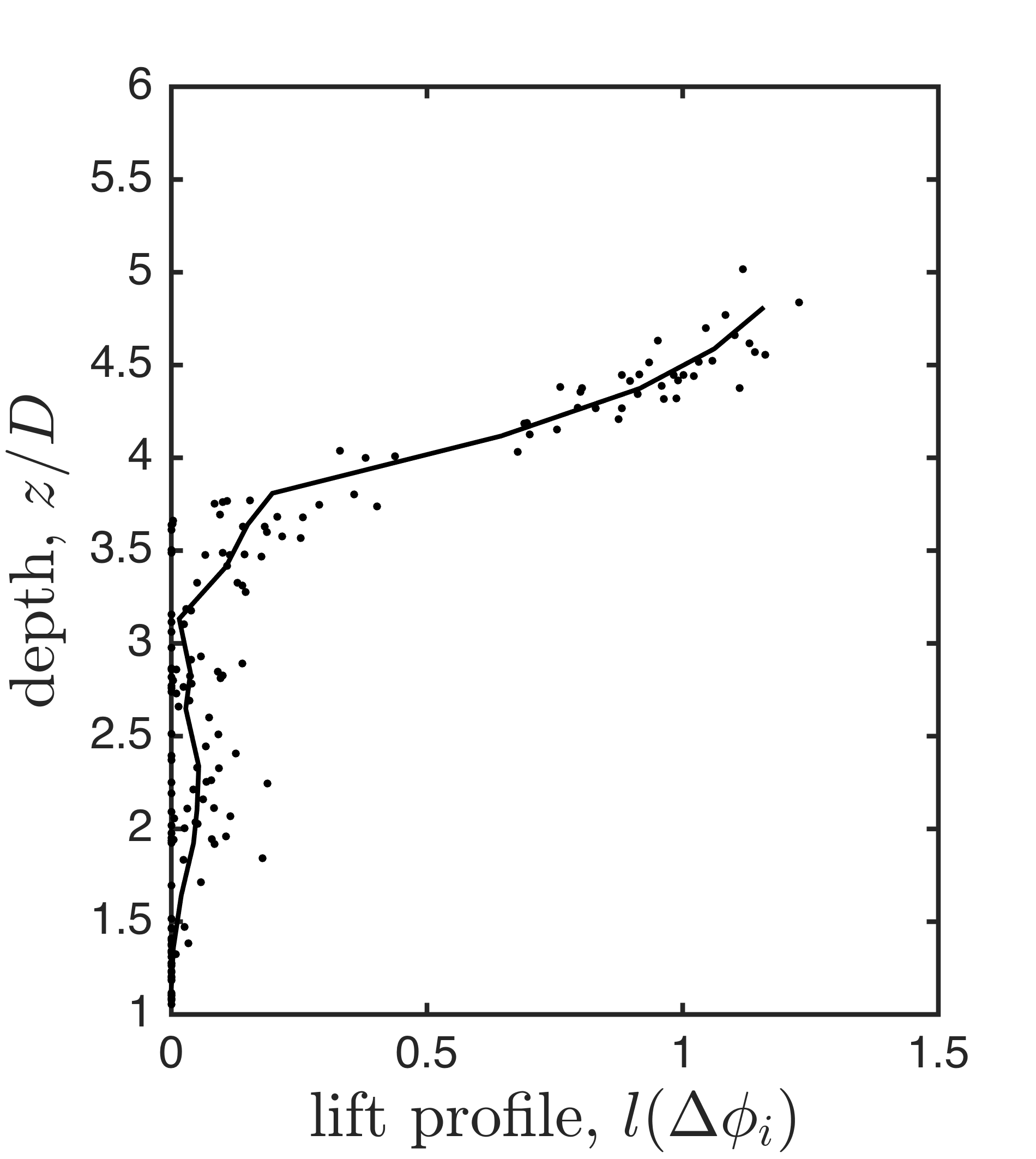}
\caption{(a) A scatter plot of the height $z/D$ ($D$ is the median grain size) of each grain above the lower boundary versus the packing fraction $\phi_i$ of a typical configuration of 200 grains in a static 3D bed. We choose $\phi_t=0.42$ to correspond with a typical packing fraction in the top layer and $\phi_m=0.7$ as the maximum packing fraction grains find in the bed (our simulations results are qualitatively insensitive to these choices). 
(b) A plot of the same data points in (a), with the fluid profile $f(\phi_i)=[\exp(-b\phi)-\exp(-b\phi_m)]/[\exp(-b\phi_t)-\exp(-b\phi_m)]$ with $b=5$ on the horizontal axis and $z/D$ on the vertical axis. The solid line shows a binned average, which roughly corresponds to the applied fluid profile, and the data points show the typical scatter, which arises from local fluctuations in $\phi_i$.
(c) A scatter plot from the same bed as in (a) and (b) of the height $z/D$ versus $\Delta \phi_i$, where $\Delta \phi_i$ is the difference between the local packing fraction calculated at the top and bottom of grain $i$ (see text for details). (d) The same data from (c), with the lift profile $l(\Delta\phi_i) = \frac{\Delta\phi_i}{0.3}$ on the horizontal axis. The solid line shows a binned average.
}
\label{fig:flow-profiles}
\end{figure*}

To simulate fluid shear, we choose a model fluid velocity profile that acts primarily on the surface grains of a static bed and that increases somewhat for mobilized grains that move above the surface. We set a characteristic fluid velocity $v_0$ at the surface of a static bed, and we multiply this velocity by a fluid profile $f(\phi_i)$, where $\phi_i$ is the local packing density at grain $i$, which yields $v_0 f(\phi_i)$ acting on grain $i$. Our simulations use the local packing fraction $\phi_i$, which varies horizontally and vertically, but our results are insensitive to horizontally averaging $\phi_i$ such that local packing fraction only varies vertically. In 2D, $f(\phi_i)=e^{-b(\phi_i-\phi_t)}$, where $b$ controls the ratio of the magnitude of the fluid flow above and inside the bed and $\phi_t = 0.5$ is the typical packing fraction of a grain at the top of a static 2D bed. $\phi_i$ is calculated in a small region with diameter $D_i+2D_l$ around the center of each grain, where $D_l$ is the diameter of the larger grains. Since $f=1$ for $\phi_i=\phi_t$, $v_0$ is roughly equal to the fluid velocity at the free granular surface. In 3D, we use a modified form, $f(\phi_i)=[\exp(-b\phi_i)-\exp(-b\phi_m)]/[\exp(-b\phi_t)-\exp(-b\phi_m)]$, as shown in Fig.~\ref{fig:flow-profiles}. $\phi_m=0.7$ approximates the maximum packing fraction in the bulk of the bed, and $\phi_t = 0.42$ is a typical packing fraction of a grain at the top of the bed. This modified form is used to sufficiently reduce the fluid velocity inside a 3D bed, where typical packing fractions are much smaller ($\approx 0.55 - 0.64$) than in 2D ($\approx 0.75 - 0.84$). For all simulations, we set $b=5$. We find that our results are insensitive to the choice of $f$, provided its magnitude is very small in the bed. Since $\phi_t$ causes a shift in our definition of the height of the bed surface (see Fig.~\ref{fig:flow-profiles}), varying this parameter corresponds to multiplying all our results for $\Theta$ by an order unity prefactor. However, our results are qualitatively unaffected by this choice, and quantitative variation is weak, provided the choice for $\phi_t$ falls within a reasonable range taken from Fig.~\ref{fig:flow-profiles} (i.e., 0.35 -- 0.5)

To add model lift forces (in 3D only), we consider the difference $\Delta \phi_i$ in packing fraction between the top and bottom of a grain (Fig.~\ref{fig:flow-profiles}). We again calculate the packing fraction in a small region with diameter $D_i+2D_l$ around the top and bottom of each grain, and $\Delta\phi_i$ is the difference between these two quantities. $\Delta\phi_i$ is small in the bulk of a bed, large at the bed surface, and small for mobilized grains above the bed. We then define a lift profile $l(\Delta\phi_i) = \frac{\Delta\phi_i}{0.3}$, where the factor 0.3 is chosen to normalize $l$ to unity at the bed surface.

We then set the fluid force $\vec{F}_f$ from Eq.~\eqref{eqn:force-law-rot} equal to the form from Eq.~\eqref{eqn:grain-dynamics},
\begin{equation}
\vec{F}_f =B_1 [v_0 f(\phi_i) \hat{x} - \vec{v}_i] + B_2 |v_0 f(\phi_i) \hat{x} - \vec{v_i}| [v_0 f(\phi_i) \hat{x} - \vec{v}_i] + B_l l(\Delta\phi_i) \hat{z},
\label{eqn:fluid-drag}
\end{equation}
where $B_1=3\pi\rho_f\nu D_i$ and $B_2 = \frac{\pi}{20}\rho_f D_i^2$, and $B_l$ is a lift coefficient. While there are two different particle diameters, we use the mean value $D$ in our dimensional analysis below. In addition to $\frac{K}{m g'}$, $\mu$, and $e_n$, which determine grain-grain interactions, Eqs.~\eqref{eqn:force-law-rot}-\eqref{eqn:fluid-drag} include four additional nondimensional numbers:
\begin{align}
\Theta &= \frac{\tau}{\Delta\rho g D} = \frac{2}{3}\left(\frac{B_1 v_0 + B_2 v_0^2}{mg'} \right)
\label{eqn:nondimen-params-Theta}
\\ {\rm Re}_* &= \sqrt{\frac{\rho_f}{\rho_g}}\frac{\tau_\nu}{\tau_\Theta} = \sqrt{3}\left(\frac{m/B_1}{\sqrt{\frac{D}{\Theta g'}}}\right) 
\label{eqn:nondimen-params-Re*}
\\ {\rm Re}_p^0 &=\frac{ v_0 D }{\nu} = \frac{60 B_2 v_0}{B_1} 
\label{eqn:nondimen-params-Re_p}\\
\frac{F_l}{F_d} &=\frac{B_l}{B_1 v_0 + B_2 v_0^2}.
\label{eqn:nondimen-params-lift}
\end{align}

$\Theta$ is the Shields parameter, where the factor $2/3$ from Eq.~\eqref{eqn:Wiberg-scaling} represents a conversion~\citep{wiberg1987} from a force ratio [i.e., in Eq.~\eqref{eqn:nondimen-params-Theta}, $B_1 v_0$, $B_2v_0^2$, and $mg'$ have units of force] to the stress ratio $\Theta=\frac{\tau}{\rho g' D}$. Assuming that the fluid stress $\tau$ acts approximately over the cross-sectional area of a sphere, $A=\frac{\pi}{4}D^2$, then $\tau A$ is the horizontal force exerted on a static grain. The gravitational stress $\rho g' D$ acts over an effective area $\frac{V}{D}=\frac{\pi}{6}D^2$, and $\rho g' D \frac{V}{D}$ is the grain weight. Thus, the explicit force ratio $\frac{B_1 v_0 + B_2 v_0^2}{mg'}$ is converted to a stress by multiplying by $\frac{V}{AD}=\frac{2}{3}$. In Eq.~\eqref{eqn:nondimen-params-Re*}, we rewrite the shear Reynolds number ${\rm Re}_* = \sqrt{\frac{\rho_f}{\rho_g}}\frac{\tau_\nu}{\tau_\Theta} $ from Eq.~\eqref{eqn:Re*} in Sec.~\ref{sec:dimensional-analysis} in terms of parameters relevant to the simulations [see Eq.~\eqref{eqn:fluid-drag} using $\tau_\nu = m/B_1$, $\tau_\Theta = \sqrt{\frac{D}{\Theta g'}}$, and $\rho_g/\rho_f = \sqrt{3}$]. These two forms in Eqs.~\eqref{eqn:Re*} and \eqref{eqn:nondimen-params-Re*} are equivalent, and both reduce to the standard form ${\rm Re}_*=\frac{u_* D}{\nu}$. Next, ${\rm Re}_p^0$ is the particle Reynolds number of a grain at the surface of the bed, which determines the relative contributions of the viscous and inertial drag terms. We emphasize that, when comparing results between a linear drag law, where $B_2=0$, and a quadratic drag law, where $B_2/B_1$ is determined by ${\rm Re}_p$ according to Eq.~\eqref{eqn:nondimen-params-Re_p}, we first assign $\Theta$ and then specify $v_0$. Simulations for linear and quadratic drag laws at the same $\Theta$ will require different values of $v_0$. This is distinct from the case where the fluid velocity is specified, in which case we would severely underestimate the stresses at high ${\rm Re}_*$. Finally, $\frac{F_l}{F_d}$ represents the characteristic ratio of lift to drag forces at the bed surface. We set this quantity to zero in all but a small number of our simulations.


To characterize the onset and cessation of bed motion in our system, we employ two protocols. To study the mobile-to-static (M-S) transition defined by $\Theta_c$, we distribute all grains randomly on a cubic lattice throughout the domain and set a constant value of $\Theta$ for a total time of roughly $10^6$ grain-grain collision times (our results are insensitive to the details of this initial condition, as long as we consider large ensembles of initial conditions where a significant fraction of grains are suspended when the model fluid flow is applied). We then observe if and when our system stops, which we define as when the maximum acceleration $a_{\rm max}<a_{\rm thresh}$ and maximum velocity $v_{\rm max}<v_{\rm thresh}$, where $a_{\rm thresh}$ is roughly one order of magnitude smaller than $g'$ and roughly three orders of magnitude smaller than typical values for a moving bed and $v_{\rm thresh}$ is roughly three orders of magnitude smaller than the fluid velocity at the surface. Our results are independent of the values of these thresholds, provided they are sufficiently small. To understand the dynamics of the static-to-mobile (S-M) transition, we begin with a static bed and slowly increase $v_0$ in small increments, corresponding to increases in stress $\Delta \Theta \leq 0.05 \Theta$ until we observe $a_{\rm max}>a_{\rm thresh}$ or $v_{\rm max}>v_{\rm thresh}$. We then keep $v_0$ constant until $a_{\rm max}<a_{\rm thresh}$ and $v_{\rm max}<v_{\rm thresh}$ or until the end of our simulation. Table~\ref{tbl:sim-protocol} gives a full list of the parameter values we explore, and we refer back to each of these settings S-1 through S-8 throughout the remainder of the manuscript.

\begin{table}
    \caption{A list of the settings for the DEM simulations presented in this article. Protocol refers to either mobile-to-static transition (M-S), where mobilized beds are allowed to search for stable configurations, or static-to-mobile transition (S-M), where we increase $\Theta$ slowly for a static bed until we observe indefinite grain motion. Dimension denotes whether the simulations are 2D or 3D. Fill height is the distance between the top of the bed and the bottom boundary. $\Theta$ is the Shields parameter. Drag law refers to linear (Lin.) or quadratic (Quad.). ${\rm Re}_p$ is the particle Reynolds number, and ${\rm Re}_*$ is the shear Reynolds number. $F_l/F_d$ is the typical ratio of lift to drag forces at the top of the bed. $e_n$ is the restitution coefficient. $\mu$ and $\mu_{\rm eff}$ are friction coefficients for the Cundall-Strack model and the grain-asperity model, respectively. We refer back to these settings by the label S-1 through S-8.}
\centering
    \vspace{0.2in}
    \begin{tabular}{|c|c|c|c|c|c|c|c|c|c|c|c|}
    \hline 
      $ $ & Protocol & Dimension & $N$ & fill height & $\Theta$  & Drag & ${\rm Re}_p$ & ${\rm Re}_*$ & ${F_l}/{F_d}$ & $e_n$ & $\mu$, $\mu_{\rm eff}$ \\ \hline  \hline

S-1 & M-S & 3D & 400 & $5D$ & 0.05 -- 0.5 & Quad. & 0.01 -- 30,000 & 0.05 -- 1,000 & 0 & 0.9 & 0 \\ 

S-2 & M-S & 3D & 400 & $5D$ & 0.05 -- 0.5 & Lin. & 0 & .03 -- 2,000 & 0 & 0.2 & 0 \\ 

S-3 & M-S & 3D & 400 & $5D$ & 0.01 -- 0.5 & Lin. & 0 & 0.05 -- 2,000 & 0 & 0.1 -- 0.9 & 0 \\ 

S-4 & M-S & 3D & 400 & $5D$ & 0.05 -- 0.5 & Quad. & 0.01 -- 30,000 & 0.05 -- 1,000 & 0 -- 3 & 0.9 & 0 \\ 

S-5 & M-S & 2D & 200 & $10D$ & 0.06 -- 1 & Lin. & 0 & 0.05 -- 1,000 & 0 & 0.1 -- 0.9 & $\mu = 10^{-4}$ -- 5 \\ 

S-6 & M-S & 2D & 200 & $10D$ & 0.06 -- 1 & Lin. & 0 & 0.05 -- 1,000 & 0 & 0.1 -- 0.9 & $\mu_{\rm eff} = 0.1$ -- 2 \\            

S-7 & S-M & 3D & 50 -- 800 & $5D$ & 0.033 -- 0.75 & Quad. & 3,000 & 100 & 0 & 0.5 & 0 \\             
         
S-8 & S-M & 2D & 50 -- 800 & $5D$ -- $40D$ & 0.033 -- 1.7 & Lin. & 0 & 10 & 0 & 0.8 & $\mu, \mu_{\rm eff} = 0.6$ \\             
            
\hline
 \end{tabular}
    \label{tbl:sim-protocol}
\end{table}

\section{Results}
\label{sec:Results}


\subsection{Summary of simulation results}
\label{sec:result-summary}
We present our results in the following way. In Sec.~\ref{sec:main-result}, we show simulations in 3D using a quadratic drag law (S-1 settings in Table~\ref{tbl:sim-protocol}). We show that $\Theta_c({\rm Re}_*)$ from this model mimics the Shields curve. Quantitative discrepancies are consistent with the relative contribution of lift forces~\citep{wiberg1987}, as well as possible contributions from turbulent fluctuations~\cite{robinson1991,diplas2008,hardy2009,schmeeckle2014} and coherent structures~\cite{adrian2007,vowinckel2016}. We also find that $\Theta_c({\rm Re}_*)$ is identical for both a quadratic drag law in 3D (S-1 settings) and a linear drag law in 3D with small $e_n$ (S-2 settings). We show these results in order to justify our argument in Sec.~\ref{sec:dimensional-analysis} that the inertial time scale $\tau_I$ plays only a secondary role, since linear drag includes only $\tau_\nu$ and not $\tau_I$. In Sec.~\ref{sec:linear-vs-quad}, we give further details of how $\Theta_c({\rm Re}_*)$ with linear drag (S-2 and S-3 settings) converges to the result from quadratic drag (S-1 settings) when $e_n$ approaches zero. In Sec.~\ref{sec:lift}, we include lift forces at the bed surface (S-4 settings), and $\Theta_c$ decreases as expected. In Sec.~\ref{sec:3D-to-2D}, we use 2D simulations (S-5 and S-6 settings) to show that friction and irregular grain shape only weakly affect the results. In Sec.~\ref{sec:weibull}, we use weakest-link statistics to show that grain motion is always initiated at $\Theta_c({\rm Re}_*)$ for large systems (using S-7 and S-8 settings). This picture suggests that fluid-sheared granular beds possess a dynamical instability for moderate to high ${\rm Re}_*$, where mobile grains are unable to stop, and that grain dynamics plays a dominant role in this regime.

\subsection{$\Theta_c$ varies with ${\rm Re}_*$-dependent grain dynamics}
\label{sec:main-result}
\begin{figure}
\centering \includegraphics[width=0.5\columnwidth]{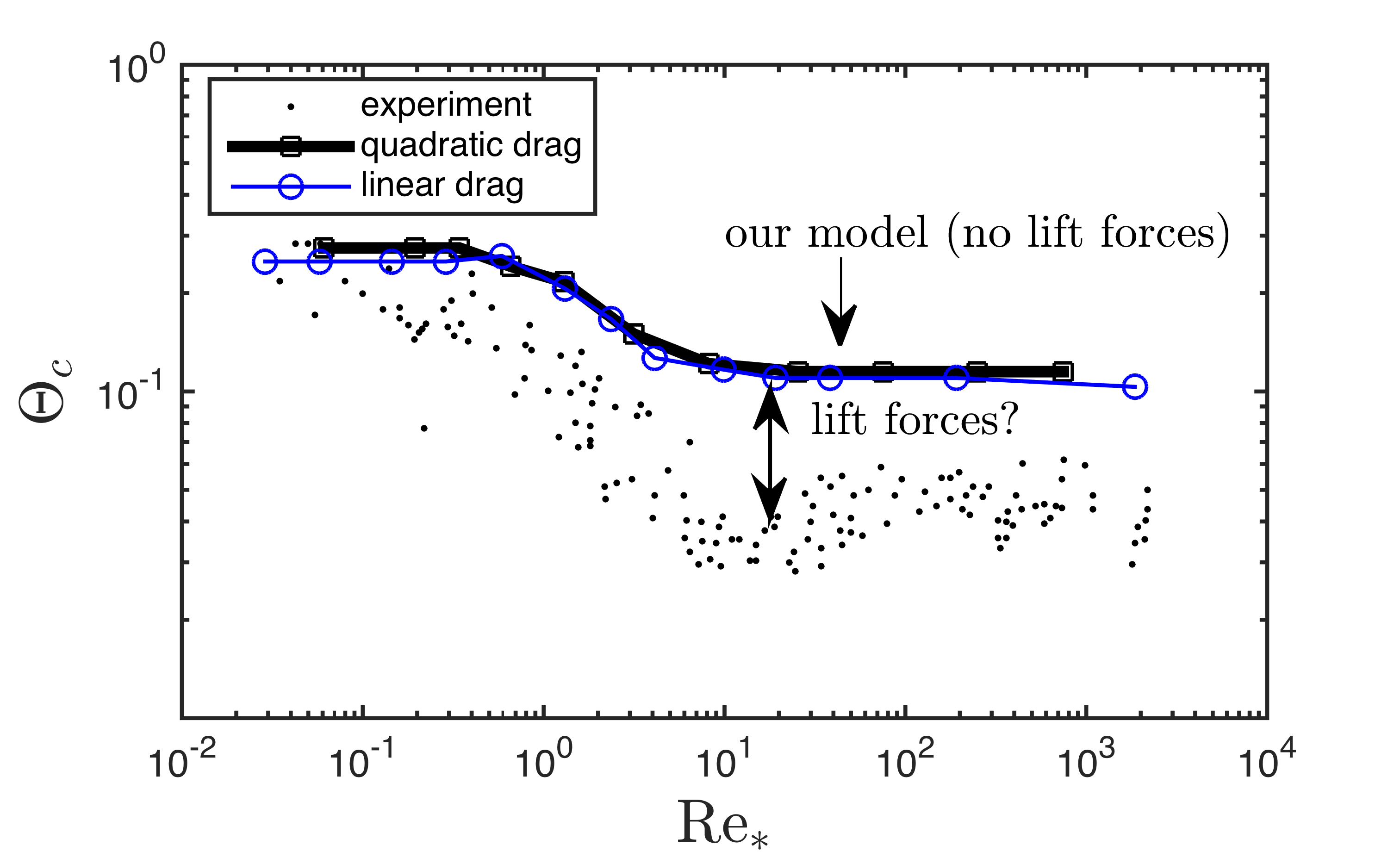}
\caption{A collection of experimental and field data (filled dots) from \citet{Dey2014} showing the variation of the minimum Shields number for grain motion $\Theta_c$ with ${\rm Re}_*$. The solid curves show the boundaries between states with and without sustained grain motion from our model in 3D, excluding lift forces. The black curve with square markers shows results for quadratic drag (S-1 settings), and the blue curve with open circles shows results for linear drag in the limit of small restitution coefficient (S-2 settings). 
}
\label{fig:Shields-curve-with-data}
\end{figure}

The main result from our work is shown in Figure~\ref{fig:Shields-curve-with-data}. The small, black dots represent experimental and field data from~\citet{Dey2014}. The black curve with square markers is the boundary $\Theta_c({\rm Re}_*)$ in 3D between systems with and without sustained grain motion from our model with S-1 settings. We find a nearly identical boundary using a linear drag law when $e_n$ is small, as shown by the solid blue curve with open circles, where $e_n=0.2$ (S-2 settings). As discussed below in Sec.~\ref{sec:linear-vs-quad}, small $e_n$ suppresses the effect of grain impact with the bed for linear drag; for quadratic drag, the inertial equilibration time scale limits the kinetic energy of mobilized grains. In both the cases of linear drag with small $e_n$ and quadratic drag, the key physics near $\Theta_c$ is the dynamics of mobilized grains between successive interactions with the bed.

\begin{figure*}
\raggedright \hspace{5 mm} (a) \hspace{50 mm} (b) \hspace{40 mm} (c) \\
\centering
\includegraphics[trim=0mm 0mm 0mm 5mm, clip, width=0.32\textwidth]{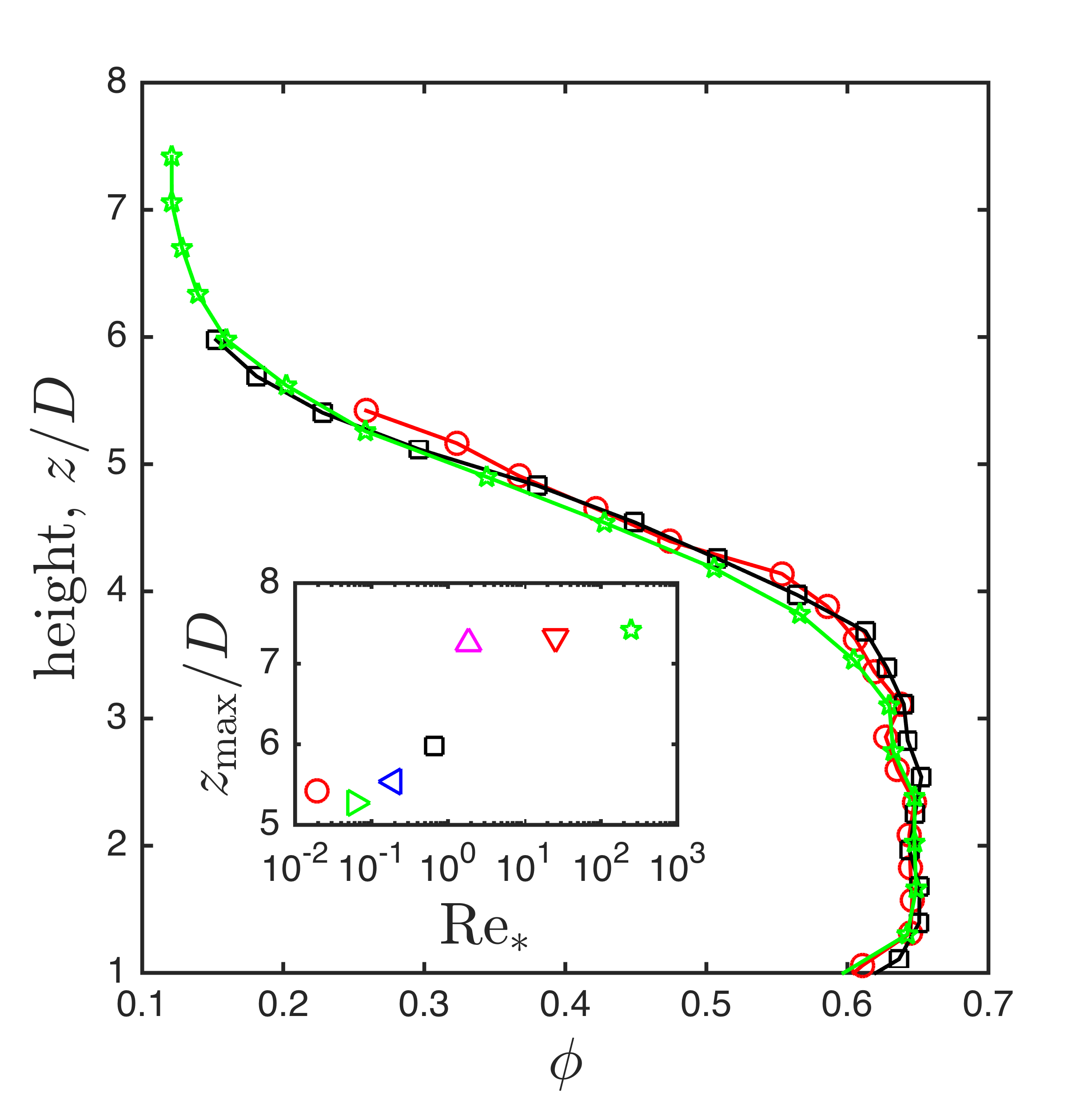}
\includegraphics[trim=0mm 0mm 0mm 5mm, clip, width=0.26\textwidth]{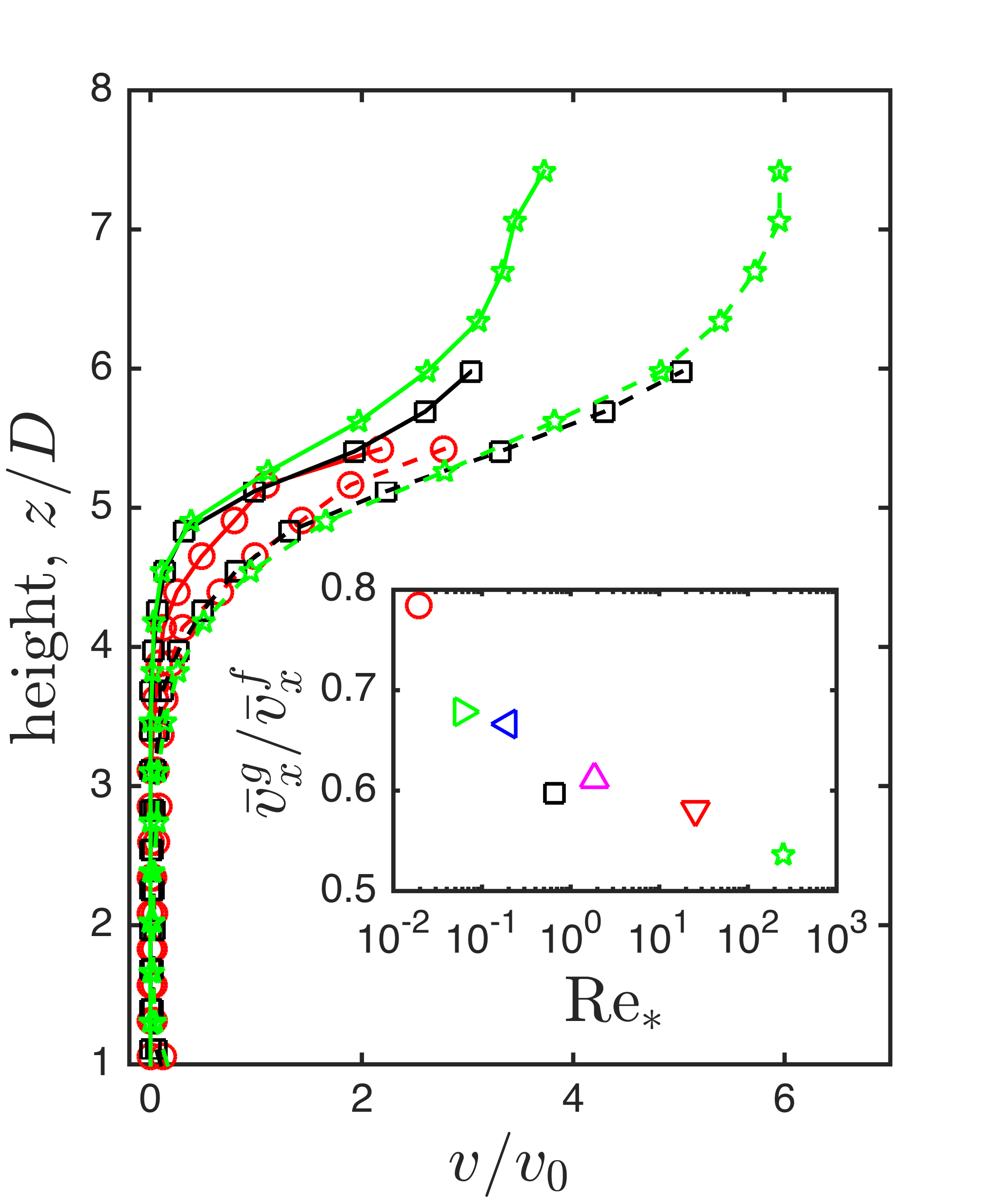}
\includegraphics[trim=0mm 0mm 0mm 5mm, clip, width=0.34\textwidth]{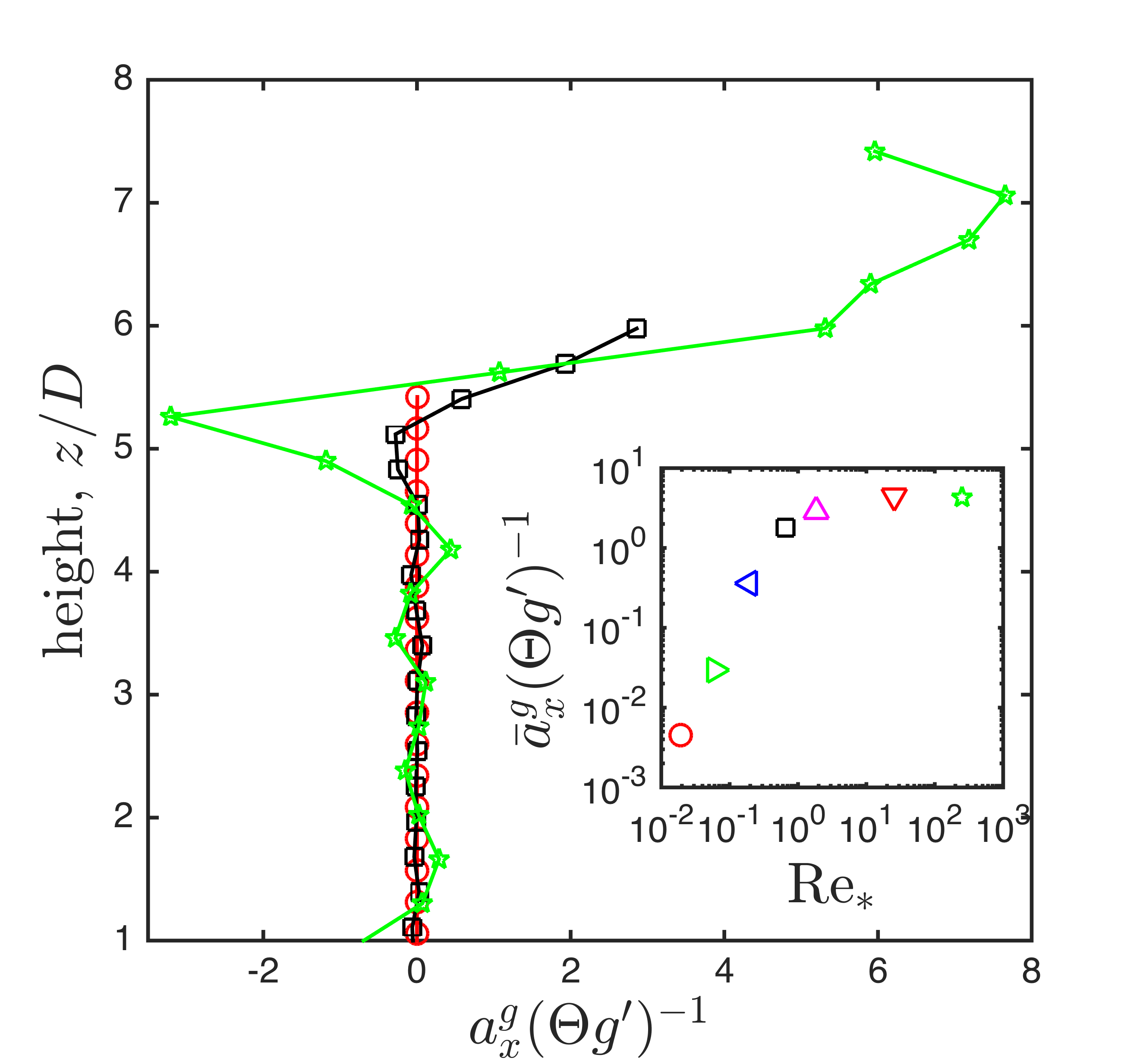}
\caption{Packing fraction, grain and fluid velocity, and grain acceleration profiles from simulations of mobilized beds using a quadratic drag law (S-1 settings) at $\Theta \approx \Theta_c({\rm Re}_*)$ and $10^{-2} \leq {\rm Re}_* \leq 10^{3}$. The profiles are obtained by time-averaging and binning by grain height $z/D$ (vertical axes) for ${\rm Re}_* \approx 0.02$ (circles), 0.7 (squares), and 250 (stars), with more values included in the insets. (a) Packing fraction $\phi$ profile for several ${\rm Re}_*$. The inset shows the maximum grain height $z_{\rm max}$ versus ${\rm Re}_*$. (b) Grain $v^g_x/v_0$ (solid lines) and fluid $v^f_x/v_0$ (dashed lines) velocity profiles. The inset shows the ratio of the average grain velocity $\bar{v}_x^g$ in the mobilized region and above to the average fluid velocity $\bar{v}_x^f$ in the same region. The mobilized region is defined as $z/D > 5.25$ and the results are qualitatively insensitive to the choice of the threshold. (c) Normalized horizontal grain acceleration $a_x^g (\Theta g')^{-1}$ profile. The inset shows the average normalized horizontal grain acceleration in the mobilized region versus ${\rm Re}_*$. } 
\label{fig:z-vs-v-a-phi}
\end{figure*}

Since we do not vary the fluid flow profile, ${\rm Re}_*$-dependent grain dynamics are solely responsible for the variation seen in Fig.~\ref{fig:Shields-curve-with-data}. To illustrate how grain dynamics vary with ${\rm Re}_*$, Fig.~\ref{fig:z-vs-v-a-phi} shows data from simulations of mobilized beds using S-1 settings at $\Theta \approx \Theta_c({\rm Re}_*)$ for $10^{-2} \leq {\rm Re}_* \leq 10^3$. Figure~\ref{fig:z-vs-v-a-phi}(a) shows height $z/D$ versus packing fraction $\phi_i$, and the inset shows the maximum height $z_{\rm max}$ that mobile grains achieve as a function of ${\rm Re}_*$. Figure~\ref{fig:z-vs-v-a-phi}(b) shows height $z/D$ versus grain velocity $v_x^g/v_0$ (solid lines) and fluid velocity $v_x^f/v_0$ (dashed lines). The inset shows the ratio of average grain velocity $\bar{v}_x^g$ in the top layer and above (i.e., $z/D>5.25$) to the average fluid velocity $\bar{v}_x^f$ in the same region. As ${\rm Re}_*$ increases, $\bar{v}_x^g/\bar{v}_x^f$ decreases, meaning that mobile grains do not equilibrate to the fluid flow. Figure~\ref{fig:z-vs-v-a-phi}(c) shows the height $z/D$ versus the normalized horizontal grain acceleration $a_g(\Theta g')^{-1}$. At small ${\rm Re}_*$, grain acceleration is negligible, and mobile grains always move with the local fluid flow. At high ${\rm Re}_*$ grains are significantly accelerated, and their momentum is lost through collisions with the bed, which is indicated by the negative acceleration peak at the bed surface for ${\rm Re}_* \approx 250$ (green stars). The inset shows the average normalized horizontal grain acceleration $\bar{a}_g(\Theta g')^{-1}$ in the top layer and above ($z/D>5.25$) as a function of ${\rm Re}_*$. As ${\rm Re}_*$ increases, average normalized horizontal grain acceleration for mobilized grains increases until it plateaus for ${\rm Re}_*>10$. Together these data show how grain dynamics vary with ${\rm Re}_*$ from viscous-dominated at low ${\rm Re}_*$ to acceleration-dominated at high ${\rm Re}_*$, as discussed in Sec.~\ref{sec:dimensional-analysis}.

Our results for $\Theta_c({\rm Re}_*)$ in Fig.~\ref{fig:Shields-curve-with-data} display plateaus at low and high ${\rm Re}_*$, denoted $\Theta_c^l$ and $\Theta_c^h$, respectively. The behavior of the Shields curve at ${\rm Re}_*< 1$ is currently an open question. Most hydraulic models~\citep{mantz1977,yalin1979,Paphitis2001} assume a decreasing trend of $\Theta_c({\rm Re}_*)$ for ${\rm Re}_*<10$. In simulations, we observe a plateau~\cite{pilotti2001,ouriemi2007} with $\Theta_c^l\approx 0.28$, which numerically agrees with~\citep{pilotti2001} as well as the data shown in Fig.~\ref{fig:Shields-curve}. We interpret this plateau as the shear force corresponding to the most geometrically stable arrangement of the bed, suggesting that configurations that can resist $\Theta>\Theta_c^l$ do not exist. As ${\rm Re}_*$ is increased, $\Theta_c$ decreases, as grain dynamics begin to transition out of the viscous regime (${\rm Re}_* \ll 1$). For ${\rm Re}_*\gg 1$, $\Theta_c$ also displays a plateau $\Theta_c^h$ that has the same value for linear and quadratic drag laws. Note that, for a linear drag law, a larger fluid velocity is required to achieve the same $\Theta$ when compared to a quadratic drag law. 

\subsection{Comparing linear and quadratic drag laws}
\label{sec:linear-vs-quad}

The quadratic drag law in Eq.~\eqref{eqn:grain-dynamics} natually includes both $\tau_\nu$ and $\tau_I$ as defined in Sec.~\ref{sec:dimensional-analysis}. However, the linear drag law (S-2 and S-3 settings) only includes $\tau_\nu$, not $\tau_I$. In Sec.~\ref{sec:dimensional-analysis}, we showed that ${\rm Re}_*\propto \tau_\nu/\tau_\Theta$, and we argued that the inertial time scale $\tau_I$ plays a secondary role since it is always longer than $\tau_\Theta$ given typical grain and fluid densities. To justify this claim and to connect to our previous work using a linear drag law~\cite{clark2015hydro}, we show results in this section from 3D simulations with linear (S-3 settings) and quadratic (S-1 settings) drag laws. We show below that $\Theta_c({\rm Re}_*)$ and the sediment transport rates are the same for linear and quadratic drag laws, provided $e_n$ is small when a linear drag law is used. If $e_n$ is not small, then mobilized grains at ${\rm Re}_* \gg 1$ are accelerated over time $\tau_\nu$ instead of being cut off by $\tau_I$, leading to hysteresis at ${\rm Re}_* \gg 1$. We again note that $\Theta$ is set by the typical force felt by surface grains, not by the velocity of the fluid. This means that when comparing linear and quadratic drag at high ${\rm Re}_*$, we use larger fluid velocities for the linear case to obtain the same value of $\Theta$, since we are neglecting the quadratic term in the drag laws shown in Eqs.~\eqref{eqn:grain-dynamics} and \eqref{eqn:fluid-drag}.

\begin{figure}
\centering  \includegraphics[width=0.5\columnwidth]{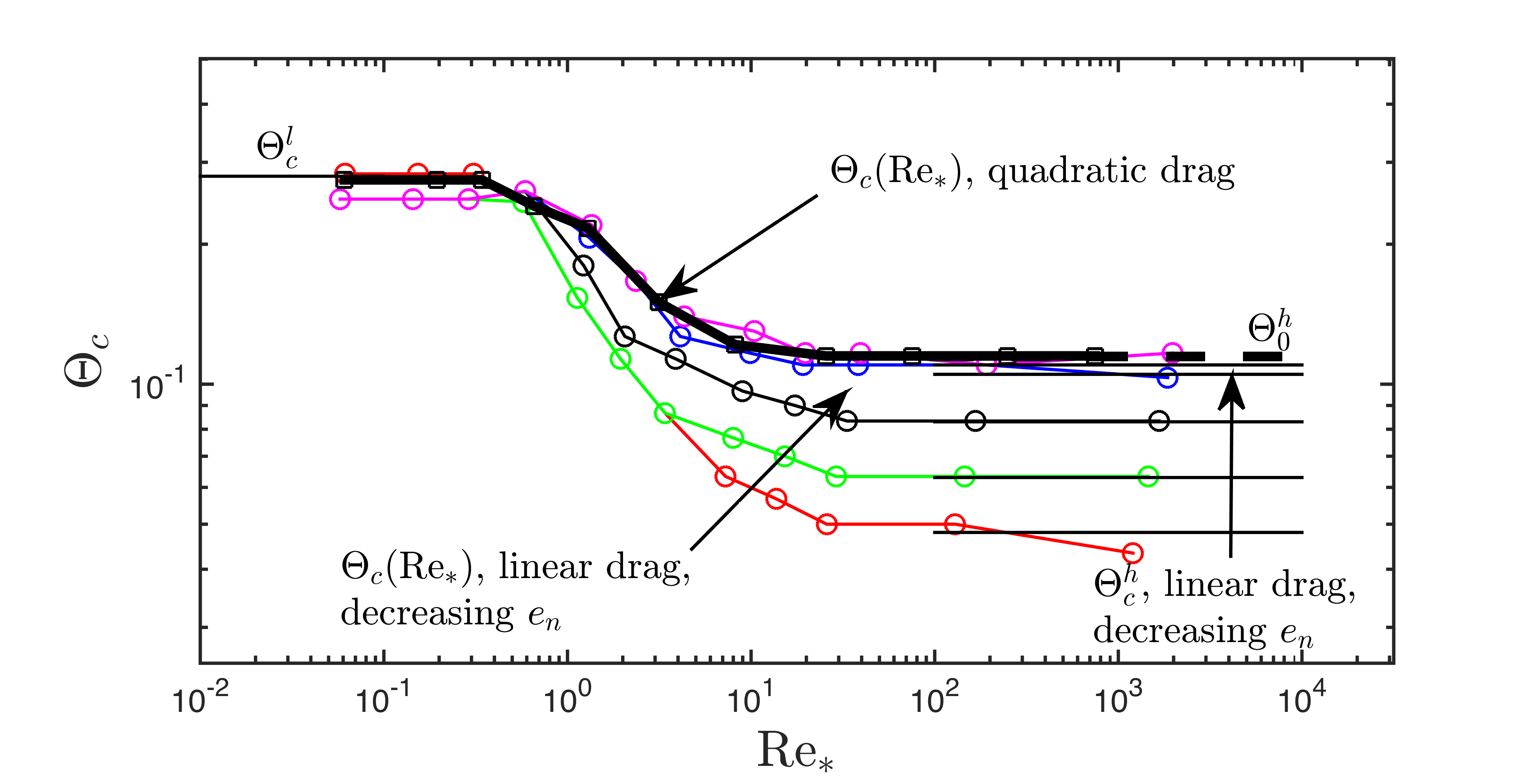}
\caption{The solid black curve with square markers is $\Theta_c({\rm Re}_*)$ with the quadratic drag law (S-1 settings). The thin curves with open circles represent $\Theta_c({\rm Re}_*)$ for linear drag (S-3 settings) with $e_n = 0.9$ (red), 0.8 (green), 0.5 (black), 0.2 (blue), and 0.1 (magenta). Plateau values $\Theta_c^l$ and $\Theta_c^h$ for the curves from a linear drag law are marked with thin black lines. The thicker, dashed line shows $\Theta_0^h$, which is the minimum $\Theta$ at ${\rm Re}_* \gg 1$ required to initiate sustained grain motion for linear and quadratic drag for all values of $e_n$ (S-1, S-2, and S-3 settings). As we discuss in Section~\ref{sec:weibull}, grain motion at ${\rm Re}_*\gg 1$ is always initiated at this boundary for large systems.}
\label{fig:Shields-curve-with-data-all}
\end{figure}

Figure~\ref{fig:Shields-curve-with-data-all} shows the boundaries between systems with and without sustained grain motion for quadratic and linear drag laws. The thick black line with square markers shows the boundary between systems with and without sustained grain motion using a quadratic drag law (S-1 settings), representing both the transitions from a mobile to a static bed as well as from a static to a mobile bed (i.e., we observe no hysteresis for a quadratic drag law). This boundary is independent of $e_n$. The thin, colored lines with circle markers show the minimum $\Theta$ required to sustain grain motion indefinitely, using a linear drag law (S-3 settings). The different colors (red, green, black, blue, magenta) represent different restitution coefficients ($e_n=0.9$, 0.8, 0.5, 0.2, and 0.1). As $e_n\rightarrow 0$, these boundaries form a single curve, with plateaus $\Theta=\Theta_c^l\approx 0.28$ and $\Theta=\Theta_c^h\approx 0.11$ at low and high ${\rm Re}_*$, respectively. Note that \citet{nino1998} showed experimentally that restitution coefficients are typically small ($e_n<0.5$) for saltating grains rebounding off a sediment bed. The dashed, black line represents $\Theta_0^h$, which is the minimum $\Theta$ required to \textit{initiate} sustained grain motion from a static bed at ${\rm Re}_*\gg 1$. As we show in Section~\ref{sec:weibull}, sustained grain motion is always initiated at this value in the large-system limit. We find that $\Theta_0^h$ is insensitive to the drag law and restitution coefficient (i.e., $\Theta_0^h$ is constant for S-1, S-2, and S-3).

\begin{figure}

\raggedright \hspace{5 mm} (a) \hspace{52 mm} (b) \hspace{52 mm} (c) 
\\ 
\centering \includegraphics[width=0.32\columnwidth]{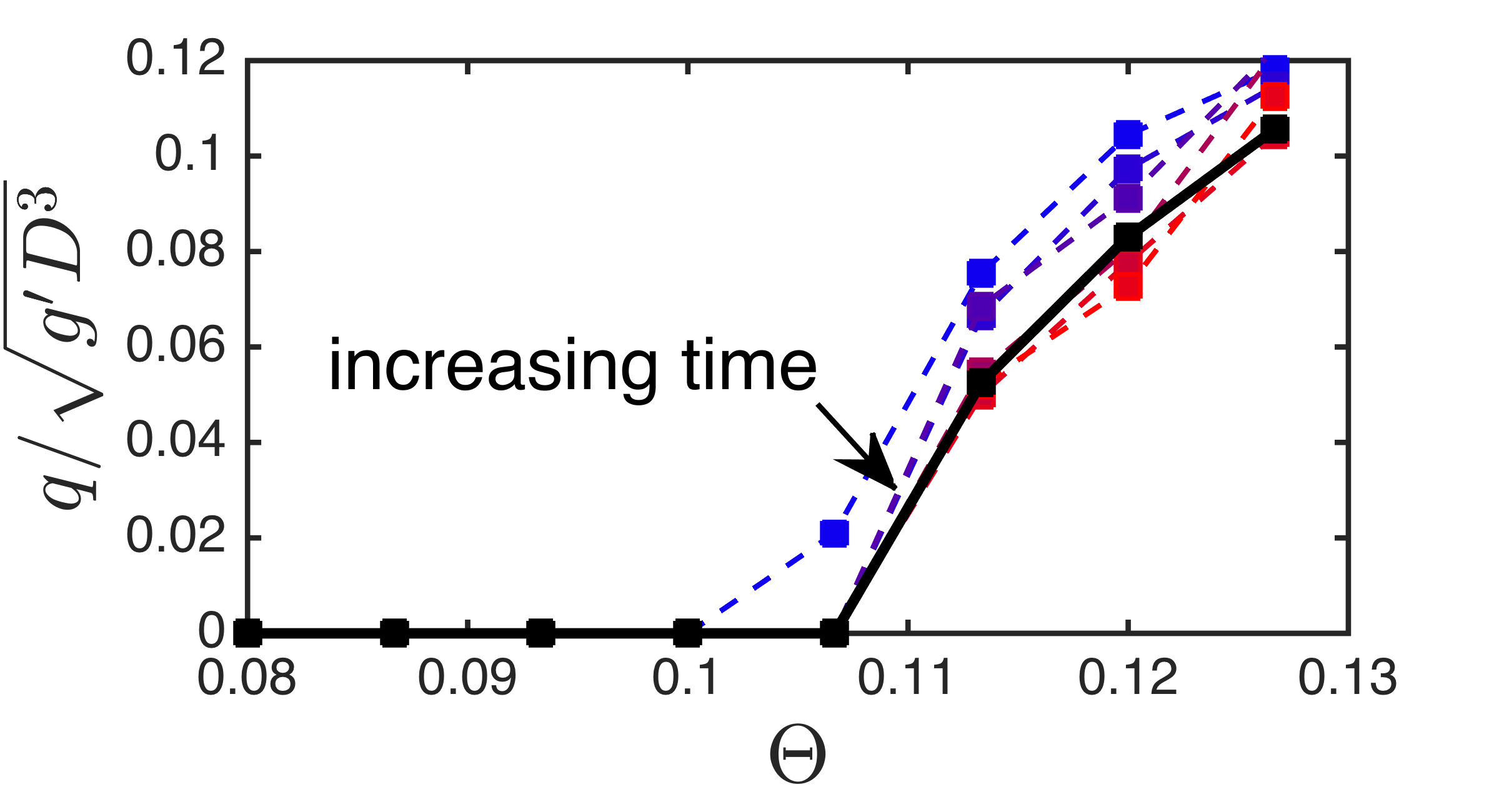}
 \includegraphics[width=0.32\columnwidth]{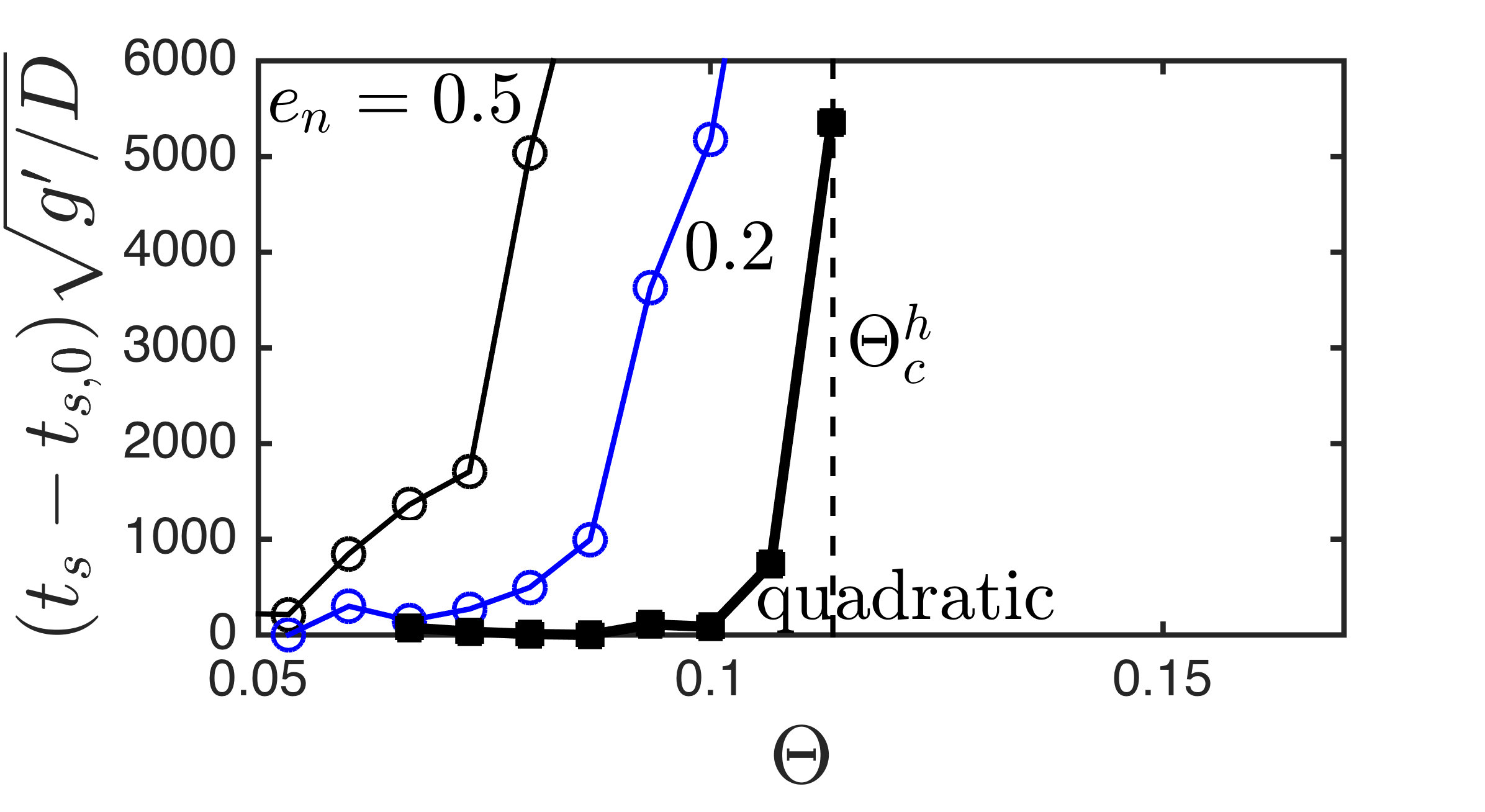} 
 \includegraphics[width=0.32\columnwidth]{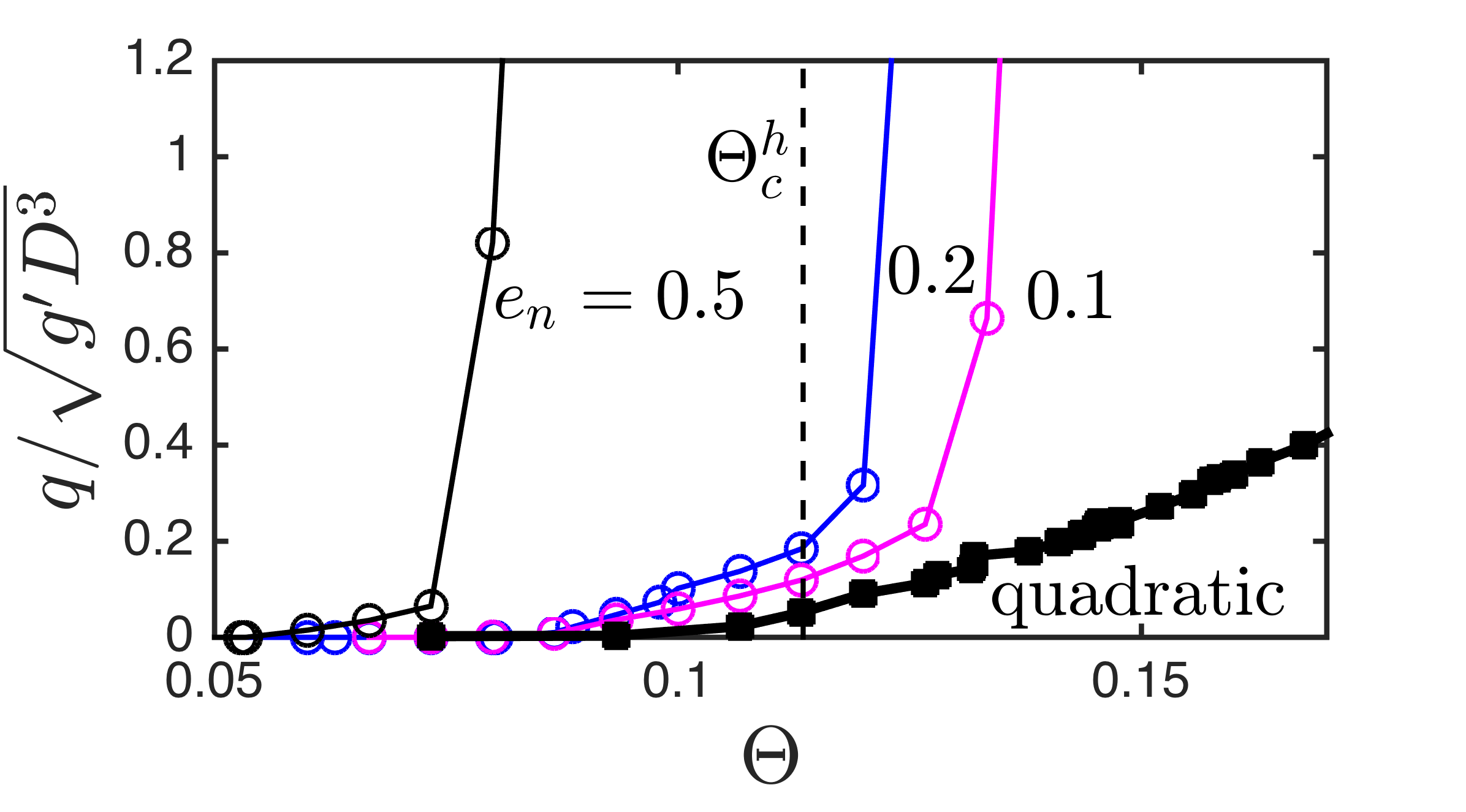}

\caption{For simulations with ${\rm Re}_*\gg 1$ using S-1 and S-3 settings, the dimensionless volumetric grain flux per cross stream width $q/\sqrt{g'D^3}$ (a) and the dimensionless stopping time $(t_s-t_0)\sqrt{g'/D}$ (b) are plotted versus $\Theta$. Data in (a) represents the ensemble average over 10 systems with quadratic drag, where the data are averaged over all systems that were still in motion. Intermediate times are given by blue (short times) to red (long times) dashed lines with square markers, and the black line represents the end of the simulation. If more than half of the simulations had stopped, then $q=0$. Data in (b) represent the mean stopping time for ensembles where grain motion ceased in at least half of the simulations. Curves show quadratic (black squares, S-1 settings) and linear (S-3 settings) drag laws with $e_n=0.5$ (black open circles) and 0.2 (blue open circles). Additionally, in (c), we compare the grain flux versus $\Theta$ for the cases of linear and quadratic drag, where we also include the flux for $e_n=0.1$ (magenta open circles). Despite the fact that the drag laws have different forms, the curve $q(\Theta)$ approaches the quadratic case near $\Theta_c$ as $e_n\rightarrow 0$.}
\label{fig:onset-highRe}
\end{figure}

When ${\rm Re}_*\ll 1$, the linear (S-2 and S-3 settings) and quadratic (S-1 settings) drag laws agree, and our simulations show identical results. Grain flux tends to zero and stopping times diverge at a critical value $\Theta_c^l \approx 0.28$. Figure~\ref{fig:onset-highRe} shows representative data for ${\rm Re}_*\gg 1$. As discussed in Section~\ref{sec:fluid-drag}, we begin with a mobilized system, where grains are suspended, and apply the model fluid flow. $\Theta_c$ is characterized by a grain discharge per unit width that tends to zero and diverging stopping times for $\Theta$ just below $\Theta_c$. We characterize the grain motion by plotting $q/\sqrt{g'D^3}$, where $q$ is the discharge per unit width. The dashed curves in Fig.~\ref{fig:onset-highRe}(a) show $q/\sqrt{g'D^3}$ at different times (blue to red represents increasing time), where each data point is obtained from an ensemble of ten simulations using S-1 settings. Each data point with a nonzero $q$ means that in at least half of the simulations the grains did not stop, and the value of $q$ represents the average of all simulations where grains were still in motion. The solid black curve represents the steady state grain flux (measured at the end of the simulation). If in at least half of the simulations, the grains stopped, we measure the average time $t_s$ that it took the grains to stop moving. Figure~\ref{fig:onset-highRe}(b) shows $(t_s-t_{s,0})\sqrt{g'/D}$ plotted versus $\Theta$ for quadratic drag (S-1 settings) and linear drag (S-3 settings). Figure~\ref{fig:onset-highRe}(c) shows $q$ versus $\Theta$ for quadratic and linear drag. We find that $q$ tends to zero and $t_s$ diverges at roughly the same value of $\Theta$. As $e_n\rightarrow 0$ for linear drag, the critical $\Theta$ approaches the value for the quadratic drag law, $\Theta=\Theta_c^h\approx 0.11$.

Thus, at ${\rm Re}_*\gg 1$, we find that using linear (with small $e_n$) and quadratic drag laws give the same value of $\Theta_c$ as well as the same dependence of $q$ versus $\Theta$. Under a linear (viscous only) drag law, grains can be accelerated to much larger speeds, since $\tau_\nu \propto {\rm Re}_*$ for all ${\rm Re}_*$, and the equilibration time is not cut off by $\tau_I$. Thus, in simulations with a linear drag law, mobile grains can deliver significant energy when they impact the bed and can rebound to large heights above the bed from elastic collisions (not lift forces) if $e_n$ is large. These effects are suppressed when $e_n$ is small and $\Theta$ is near $\Theta_c$, and the behavior of $q(\Theta)$ for linear drag approaches the $q(\Theta)$ curve for quadratic drag as $e_n\rightarrow 0$. In this case, the trajectory of mobilized grains is confined to positions near the bed, and viscous and inertial drag laws yield the same behavior very close to $\Theta_c$.

\subsection{Including model lift forces}
\label{sec:lift}
The results of our model do not quantitatively capture the global minimum in the Shields curve at ${\rm Re}_*\approx 10$ and overestimate $\Theta_c$ by roughly a factor two at high ${\rm Re}_*$. However, in this model we neglected lift forces and turbulence, and thus we expect to overpredict $\Theta_c$ in this regime. For instance, the calculation from~\citet{wiberg1987} discussed in Sec.~\ref{sec:prior-descriptions} and shown in Fig.~\ref{fig:Shields-curve} could be combined with our model to capture ${\rm Re}_*$-dependent lift forces. To demonstrate the viability of this approach, in this section we show that lift forces decrease $\Theta_c$ in a way that is quantitatively consistent with Eq.~\eqref{eqn:Wiberg-scaling}.

\begin{figure}
\centering \includegraphics[width=0.5\columnwidth]{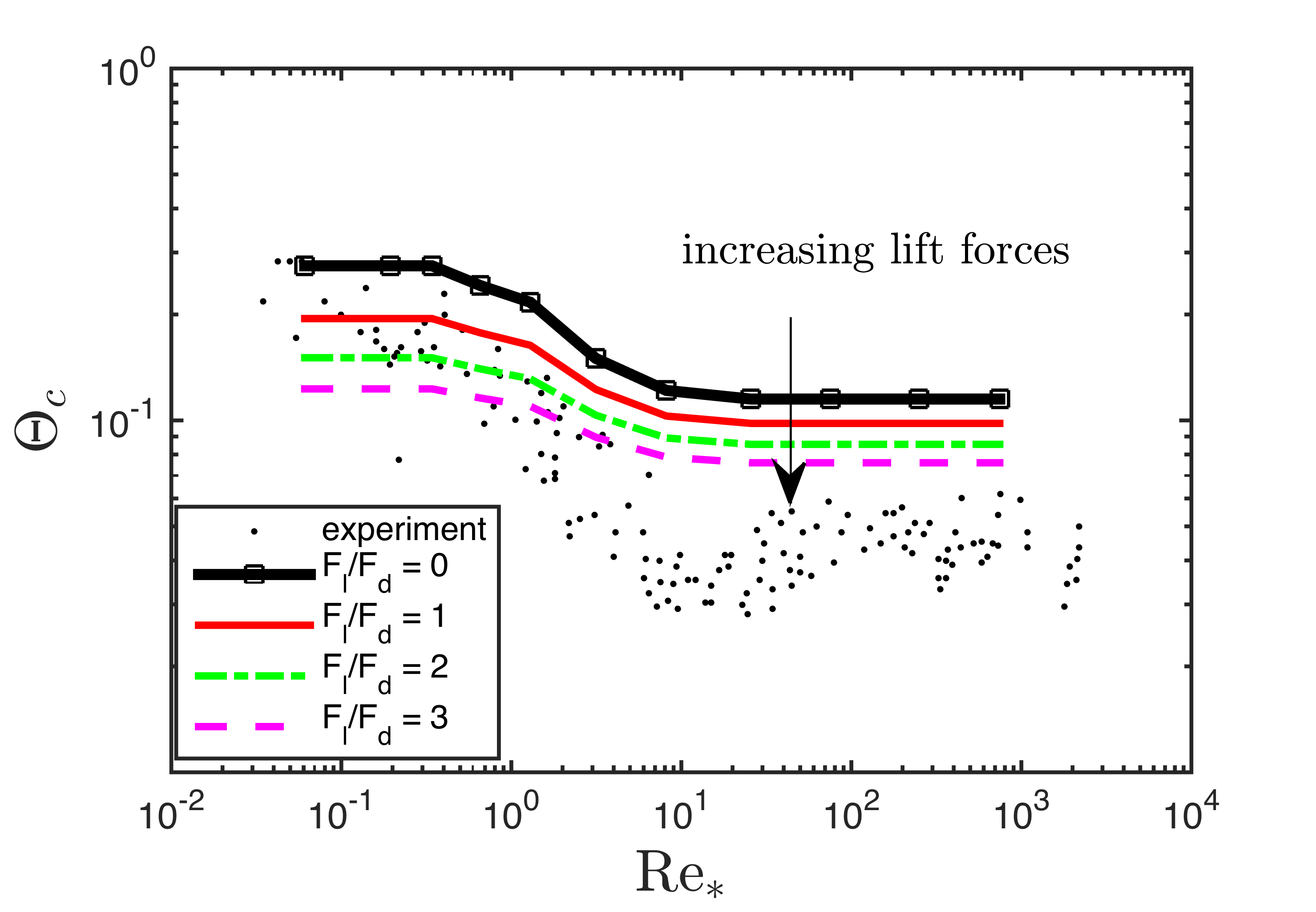}
\caption{We plot $\Theta_{c}\left({\rm Re}_*,\frac{F_l}{F_d}\right)$ versus ${\rm Re}_*$ with $\frac{F_l}{F_d}=0$, 1, 2, and 3. As $\frac{F_l}{F_d}$ is increased, $\Theta_{c}\left({\rm Re}_*,\frac{F_l}{F_d}\right)$ decreases according to Eq.~\eqref{eqn:lift-sim}. Quantitatively capturing the Shields curve would require an ${\rm Re}_*$-dependent lift force, as shown in Fig.~\ref{fig:Shields-curve} from~\cite{wiberg1987}.}
\label{fig:Shields-curve-with-lift}
\end{figure}

Figure~\ref{fig:Shields-curve-with-lift} shows the results of 3D simulations with quadratic drag and constant lift forces (S-4 settings) which are not ${\rm Re}_*$-dependent. We vary $\frac{F_l}{F_d}$ from 0 to 3, and the corresponding curves $\Theta_{c}\left({\rm Re}_*,\frac{F_l}{F_d}\right)$ decrease, as expected. The magnitude of these curves follows the scaling shown in Eq.~\eqref{eqn:Wiberg-scaling}, but where $\cot \psi$ is replaced by $\frac{3}{2}\Theta_{c}\left({\rm Re}_*,0\right)$, which is the ratio of horizontal to vertical forces required in simulations at varying ${\rm Re}_*$:
\begin{equation}
\Theta_{c}\left({\rm Re}_*,\frac{F_l}{F_d}\right) = \Theta_{c}\left({\rm Re}_*,0\right)\frac{1}{1+ \frac{3}{2}\Theta_{c}\left({\rm Re}_*,0\right)\frac{F_l}{F_d}}.
\label{eqn:lift-sim}
\end{equation}
We reiterate that the model lift forces we include have constant $\frac{F_l}{F_d}$, whereas physical systems likely have lift forces that vary strongly with ${\rm Re}_*$. It is possible that a scaling function $\frac{1}{1+ \frac{3}{2}\Theta_{c}\left({\rm Re}_*,0\right)\frac{F_l}{F_d}}$ that is directly calculated from the ${\rm Re}_*$-dependent fluid forces could by combined with our results to quantitatively recapitulate the Shields curve over the full range of ${\rm Re}_*$.

\subsection{Friction, shape, and dimension}
\label{sec:3D-to-2D}

\begin{figure}
\centering \includegraphics[width=0.5\columnwidth]{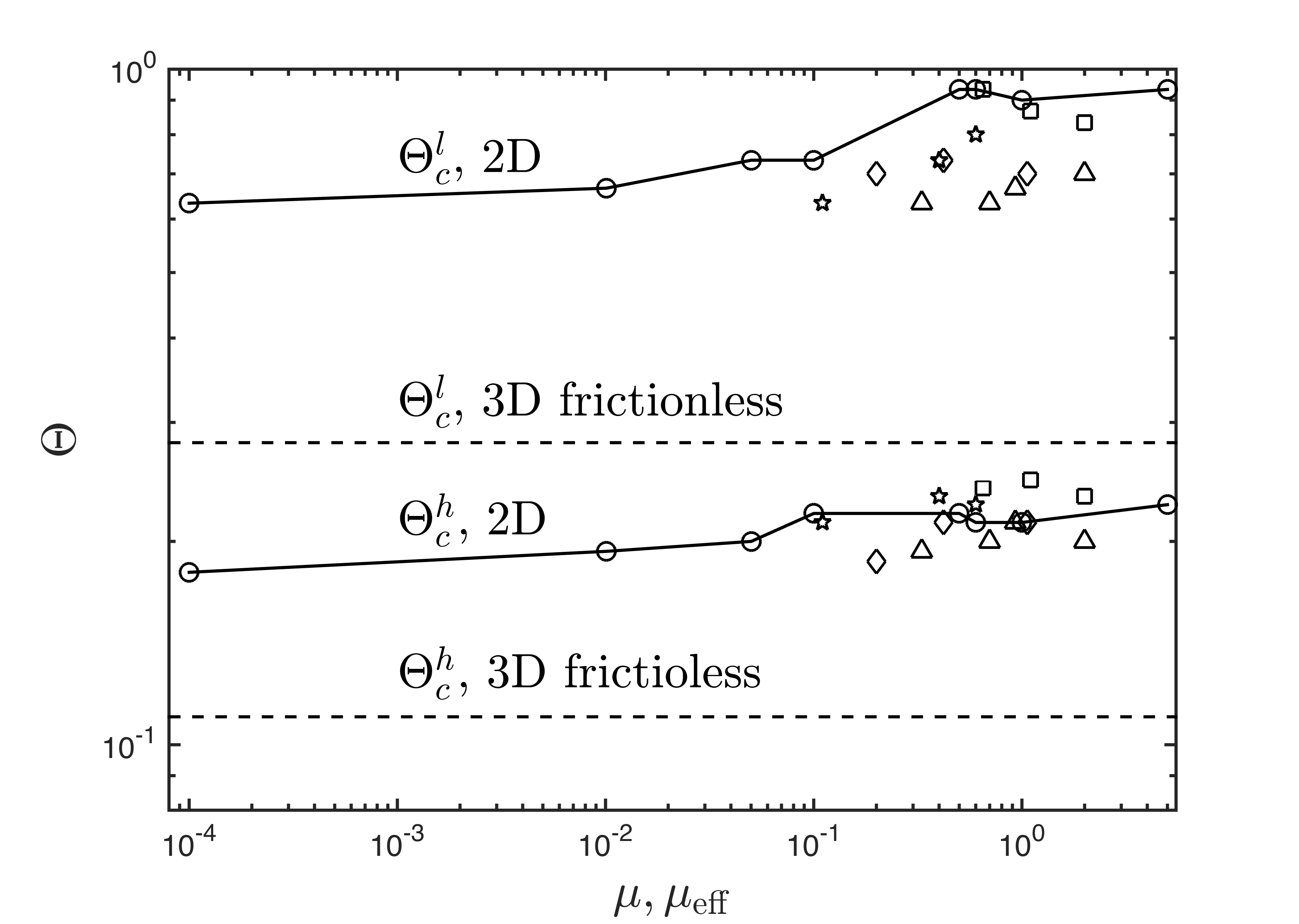}
\caption{The plateau values $\Theta_c^l$ and $\Theta_c^h=\Theta_0^h$ (for quadratic drag or small $e_n$) are plotted as a function of friction coefficient $\mu$ or $\mu_{\rm eff}$. Solid lines with open circles show data for disks with Cundall-Strack friction (S-5 settings). Other open symbols correspond to irregularly shaped grains (S-6 settings) with $n=2$ (diamonds), 3 (triangles), 4 (squares), and 5 (stars).}
\label{fig:Th-plateaus}
\end{figure}

In this section, we use 2D simulations (S-5 and S-6 settings) to show that $\Theta_c({\rm Re}_*)$ is only weakly dependent on friction and irregular grain shape. These results are shown in Fig.~\ref{fig:Th-plateaus}. In 2D, the numerical values for $\Theta_c$ increase by roughly a factor of two. We interpret the increase in $\Theta_c$ from 3D to 2D to follow from the fact that the energy landscape of a 2D bed is more difficult for grains to navigate, as grains must roll over obstacles instead of traversing the low points between impeding grains. This raises important questions about whether and how quasi-2D calculations of pocket angles relate to 3D systems. 

We perform simulations in 2D with a linear drag law where grain-grain interactions include Cundall-Strack friction (S-5 settings) or geometrical friction from irregular grain shape (S-6 settings). We find hysteresis that vanishes in the limit of small $e_n$, as in 3D (Fig.~\ref{fig:Shields-curve-with-data-all}), and we plot the small $e_n$ values in Fig.~\ref{fig:Th-plateaus}, which correspond to the result with a quadratic drag law. Figure~\ref{fig:Th-plateaus} shows the values of the plateaus $\Theta_c^l$ and $\Theta_c^h$ as friction is varied. Circles represent disks with Cundall-Strack friction, with a friction coefficient $\mu$. Other symbols correspond to $n-$mers with $n=2$ (diamonds), 3 (triangles), 4 (squares), and 5 (stars). Both $\Theta_c^l$ and $\Theta_c^h$ increase by less than 50\% with increasing friction, which is still comparable to the scatter in the experimental and field data shown in Fig.~\ref{fig:Shields-curve}. We note that~\citet{joseph2004} found a friction coefficient of $\mu \approx 0.15$ for submerged spheres, and we observe almost no variation in $\Theta_c^l$ and $\Theta_c^h$ for friction coefficients at or below this measured experimental value.

\subsection{Onset of grain motion}
\label{sec:weibull}

Thus far, we have argued that grain dynamics at varying ${\rm Re}_*$ play a dominant role in determining $\Theta_c({\rm Re}_*)$, as opposed to static force and torque balance of individual grains in typical pocket geometries. However, the onset of grain motion must follow from a breakdown of force and torque balance on one or more grains. In this section, we show data (using S-7 and S-8 settings) for the initiation of sustained grain motion (i.e., the static-to-mobile transition), which, in the limit of large systems, always occurs at the same dynamical boundary $\Theta_c({\rm Re}_*)$ that denotes the minimum applied fluid stress at which mobilized systems are unable to stop. We show that local motion in one of many uncorrelated subsystems leads to global sustained grain motion using Weibullian weakest-link statistics~\citep{weibull1939,weibull1951}. Additionally, we find that the characteristic time $t_m$ for the bed to fully mobilize diverges at $\Theta_c$, consistent with a dynamical instability at $\Theta=\Theta_c$ that is activated by a single mobilized region.

If we consider the bed to be a composite system of $M$ \textit{uncorrelated} subsystems that begin to move if any of the subsystems move at $\Theta>\Theta_c$ (when grain motion will be sustained indefinitely), then the cumulative distribution $C_M(\Theta)$ for the initiation of grain motion in the collective system is related to that of a single subsystem $C(\Theta)$ by
\begin{equation}
1-C_M(\Theta)=\left[1-C(\Theta)\right]^M.
\label{eqn:Weibull}
\end{equation} 
By assuming a Weibull distribution for $C(\Theta)$~\citep{weibull1939,weibull1951,franklin2014}
\begin{equation}
C(\Theta)=1-\exp\left[\left(\frac{\Theta-\Theta_c}{\beta}\right)^{\alpha}\right],
\label{eqn:Weibull-CDF}
\end{equation}
then $C_M(\Theta)$ in Eq.~\eqref{eqn:Weibull} has the same form with $\alpha_M=\alpha$ and $\beta_M=\beta M^{-1/\alpha}$. As in our previous study~\citep{clark2015hydro}, we find that this scaling holds for all systems we have considered here. This means that Eq.~\eqref{eqn:Weibull} applies, confirming that global grain motion is initiated by a single member of a collection of uncorrelated subsystems (i.e., local pockets). In the limit of large system size, we find that grain motion is always initiated at $\Theta_c$. 

\begin{figure}
\raggedright \hspace{5mm} (a) \hspace{60mm} (b) \hspace{50mm} (c) 
\\ 
\centering \includegraphics[width=0.38\textwidth]{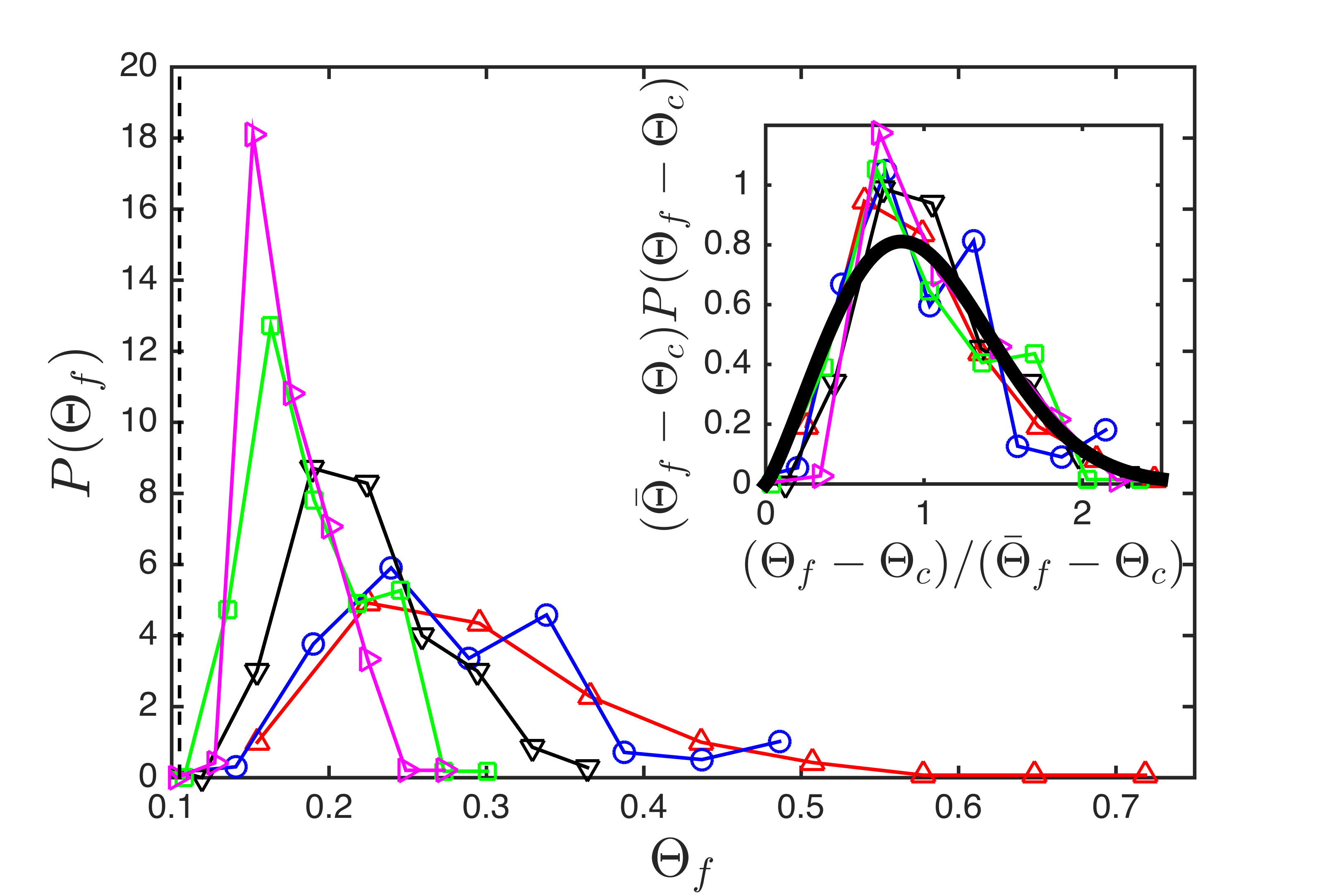} 
\includegraphics[width=0.27\textwidth]{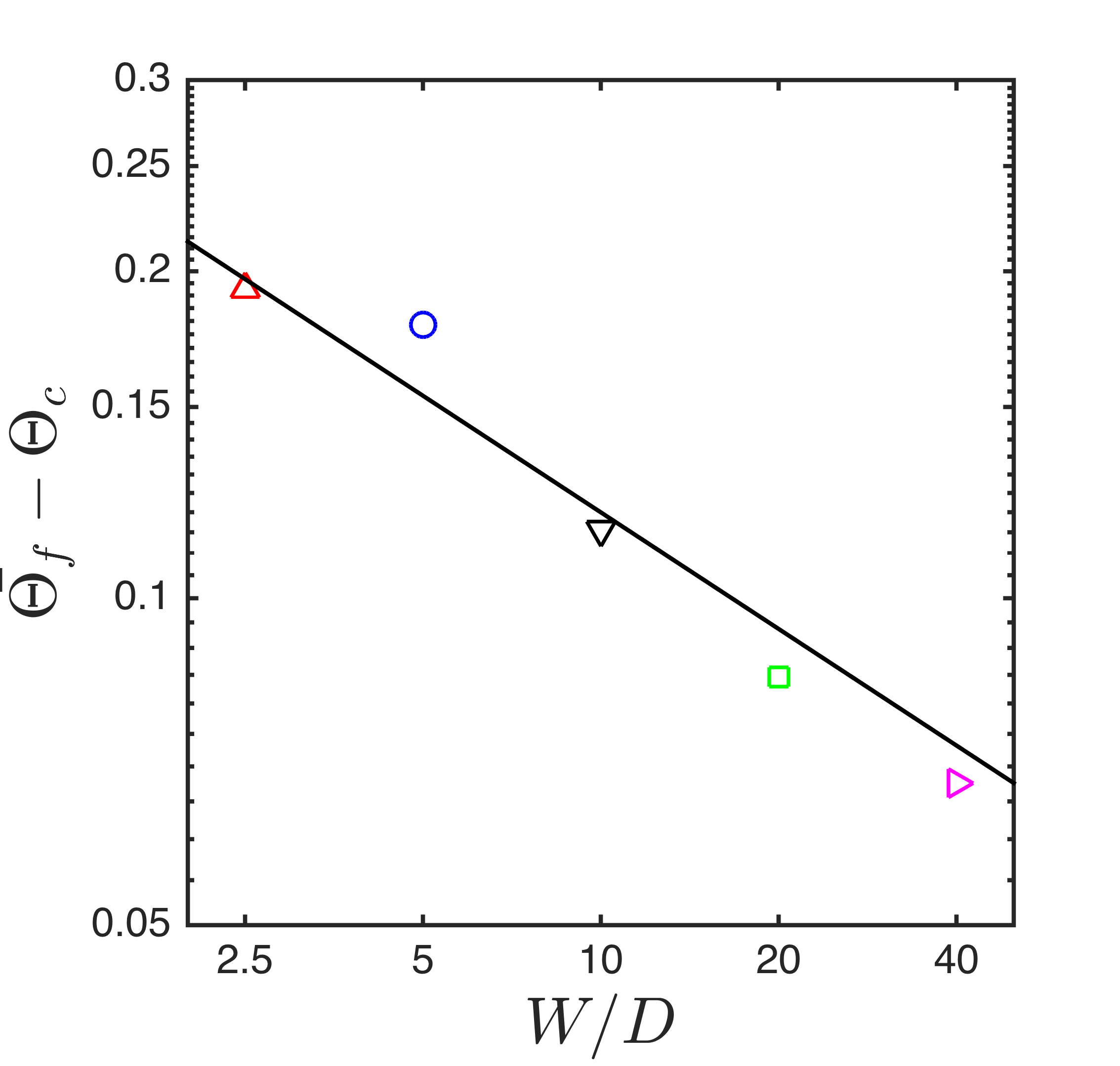} 
\includegraphics[width=0.31\textwidth]{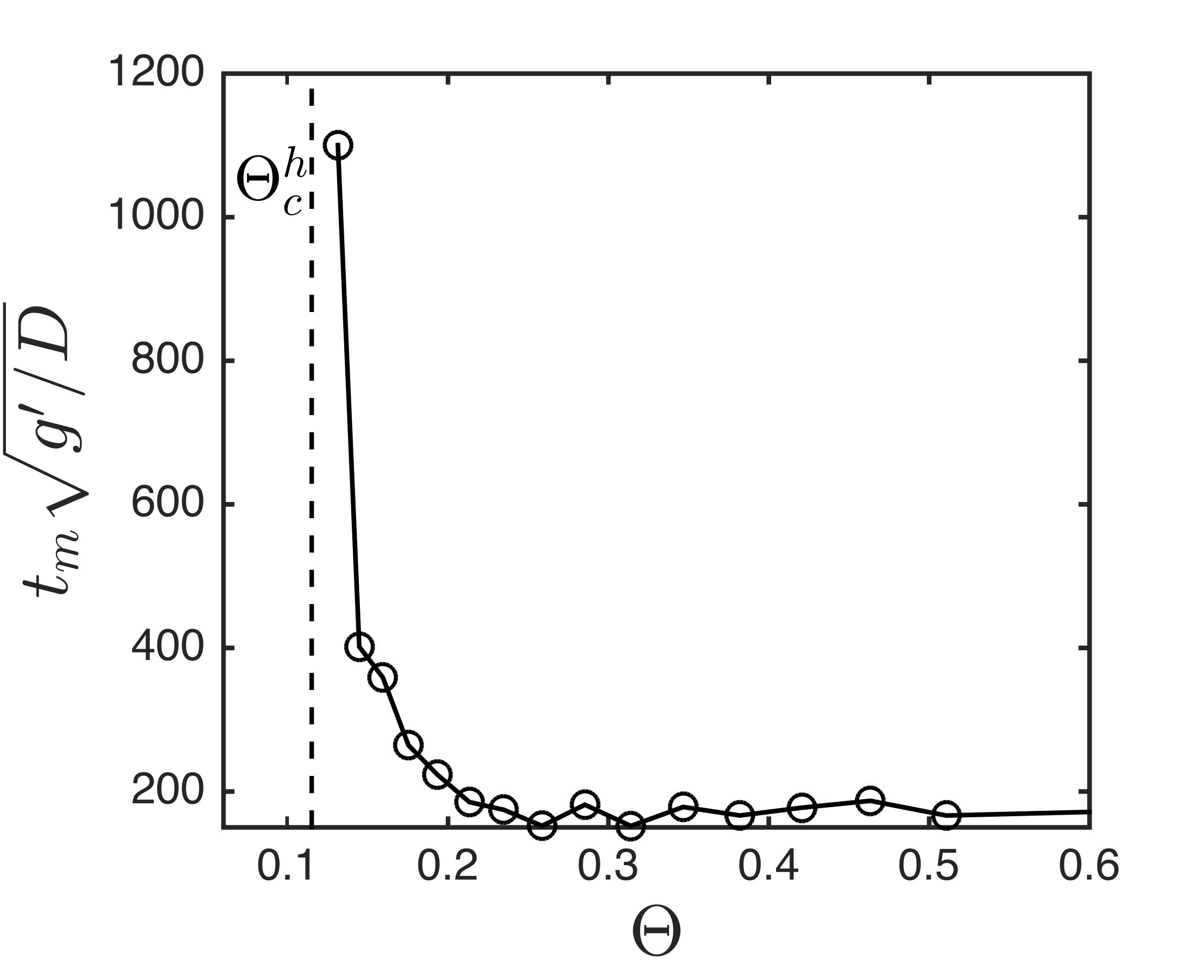}
\caption{(a) The probability distributions $P(\Theta_f)$ of the Shields number $\Theta_f$ required to initiate sustained grain motion in an initially static 3D system as $\Theta$ is slowly increased at high ${\rm Re}_*$, computed with a quadratic drag law (S-7 settings). The vertical dashed line represents $\Theta = \Theta_c^h$. These ensembles, consisting of 200 simulations at each system size, were prepared with $\Theta=0.033<\Theta_c$, a fill height $5D$, a cross-stream width of $4D$, and a stream-wise distance $W/D$, which we vary between 2.5 and 40. The inset shows that these distributions collapse when rescaled by $\bar{\Theta}_f-\Theta_c^h$, and the thick black line shows a Weibull distribution with shape parameter $\alpha = 2.2$. (b) The mean excess stress $\bar{\Theta}_f-\Theta_c^h$ for the initiation of sustained grain motion decreases as a power law with exponent $-1/\alpha$, in accordance with Eqs.~\eqref{eqn:Weibull} and \eqref{eqn:Weibull-CDF}. (c) The mean normalized mobilization time $t_m\sqrt{g'/D}$, binned and averaged over $\Theta$. $t_m$ is defined as the time required for the grain flux $q$ to rise from zero to the steady-state value.}
\label{fig:size-scaling-3D}
\end{figure}

For a given system with static grains, we slowly increase $\Theta$ until sustained grain motion occurs at $\Theta = \Theta_f > \Theta_c$. Figure~\ref{fig:size-scaling-3D}(a) shows the distributions of the excess stress $\Theta_f-\Theta_c$ required to initiate sustained grain motion in ensembles of static 3D quadratic-drag systems (S-7 settings) as $\Theta$ is slowly increased. These ensembles are prepared at $\Theta\approx 0.067 < \Theta_c$ with a fill height of $5D$ and a cross stream width of $4D$, and we vary the stream-wise distance $W/D$ between 2.5 and 40. These distributions collapse when rescaled by their mean, as shown in the inset, with a shape parameter $\alpha\approx 2.2$. This value is comparable to that measured in the frictionless, elastic case, where we found $\alpha\approx 2.6$~\citep{clark2015hydro}. The scaling of mean excess stress $\bar{\Theta}_f-\Theta_c$, as shown in Figure~\ref{fig:size-scaling-3D}(b), is consistent with a power law scaling with exponent $-1/\alpha$, confirming Eqs.~\eqref{eqn:Weibull} and \eqref{eqn:Weibull-CDF}. For systems that fail, we measure the mobilization time $t_m$, which we define as the time required for the grain flux $q$ to go from zero to the steady-state value. This quantity is plotted versus $\Theta$ in Fig.~\ref{fig:size-scaling-3D}(c), and it diverges at $\Theta_c$, like the divergence of $t_s$ shown in Fig.~\ref{fig:onset-highRe}.

\begin{figure}
\raggedright \hspace{10 mm} (a) \hspace{78 mm} (b) \\
\centering
\includegraphics[width=0.48\textwidth]{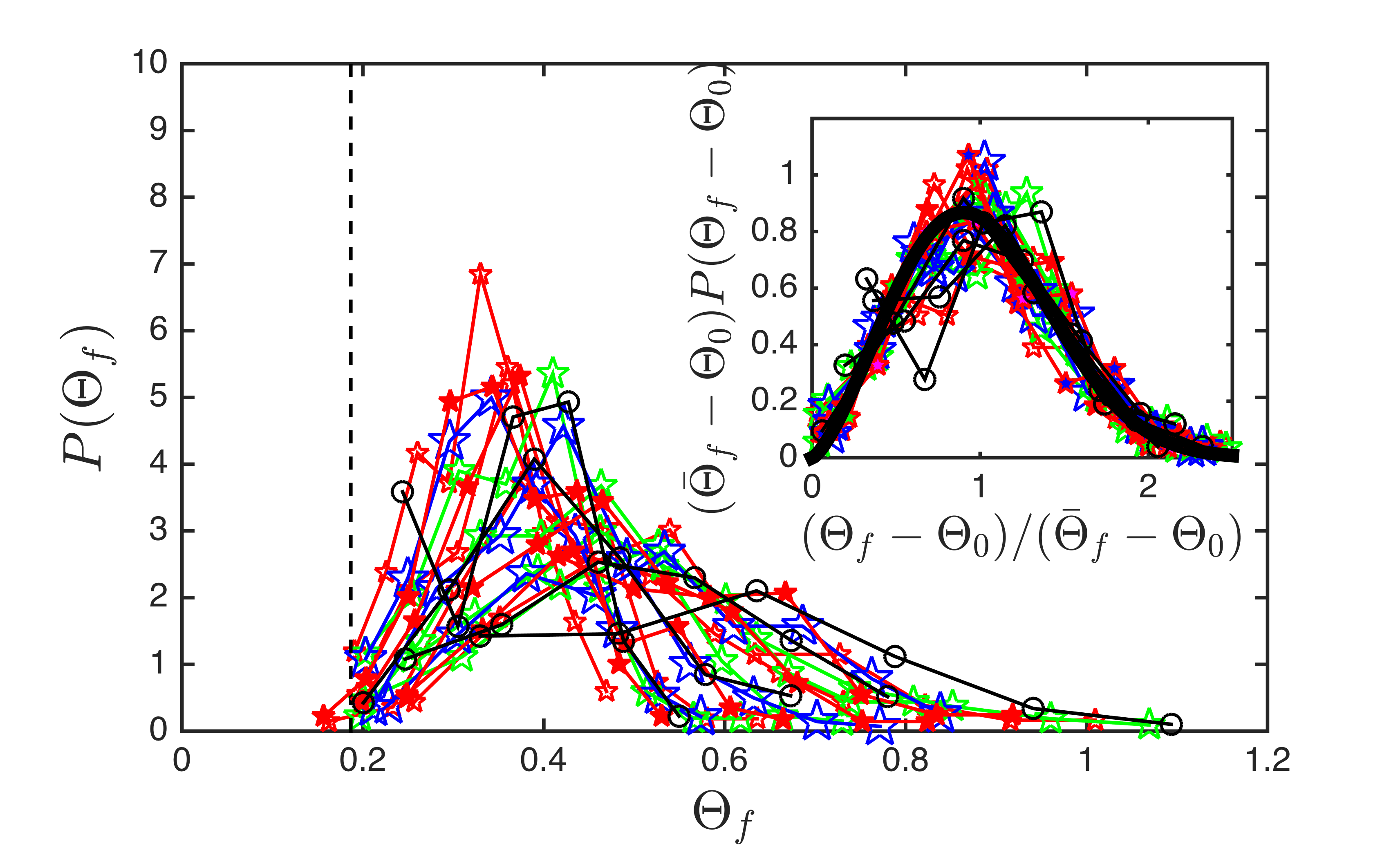}
\includegraphics[width=0.48\textwidth]{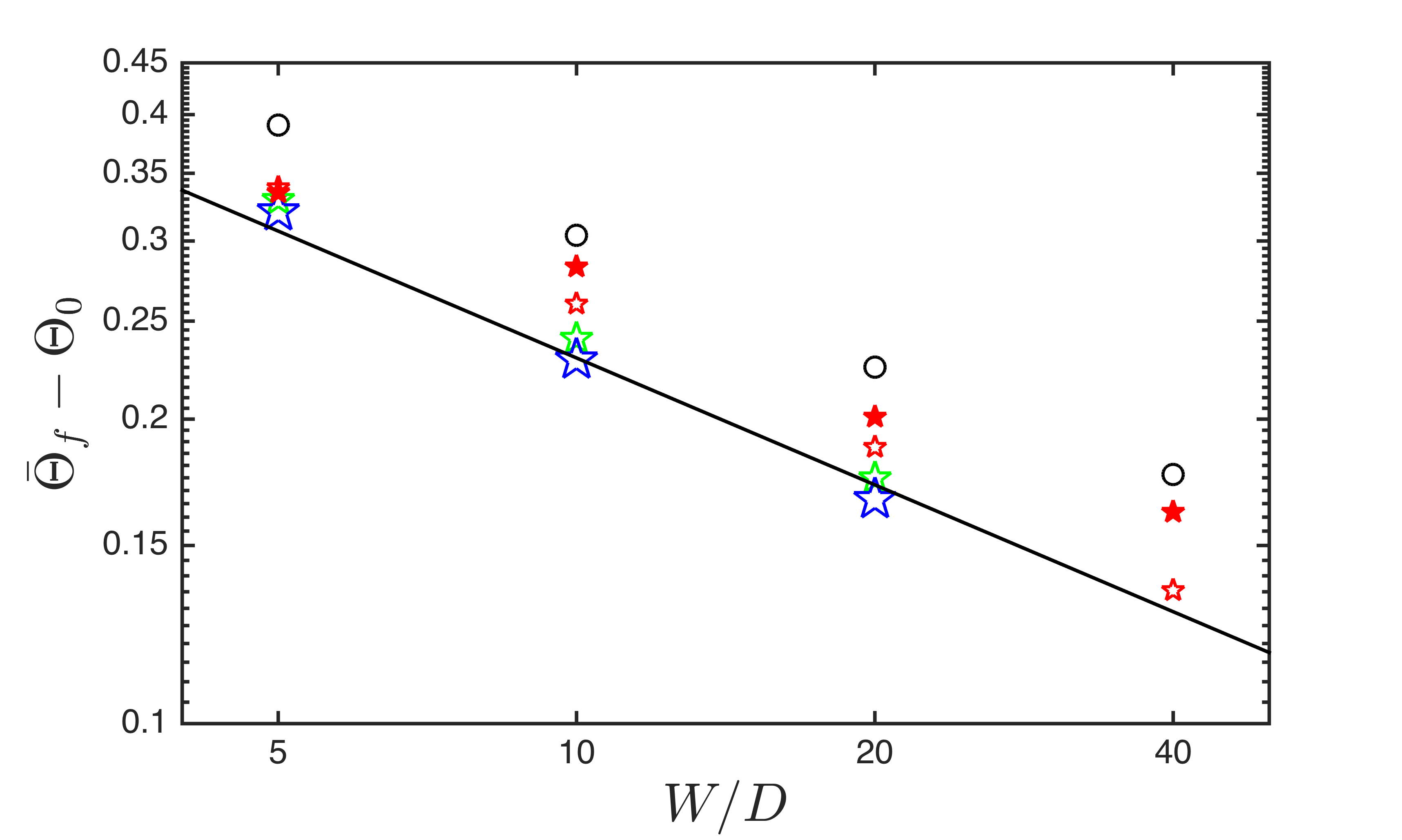}
\caption{(a) The probability distributions $P(\Theta_f)$ of the Shields number $\Theta_f$ where sustained grain motion is initiated ensembles of 2D, frictional, static systems, computed with a linear drag law (S-8 settings). The dashed vertical line represents $\Theta_0$, which is equivalent to $\Theta_c$ when $e_n$ is small. Open symbols correspond to systems that settled at $\Theta=0.033$: disks with Cundall-Strack friction with $\mu=0.6$ and fill height $10D$ (open black circles), as well as grain clusters (5-mers with $\mu_{\rm eff}=0.6$) with fill heights of $10D$ (open red stars), $20D$ (open green stars), and $40D$ (open blue stars), where the size of these symbols increases with fill height. The inset shows that these distributions collapse when rescaled by $\bar{\Theta}_f-\Theta_0$, and the thick black line shows a Weibull distribution with shape parameter $\alpha = 2.4$. (b) The mean excess stress for the initiation of grain motion $\bar{\Theta}_f-\Theta_0$ decreases as $W^{-1/\alpha}$, in accordance with the behavior of Eqs.~\eqref{eqn:Weibull} and \eqref{eqn:Weibull-CDF}.}
\label{fig:size-scaling-bumpy}
\end{figure}

Figure~\ref{fig:size-scaling-bumpy} shows the same data, but for 2D systems using Cundall-Strack friction with $\mu=0.6$ and the grain-asperity model with $\mu_{\rm eff}=0.6$ (S-8 settings). Our results are again consistent with Eqs.~\eqref{eqn:Weibull} and \eqref{eqn:Weibull-CDF} with $\alpha\approx 2.4$. We note that the primary difference between the frictionless case~\citep{clark2015hydro} and the frictional results shown here is in the effective system size $M_{\rm eff}= W_{\rm eff}H_{\rm eff}$. In systems with tangential forces, where $W$ is larger than a few grains, we find that $M_{\rm eff} = W$, and the system height is nearly irrelevant. For frictionless disks, $H_{\rm eff}$ is calculated by integrating the probability of the initiation of grain motion over the depth of the system, which is equal to the fluid force profile. Thus, friction strongly suppresses the initiation of surface grain motion from rearrangements below the surface.

We also note that the initiation of grain motion in systems with tangential forces depends on preparation history in a way that is different from frictionless systems. Specifically, systems with tangential contact forces that settle at a larger value of $\Theta$ tend to fail at larger values of $\Theta_f$. In frictionless simulations, we found no variation of the statistics of $\Theta_f$ with the value of $\Theta$ at which the system settled. The open stars shown in Fig.~\ref{fig:size-scaling-bumpy} settled at $\Theta=0.033$, whereas the filled red stars settled at $\Theta=0.067$. Settling at a larger value of $\Theta$ makes these systems statistically stronger on average but does not affect $\Theta_c$.

\section{Discussion}
\label{sec:discussion}

In this manuscript, we introduced a physical mechanism for how ${\rm Re}_*$-dependent grain dynamics can affect the critical applied fluid stress $\Theta_c$ required to sustain permanent grain motion. Using numerical simulations, we showed that the minimum dimensionless shear force to maintain grain motion can be described by a function $\Theta_c({\rm Re}_*)$ that consists of two distinct regimes. At ${\rm Re}_*<1$, grain dynamics are viscous dominated, and grains are not significantly accelerated between interactions with the bed. This means that grains are unlikely to bounce over a stable surface configuration or disrupt existing pockets during interactions with surface grains. In this regime, more geometrical configurations are available to grains as they search for stability, and grains find a state that is stable to a maximum Shields number of $\Theta_c^l \approx 0.28$. At ${\rm Re}_*>10$, grains are accelerated significantly between interactions with the bed. Based on the physical reasoning given in Section~\ref{sec:dimensional-analysis}, this makes some geometrical configurations inaccessible to the grains. This picture is confirmed by our numerical results, where grains that are sheared by the exact same fluid flow at ${\rm Re}_*>10$ are unable to find the configurations that are stable in the region $\Theta_c^h < \Theta < \Theta_c^l$, where $\Theta_c^h \approx 0.11$. 

These results suggest that the most common physical picture of the onset of sediment transport, namely the conditions at which static equilibrium is violated for surface grains, should be updated to include grain reorganization dynamics. The divergence of the transition times $t_s$ and $t_m$ at $\Theta=\Theta_c$ suggest the existence of a dynamical instability. Theories that account for when mobile grains can stop may be more successful than simply focusing on when static grains can first move. The roughness and geometry of the granular bed must play a role, but its role should be expanded from a focus on pocket angles of surface grains to a broader picture that describes the dynamics of grains as they traverse a rough, granular bed. In our theoretical analysis, we include this effect by assuming that grains will collide with the bed after moving roughly one grain diameter. This picture is supported by the data shown in Fig.~\ref{fig:z-vs-v-a-phi}.

Our results also suggest that previous approaches, which analyzed a single representative pocket geometry of a surface grain, could possibly be improved by accounting for the distribution of pocket geometries. Such an approach is similar in spirit to the results presented in Section~\ref{sec:weibull}, where for large systems we find that grain motion will always be initiated somewhere in the system once $\Theta$ exceeds $\Theta_c$. We argue that the grain dynamics that follow the initial force imbalance are important. Unsteadiness in the fluid flow, from turbulence or other external sources of fluctuations, can initiate grain motion, which will then be sustained only above the dynamical boundary $\Theta_c$. Our simulations do not include explicit temporal fluctuations, and there are likely important differences that arise when turbulence or other unsteadiness in the fluid stress is included, such as intermittency and fluctuations in the grain dynamics near $\Theta_c$~\citep{gonzalez2016}.

Finally, the boundaries $\Theta_c({\rm Re}_*)$ for simulations with linear and quadratic drag laws show strong agreement. At ${\rm Re}_*\ll 1$, these two approaches are expected to agree, since Stokes drag dominates in both cases. However, at ${\rm Re}_*\gg 1$, such good agreement is quite surprising. This means that, with regard to how grains search for stable configurations to an applied shear force, the form of the drag law is not important. The only relevant parameters are the shear stress $\Theta$, the bed collision time $\tau_\Theta$, and the characteristic time for a grain to equilibrate to the flow, which is given either by $\tau_\nu$ or $\tau_I$. For ${\rm Re}_*\gg 1$, the fluid equilibration time scales are always longer than the bed collision time scale. Thus, the minimum value of $\Theta$ to initiate sustained grain motion from a static bed is the same for both drag laws. For a linear drag law, we note that large restitution coefficients cause grains to bounce up into the fluid flow and accelerate in a way that is physically unreasonable in subaqueous sediment transport. This is the cause of the hysteresis shown in Fig.~\ref{fig:Shields-curve-with-data-all} and in Fig. 2 of our prior work~\cite{clark2015hydro}. When $e_n$ is decreased, these effects are suppressed and the hysteresis vanishes. For linear drag with small $e_n$ and quadratic drag for all $e_n$, the onset and cessation of grain motion occur at the same boundary, since the dominant physics relates to surface grain dynamics. However, we note that for the case of Aeolian sediment transport, the sediment grains are much denser than air, $\rho_g/\rho_f \sim 2000$, which yields $\tau_I / \tau_\Theta \sim 40-50$. This means saltating grains are substantially accelerated, and grain-grain collisions can be relatively elastic upon collision with the bed. Under these conditions, our results predict hysteresis, where motion is sustained by fast moving grains colliding with the bed~\citep{mitha1986,carneiro2011} as in our simulations with a linear drag law and large $e_n$.

Future work will focus on microstructural differences between contact geometries in the different regimes of ${\rm Re}_*$. Our results show that grains at ${\rm Re}_*> 10$ are unable to find configurations that are stable to an identical applied shear force profile at ${\rm Re}_*< 1$. Understanding the structure of these grain configurations may clarify why they are inaccessible at high ${\rm Re}_*$.

\appendix
\section{Frictional forces: Geometrical asperity and Cundall-Strack model}
\label{app:friction}

\begin{figure*}
\raggedright \hspace{10 mm} (a) \hspace{50 mm} (b) \hspace{50 mm} (c) \\ \centering
\includegraphics[trim=10mm 0mm 10mm 0mm, clip, width=0.3\textwidth]{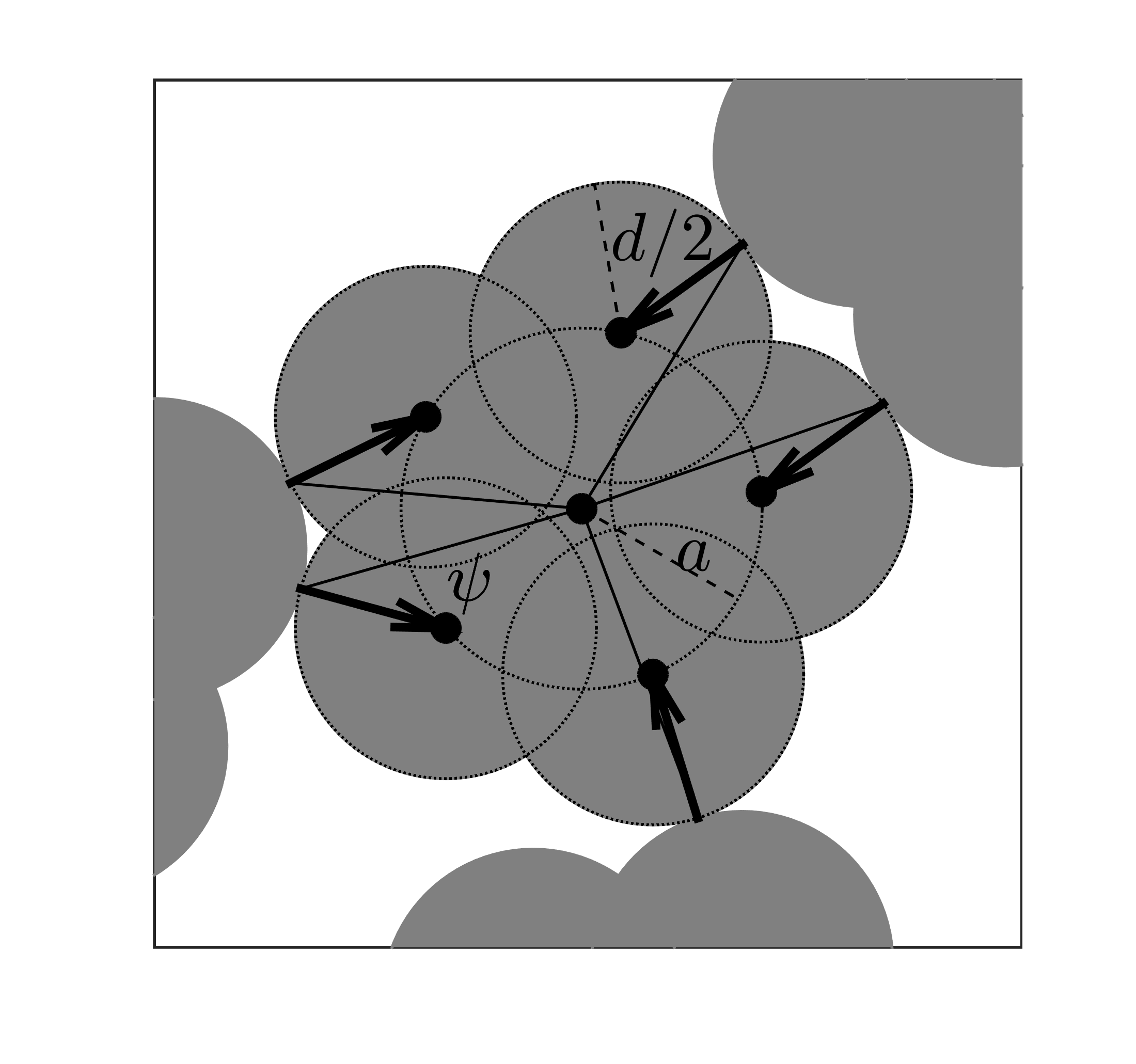}
\includegraphics[trim=10mm 0mm 10mm 0mm, clip, width=0.3\textwidth]{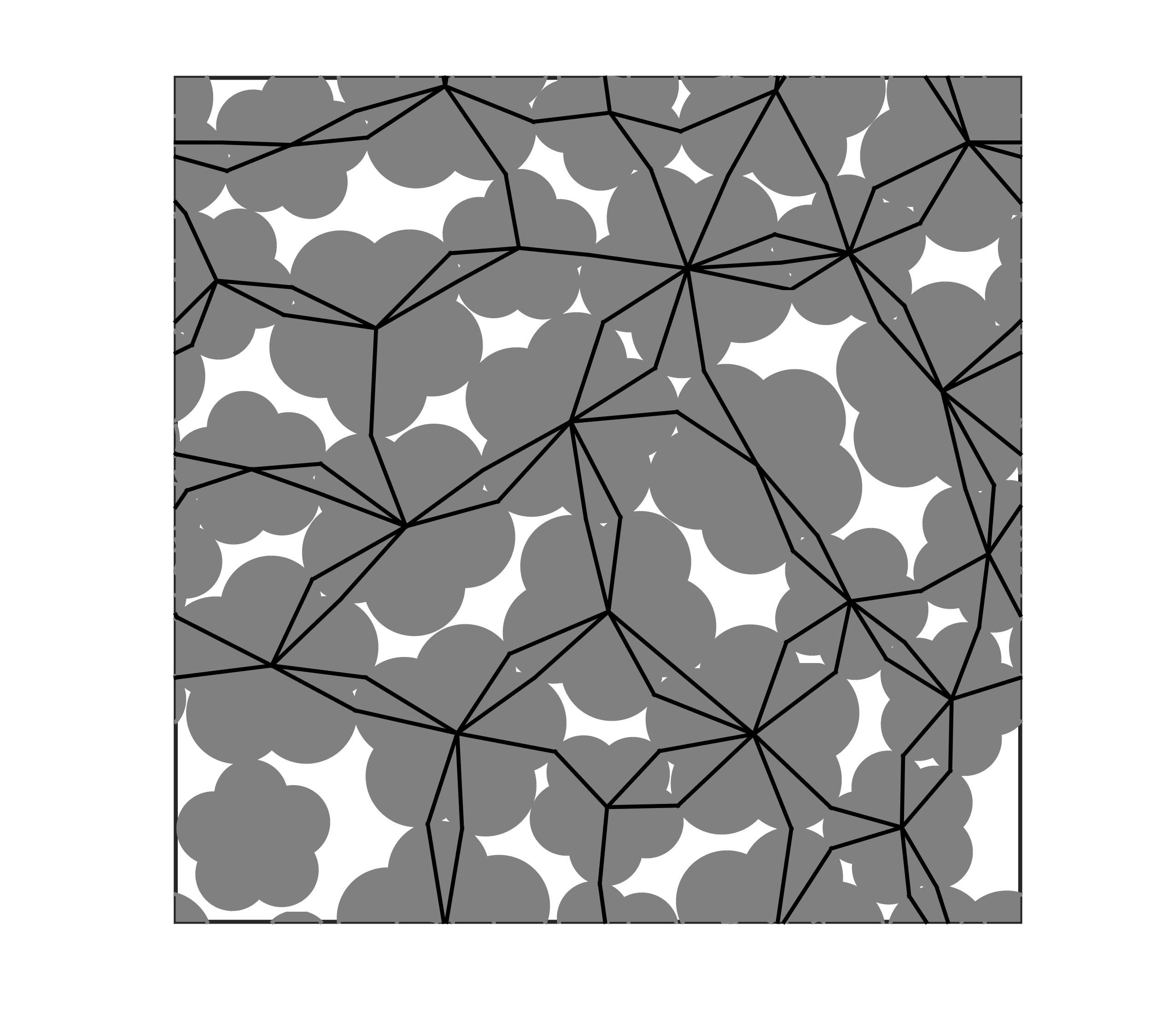}
\includegraphics[trim=10mm 0mm 10mm 0mm, clip, width=0.3\textwidth]{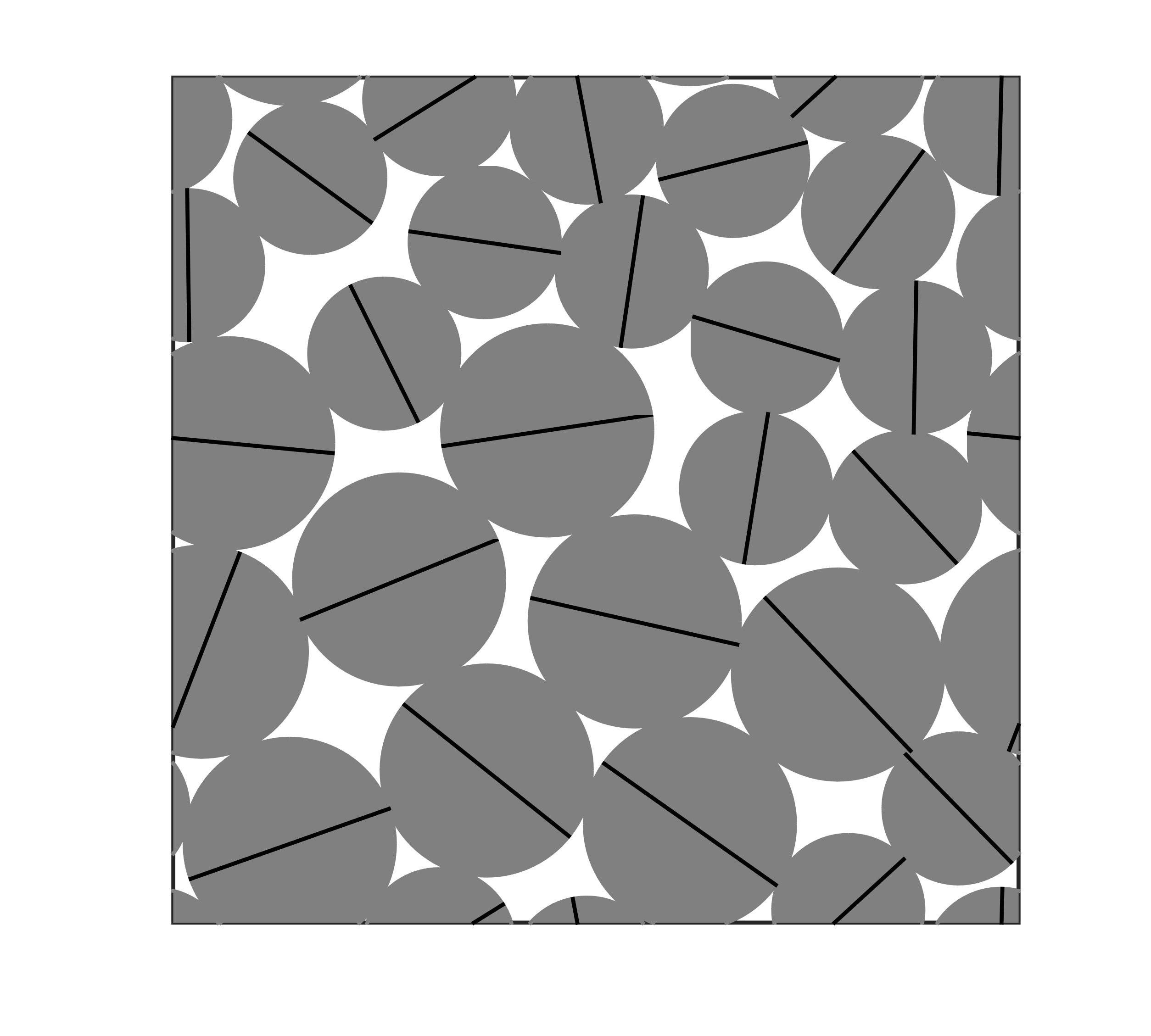}
\\  \raggedright (d) \hspace{80 mm} (e) \\
\centering \includegraphics[width=0.48\textwidth]{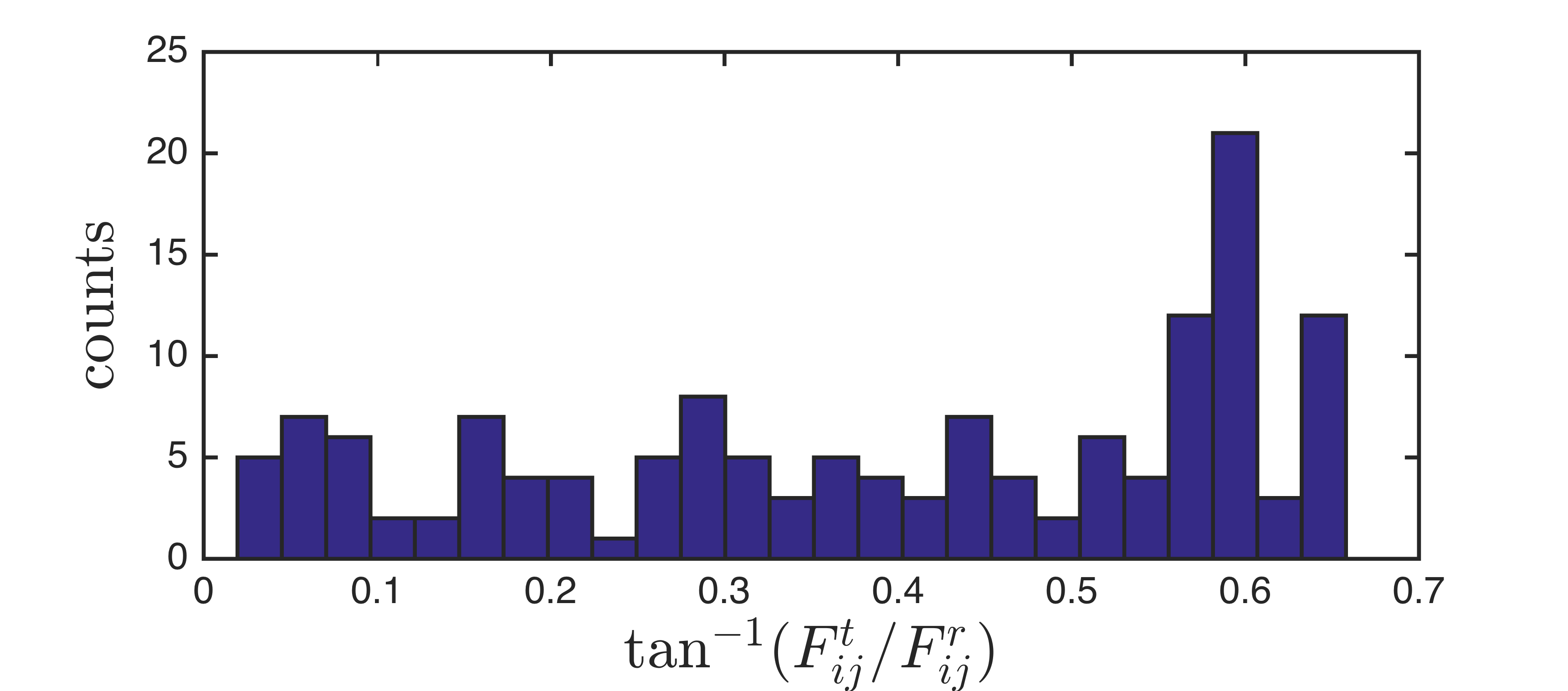}
\centering \includegraphics[width=0.48\textwidth]{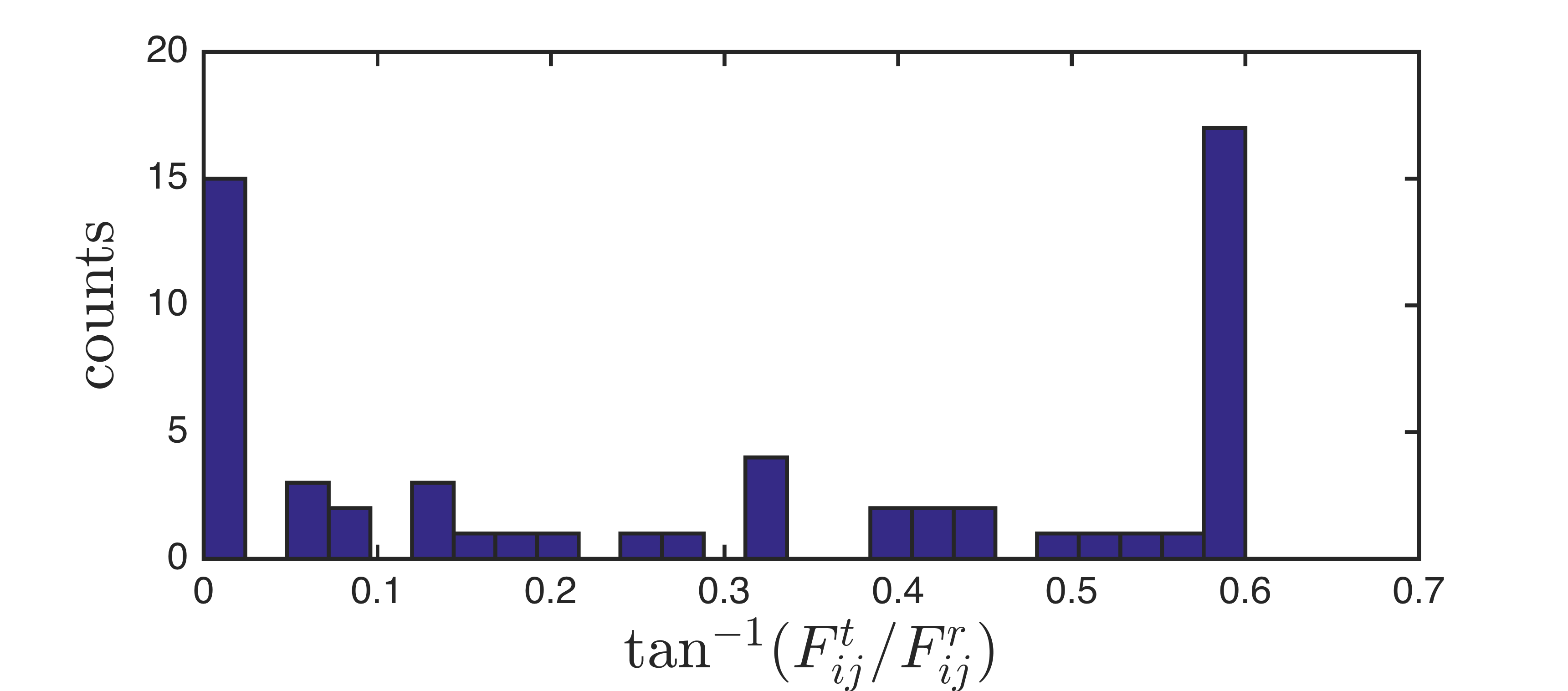}
\caption{(a) A depiction of the grain-asperity model \citep{Buchholtz1994,papanikolaou2013}. Grain clusters are composed of $n$ frictionless disk-shaped asperities, each with diameter $d$, with their centers regularly spaced on a circle of radius $a$. Arrows show the direction of contact forces, and solid lines connect the contact point to the center of the cluster. The angle $\psi$ between the arrows and solid lines sets the ratio of tangential $F^t_{ij}$ and normal $F^r_{ij}$ components of the contact forces, and thus the maximum ratio $\max(F^t_{ij}/F^r_{ij})=\mu_{\rm eff}$ is set purely by geometry. (b,c) Packings constructed from (b) grain clusters and (c) disk-shaped grains with Cundall-Strack friction~\citep{cundall79}, both containing 25 grains, using an athermal packing generation protocol~\citep{gao2006}. Grain clusters shown here have $n=5$ and $a/d=0.6$, which gives $\mu_{\rm eff}=0.6$ for two grains of the same size. The center grain in panel (a) is contacting the grain to its left with $F^t_{ij}/F^r_{ij}=\mu_{\rm eff}$. The disk-shaped grains with Cundall-Strack friction have $\mu=0.6$ to match the grain clusters. Panels (d) and (e) show histograms of the ratio $F^t_{ij}/F^n_{ij}$ for all contact forces in (b) and (c), respectively.}
\label{fig:bumpy-packing}
\end{figure*}

In the grain-asperity model, shown in Fig.~\ref{fig:bumpy-packing}(a)-(b) and Fig.~\ref{fig:bumpy-flow}(b), we use clusters of $n$ disks of a fixed size $d$. The centers of the disks lie on a circle of radius $a$, spaced at angular intervals of $2\pi/n$. The non-overlapping area $A_i$ of each cluster is calculated and the effective diameter for use in setting the fluid drag force is $D_i = \sqrt{4 A_i/\pi}$. We assume that small disks on each grain cluster interact via purely repulsive linear spring forces. These forces do not generally act through the center of mass of the cluster, and therefore generate torques. This means that the sum over $j$ in Eqs.~\eqref{eqn:force-law-rot} and \eqref{eqn:torque} now includes multiple contacts between clusters $i$ and $j$. In this way, macroscopic geometrical friction is introduced via the asperities, as is the case in natural systems, where grains of sand or gravel are almost never spherical. 

In the second approach, shown in Figs.~\ref{fig:bumpy-packing}(c) and~\ref{fig:bumpy-flow}(c), grains are represented by disks that interact via Cundall-Strack friction~\citep{cundall79}, which approximates microscopic friction through the use of linear tangential springs at intergrain contacts with tangential force $F^t_{ij} =-K_t u^t_{ij}$, where $K_t=K/3$ and $u^t_{ij}$ is the relative displacement of the point of contact between grains $i$ and $j$. At each contact, we enforce the Coulomb sliding condition, $F^t_{ij}\leq \mu F^r_{ij}$, where $\mu$ is the static friction coefficient. When $F^t_{ij}$ exceeds $\mu F^r_{ij}$, we set $u^t_{ij}=\mu F^r_{ij}/K_t$, and the grains slide relative to each other.

For the case of the grain-asperity model, the value of $F^t_{ij}/F^r_{ij}$ is determined by local geometry at the points of contact, with $F^t_{ij}/F^r_{ij} \leq \mu_{\rm eff}$, where $\mu_{\rm eff}$ corresponds to the case of an asperity from one cluster contacting two asperities from a different cluster. For the Cundall-Strack model, the value of $F^t_{ij}/F^r_{ij}$ depends on the history of the contact, i.e., the accumulated tangential displacement $u^t_{ij}$. Figure~\ref{fig:bumpy-packing} shows a comparison of packings generated with $\mu = \mu_{\rm eff} = 0.6$. The distributions of $F^t_{ij}/F^r_{ij}$ are similar for the two models; both have a maximum near 0.6 and a broad distribution below the maximum. Note that $\mu_{\rm eff}$ for contacts between two clusters of the same size is 0.6, whereas $\mu_{\rm eff}$ can be larger for an asperity from a small grain that is in contact with a asperities on a large grain.

\begin{table}
    \caption{A list of the characteristics for different simulations of the grain-aspertiy model, where $n$ is the number of disk-shaped asperities per grain, $a/d$ is the distance from the center of each disk-shaped asperity to the center of the grain, and $\mu_{\rm eff}$ is the maximum ratio of $F^t_{ij}/F^n_{ij}$.}
\centering
    \vspace{0.2in}
    \begin{tabular}{ccc}
    \hline \hline
      $n$ & $a/d$ & $\mu_{\rm eff}$ \\ \hline 
             2 & 0.1 & 0.2 \\ 
             2 & 0.2 & 0.42 \\ 
             2 & 0.4 & 1.06 \\ 
             3 & 0.2 & 0.33 \\ 
             3 & 0.4 & 0.7 \\ 
             3 & 0.5 & 0.93 \\ 
             3 & 0.75 & 2 \\ 
             4 & 0.5 & 0.65 \\ 
             4 & 0.75 & 1.1 \\ 
             4 & 1 & 2 \\ 
             5 & 0.1 & 0.11 \\ 
             5 & 0.4 & 0.4 \\ 
             5 & 0.6 & 0.6 \\ \hline 
 \end{tabular}
    \label{tbl:bumpy-grain-props}
\end{table}

\begin{acknowledgments}
This work was supported by the US Army Research Office under Grant No.~W911NF-14-1-0005 (A.H.C., N.T.O., C.S.O.) and by the National Science Foundation (NSF) Grant No.~CBET-0968013 (M.D.S.). This work also benefited from the facilities and staff of the Yale University Faculty of Arts and Sciences High Performance Computing Center and the NSF (Grant No. CNS-0821132) that, in part, funded acquisition of the computational facilities. We thank Arshad Kudrolli for his help in understanding various aspects of this problem, and we thank Michael Loewenberg for helpful discussions.
\end{acknowledgments}

\bibliography{erosion_sim2}

\begin{thebibliography}{76}%
\makeatletter
\providecommand \@ifxundefined [1]{%
 \@ifx{#1\undefined}
}%
\providecommand \@ifnum [1]{%
 \ifnum #1\expandafter \@firstoftwo
 \else \expandafter \@secondoftwo
 \fi
}%
\providecommand \@ifx [1]{%
 \ifx #1\expandafter \@firstoftwo
 \else \expandafter \@secondoftwo
 \fi
}%
\providecommand \natexlab [1]{#1}%
\providecommand \enquote  [1]{``#1''}%
\providecommand \bibnamefont  [1]{#1}%
\providecommand \bibfnamefont [1]{#1}%
\providecommand \citenamefont [1]{#1}%
\providecommand \href@noop [0]{\@secondoftwo}%
\providecommand \href [0]{\begingroup \@sanitize@url \@href}%
\providecommand \@href[1]{\@@startlink{#1}\@@href}%
\providecommand \@@href[1]{\endgroup#1\@@endlink}%
\providecommand \@sanitize@url [0]{\catcode `\\12\catcode `\$12\catcode
  `\&12\catcode `\#12\catcode `\^12\catcode `\_12\catcode `\%12\relax}%
\providecommand \@@startlink[1]{}%
\providecommand \@@endlink[0]{}%
\providecommand \url  [0]{\begingroup\@sanitize@url \@url }%
\providecommand \@url [1]{\endgroup\@href {#1}{\urlprefix }}%
\providecommand \urlprefix  [0]{URL }%
\providecommand \Eprint [0]{\href }%
\providecommand \doibase [0]{http://dx.doi.org/}%
\providecommand \selectlanguage [0]{\@gobble}%
\providecommand \bibinfo  [0]{\@secondoftwo}%
\providecommand \bibfield  [0]{\@secondoftwo}%
\providecommand \translation [1]{[#1]}%
\providecommand \BibitemOpen [0]{}%
\providecommand \bibitemStop [0]{}%
\providecommand \bibitemNoStop [0]{.\EOS\space}%
\providecommand \EOS [0]{\spacefactor3000\relax}%
\providecommand \BibitemShut  [1]{\csname bibitem#1\endcsname}%
\let\auto@bib@innerbib\@empty
\bibitem [{\citenamefont {Knisel}(1980)}]{knisel1980}%
  \BibitemOpen
  \bibfield  {author} {\bibinfo {author} {\bibfnamefont {W.~G.}\ \bibnamefont
  {Knisel}},\ }\bibfield  {title} {\enquote {\bibinfo {title} {Creams: A
  field-scale model for chemicals, runoff and erosion from agricultural
  management systems.}}\ }\href@noop {} {\bibfield  {journal} {\bibinfo
  {journal} {USDA Conservation Research Report}\ } (\bibinfo {year}
  {1980})}\BibitemShut {NoStop}%
\bibitem [{\citenamefont {Walling}(1983)}]{Walling1983}%
  \BibitemOpen
  \bibfield  {author} {\bibinfo {author} {\bibfnamefont {D.~E.}\ \bibnamefont
  {Walling}},\ }\bibfield  {title} {\enquote {\bibinfo {title} {The sediment
  delivery problem},}\ }\href@noop {} {\bibfield  {journal} {\bibinfo
  {journal} {J. Hydrol.}\ }\textbf {\bibinfo {volume} {65}},\ \bibinfo {pages}
  {209 -- 237} (\bibinfo {year} {1983})}\BibitemShut {NoStop}%
\bibitem [{\citenamefont {Renard}\ \emph {et~al.}(1997)\citenamefont {Renard},
  \citenamefont {Foster}, \citenamefont {Weesies}, \citenamefont {McCool},\
  and\ \citenamefont {Yoder}}]{renard1997}%
  \BibitemOpen
  \bibfield  {author} {\bibinfo {author} {\bibfnamefont {K.~G.}\ \bibnamefont
  {Renard}}, \bibinfo {author} {\bibfnamefont {G.~R.}\ \bibnamefont {Foster}},
  \bibinfo {author} {\bibfnamefont {G.~A.}\ \bibnamefont {Weesies}}, \bibinfo
  {author} {\bibfnamefont {D.~K.}\ \bibnamefont {McCool}}, \ and\ \bibinfo
  {author} {\bibfnamefont {D.~C.}\ \bibnamefont {Yoder}},\ }\href@noop {}
  {\emph {\bibinfo {title} {Predicting soil erosion by water: {A} guide to
  conservation planning with the Revised Universal Soil Loss Equation
  (RUSLE)}}},\ Vol.\ \bibinfo {volume} {703}\ (\bibinfo  {publisher} {United
  States Department of Agriculture Washington, DC},\ \bibinfo {year}
  {1997})\BibitemShut {NoStop}%
\bibitem [{\citenamefont {Dey}(2014)}]{Dey2014}%
  \BibitemOpen
  \bibfield  {author} {\bibinfo {author} {\bibfnamefont {S.}~\bibnamefont
  {Dey}},\ }\enquote {\bibinfo {title} {Fluvial hydrodynamics: Hydrodynamic and
  sediment transport phenomena},}\ \ (\bibinfo  {publisher} {Springer Berlin
  Heidelberg},\ \bibinfo {address} {Berlin, Heidelberg},\ \bibinfo {year}
  {2014})\ Chap.\ \bibinfo {chapter} {Sediment Threshold}, pp.\ \bibinfo
  {pages} {189--259}\BibitemShut {NoStop}%
\bibitem [{\citenamefont {Buffington}\ and\ \citenamefont
  {Montgomery}(1997)}]{buffington1997}%
  \BibitemOpen
  \bibfield  {author} {\bibinfo {author} {\bibfnamefont {J.~M.}\ \bibnamefont
  {Buffington}}\ and\ \bibinfo {author} {\bibfnamefont {D.~R.}\ \bibnamefont
  {Montgomery}},\ }\bibfield  {title} {\enquote {\bibinfo {title} {A systematic
  analysis of eight decades of incipient motion studies, with special reference
  to gravel-bedded rivers},}\ }\href@noop {} {\bibfield  {journal} {\bibinfo
  {journal} {Water Resour. Res.}\ }\textbf {\bibinfo {volume} {33}},\ \bibinfo
  {pages} {1993--2029} (\bibinfo {year} {1997})}\BibitemShut {NoStop}%
\bibitem [{\citenamefont {Xu}\ and\ \citenamefont {O'Hern}(2006)}]{xu2006}%
  \BibitemOpen
  \bibfield  {author} {\bibinfo {author} {\bibfnamefont {N.}~\bibnamefont
  {Xu}}\ and\ \bibinfo {author} {\bibfnamefont {C.~S.}\ \bibnamefont
  {O'Hern}},\ }\bibfield  {title} {\enquote {\bibinfo {title} {Measurements of
  the yield stress in frictionless granular systems},}\ }\href@noop {}
  {\bibfield  {journal} {\bibinfo  {journal} {Phys. Rev. E}\ }\textbf {\bibinfo
  {volume} {73}},\ \bibinfo {eid} {061303} (\bibinfo {year}
  {2006})}\BibitemShut {NoStop}%
\bibitem [{\citenamefont {Barrett}(1980)}]{barrett1980}%
  \BibitemOpen
  \bibfield  {author} {\bibinfo {author} {\bibfnamefont {P.~J.}\ \bibnamefont
  {Barrett}},\ }\bibfield  {title} {\enquote {\bibinfo {title} {The shape of
  rock particles, a critical review},}\ }\href@noop {} {\bibfield  {journal}
  {\bibinfo  {journal} {Sedimentology}\ }\textbf {\bibinfo {volume} {27}},\
  \bibinfo {pages} {291--303} (\bibinfo {year} {1980})}\BibitemShut {NoStop}%
\bibitem [{\citenamefont {Schumm}(1960)}]{schumm1960}%
  \BibitemOpen
  \bibfield  {author} {\bibinfo {author} {\bibfnamefont {S.~A.}\ \bibnamefont
  {Schumm}},\ }\bibfield  {title} {\enquote {\bibinfo {title} {The shape of
  alluvial channels in relation to sediment type},}\ }\href@noop {} {\bibfield
  {journal} {\bibinfo  {journal} {U.S. Geol. Surv. Prof. Pap.}\ }\textbf
  {\bibinfo {volume} {352B}},\ \bibinfo {pages} {17--30} (\bibinfo {year}
  {1960})}\BibitemShut {NoStop}%
\bibitem [{\citenamefont {Parker}(1979)}]{parker1979}%
  \BibitemOpen
  \bibfield  {author} {\bibinfo {author} {\bibfnamefont {G.}~\bibnamefont
  {Parker}},\ }\bibfield  {title} {\enquote {\bibinfo {title} {Hydraulic
  geometry of active gravel rivers},}\ }\href@noop {} {\bibfield  {journal}
  {\bibinfo  {journal} {J. Hydr. Div.}\ }\textbf {\bibinfo {volume} {105}},\
  \bibinfo {pages} {1185--1201} (\bibinfo {year} {1979})}\BibitemShut {NoStop}%
\bibitem [{\citenamefont {Nield}\ and\ \citenamefont
  {Bejan}(2013)}]{nield2013}%
  \BibitemOpen
  \bibfield  {author} {\bibinfo {author} {\bibfnamefont {D.~A.}\ \bibnamefont
  {Nield}}\ and\ \bibinfo {author} {\bibfnamefont {A.}~\bibnamefont {Bejan}},\
  }\href@noop {} {\emph {\bibinfo {title} {Mechanics of Fluid Flow Through a
  Porous Medium}}}\ (\bibinfo  {publisher} {Springer},\ \bibinfo {year}
  {2013})\BibitemShut {NoStop}%
\bibitem [{\citenamefont {Gilbert}\ and\ \citenamefont
  {Murphy}(1914)}]{gilbert1914}%
  \BibitemOpen
  \bibfield  {author} {\bibinfo {author} {\bibfnamefont {G.~K.}\ \bibnamefont
  {Gilbert}}\ and\ \bibinfo {author} {\bibfnamefont {E.~C.}\ \bibnamefont
  {Murphy}},\ }\href@noop {} {\emph {\bibinfo {title} {The transportation of
  debris by running water}}},\ \bibinfo {number} {86}\ (\bibinfo  {publisher}
  {US Government Printing Office},\ \bibinfo {year} {1914})\BibitemShut
  {NoStop}%
\bibitem [{\citenamefont {Casey}(1935)}]{casey1935}%
  \BibitemOpen
  \bibfield  {author} {\bibinfo {author} {\bibfnamefont {H.J.}\ \bibnamefont
  {Casey}},\ }\emph {\bibinfo {title} {{\"{U}ber die geschiebebewegung}}},\
  \href@noop {} {Ph.D. thesis},\ \bibinfo  {school} {Teknikal
  Hochschule-Scharlottenburg, Berlin, Germany} (\bibinfo {year}
  {1935})\BibitemShut {NoStop}%
\bibitem [{\citenamefont {Shields}(1936)}]{shields1936}%
  \BibitemOpen
  \bibfield  {author} {\bibinfo {author} {\bibfnamefont {A.}~\bibnamefont
  {Shields}},\ }\bibfield  {title} {\enquote {\bibinfo {title} {Anwendung der
  {\" a}hnlichkeitsmechanik und der {T}urbulenzforschung auf die
  {G}eschiebebewegung},}\ }in\ \href@noop {} {\emph {\bibinfo {booktitle}
  {Mitteilungen der Preussischen Versuchsanstalt f{\" u}r Wasserbau und
  Schiffbau}}},\ Vol.~\bibinfo {volume} {26}\ (\bibinfo {year}
  {1936})\BibitemShut {NoStop}%
\bibitem [{\citenamefont {USWES}(1936)}]{USWES1936}%
  \BibitemOpen
  \bibfield  {author} {\bibinfo {author} {\bibnamefont {USWES}},\ }\bibfield
  {title} {\enquote {\bibinfo {title} {Flume tests made to develop a synthetic
  sand which will not form ripples when used in movable bed models},}\
  }\href@noop {} {\bibfield  {journal} {\bibinfo  {journal} {Tech. Memo. 99-1,
  United States Waterways Experiment Station, Vicksburg, Mississippi}\ }
  (\bibinfo {year} {1936})}\BibitemShut {NoStop}%
\bibitem [{\citenamefont {White}(1940)}]{white1940}%
  \BibitemOpen
  \bibfield  {author} {\bibinfo {author} {\bibfnamefont {C.~M.}\ \bibnamefont
  {White}},\ }\bibfield  {title} {\enquote {\bibinfo {title} {The equilibrium
  of grains on the bed of a stream},}\ }\href@noop {} {\bibfield  {journal}
  {\bibinfo  {journal} {Proc. R. Soc. London, Ser. A}\ }\textbf {\bibinfo
  {volume} {174}},\ \bibinfo {pages} {322--338} (\bibinfo {year}
  {1940})}\BibitemShut {NoStop}%
\bibitem [{\citenamefont {Vanoni}(1946)}]{vanoni1946}%
  \BibitemOpen
  \bibfield  {author} {\bibinfo {author} {\bibfnamefont {V.~A.}\ \bibnamefont
  {Vanoni}},\ }\href@noop {} {\bibfield  {journal} {\bibinfo  {journal}
  {Transportation by Water Suspension, ASCE}\ }\textbf {\bibinfo {volume}
  {3}},\ \bibinfo {pages} {67} (\bibinfo {year} {1946})}\BibitemShut {NoStop}%
\bibitem [{\citenamefont {Meyer-Peter}\ and\ \citenamefont
  {M{\"u}ller}(1948)}]{meyer1948}%
  \BibitemOpen
  \bibfield  {author} {\bibinfo {author} {\bibfnamefont {E.}~\bibnamefont
  {Meyer-Peter}}\ and\ \bibinfo {author} {\bibfnamefont {R.}~\bibnamefont
  {M{\"u}ller}},\ }\bibfield  {title} {\enquote {\bibinfo {title} {Formulas for
  bed-load transport},}\ \ }(\bibinfo {organization} {IAHR},\ \bibinfo {year}
  {1948})\BibitemShut {NoStop}%
\bibitem [{\citenamefont {Neill}(1967)}]{neill1967}%
  \BibitemOpen
  \bibfield  {author} {\bibinfo {author} {\bibfnamefont {C.~R.}\ \bibnamefont
  {Neill}},\ }\bibfield  {title} {\enquote {\bibinfo {title} {Mean velocity
  criterion for scour of course uniform bed material},}\ \ }(\bibinfo
  {organization} {IAHR},\ \bibinfo {year} {1967})\BibitemShut {NoStop}%
\bibitem [{\citenamefont {Grass}(1970)}]{grass1970}%
  \BibitemOpen
  \bibfield  {author} {\bibinfo {author} {\bibfnamefont {A.~J.}\ \bibnamefont
  {Grass}},\ }\bibfield  {title} {\enquote {\bibinfo {title} {Initial
  instability of fine bed sand},}\ }\href@noop {} {\bibfield  {journal}
  {\bibinfo  {journal} {J. Fluid Mech.}\ }\textbf {\bibinfo {volume} {50}},\
  \bibinfo {pages} {619} (\bibinfo {year} {1970})}\BibitemShut {NoStop}%
\bibitem [{\citenamefont {White}(1970)}]{white1970}%
  \BibitemOpen
  \bibfield  {author} {\bibinfo {author} {\bibfnamefont {S.~J.}\ \bibnamefont
  {White}},\ }\bibfield  {title} {\enquote {\bibinfo {title} {Plane bed
  thresholds of fine grained sediments},}\ }\href@noop {} {\bibfield  {journal}
  {\bibinfo  {journal} {Nature}\ }\textbf {\bibinfo {volume} {228}},\ \bibinfo
  {pages} {152--153} (\bibinfo {year} {1970})}\BibitemShut {NoStop}%
\bibitem [{\citenamefont {Karahan}(1975)}]{karahan1975}%
  \BibitemOpen
  \bibfield  {author} {\bibinfo {author} {\bibfnamefont {E.}~\bibnamefont
  {Karahan}},\ }\emph {\bibinfo {title} {{Initiation of motion for uniform and
  nonuniform materials}}},\ \href@noop {} {Ph.D. thesis},\ \bibinfo  {school}
  {Technical University, Istanbul, Turkey,} (\bibinfo {year}
  {1975})\BibitemShut {NoStop}%
\bibitem [{\citenamefont {Mantz}(1977)}]{mantz1977}%
  \BibitemOpen
  \bibfield  {author} {\bibinfo {author} {\bibfnamefont {P.~A.}\ \bibnamefont
  {Mantz}},\ }\bibfield  {title} {\enquote {\bibinfo {title} {Incipient
  transport of fine grains and flakes by fluids -- extended {S}hields
  diagram},}\ }\href@noop {} {\bibfield  {journal} {\bibinfo  {journal} {J.
  Hydr. Div.}\ }\textbf {\bibinfo {volume} {103}},\ \bibinfo {pages} {601--615}
  (\bibinfo {year} {1977})}\BibitemShut {NoStop}%
\bibitem [{\citenamefont {Yalin}\ and\ \citenamefont
  {Karahan}(1979)}]{yalin1979}%
  \BibitemOpen
  \bibfield  {author} {\bibinfo {author} {\bibfnamefont {M.~S.}\ \bibnamefont
  {Yalin}}\ and\ \bibinfo {author} {\bibfnamefont {E.}~\bibnamefont
  {Karahan}},\ }\bibfield  {title} {\enquote {\bibinfo {title} {Inception of
  sediment transport},}\ }\href@noop {} {\bibfield  {journal} {\bibinfo
  {journal} {J. Hydr. Div.}\ }\textbf {\bibinfo {volume} {105}},\ \bibinfo
  {pages} {1433--1443} (\bibinfo {year} {1979})}\BibitemShut {NoStop}%
\bibitem [{\citenamefont {Wiberg}\ and\ \citenamefont
  {Smith}(1987)}]{wiberg1987}%
  \BibitemOpen
  \bibfield  {author} {\bibinfo {author} {\bibfnamefont {P.~L.}\ \bibnamefont
  {Wiberg}}\ and\ \bibinfo {author} {\bibfnamefont {J.~D.}\ \bibnamefont
  {Smith}},\ }\bibfield  {title} {\enquote {\bibinfo {title} {Calculations of
  the critical shear stress for motion of uniform and heterogeneous
  sediments},}\ }\href@noop {} {\bibfield  {journal} {\bibinfo  {journal}
  {Water Resour. Res.}\ }\textbf {\bibinfo {volume} {23}},\ \bibinfo {pages}
  {1471--1480} (\bibinfo {year} {1987})}\BibitemShut {NoStop}%
\bibitem [{\citenamefont {Miller}\ \emph {et~al.}(1977)\citenamefont {Miller},
  \citenamefont {McCave},\ and\ \citenamefont {Komar}}]{miller1977}%
  \BibitemOpen
  \bibfield  {author} {\bibinfo {author} {\bibfnamefont {M.~C.}\ \bibnamefont
  {Miller}}, \bibinfo {author} {\bibfnamefont {I.~N.}\ \bibnamefont {McCave}},
  \ and\ \bibinfo {author} {\bibfnamefont {P.~D.}\ \bibnamefont {Komar}},\
  }\bibfield  {title} {\enquote {\bibinfo {title} {Threshold of sediment motion
  under unidirectional currents},}\ }\href@noop {} {\bibfield  {journal}
  {\bibinfo  {journal} {Sedimentology}\ }\textbf {\bibinfo {volume} {24}},\
  \bibinfo {pages} {507--527} (\bibinfo {year} {1977})}\BibitemShut {NoStop}%
\bibitem [{\citenamefont {Yalin}(1972)}]{yalin1972}%
  \BibitemOpen
  \bibfield  {author} {\bibinfo {author} {\bibfnamefont {M.~S.}\ \bibnamefont
  {Yalin}},\ }\href@noop {} {\emph {\bibinfo {title} {Mechanics of sediment
  transport}}}\ (\bibinfo  {publisher} {Pergamon Press},\ \bibinfo {year}
  {1972})\BibitemShut {NoStop}%
\bibitem [{\citenamefont {Van~Rijn}(1993)}]{vanrijn1993}%
  \BibitemOpen
  \bibfield  {author} {\bibinfo {author} {\bibfnamefont {L.~C.}\ \bibnamefont
  {Van~Rijn}},\ }\href@noop {} {\emph {\bibinfo {title} {Principles of sediment
  transport in rivers, estuaries and coastal seas}}},\ Vol.\ \bibinfo {volume}
  {1006}\ (\bibinfo  {publisher} {Aqua publications Amsterdam},\ \bibinfo
  {year} {1993})\BibitemShut {NoStop}%
\bibitem [{\citenamefont {Paphitis}(2001)}]{Paphitis2001}%
  \BibitemOpen
  \bibfield  {author} {\bibinfo {author} {\bibfnamefont {D.}~\bibnamefont
  {Paphitis}},\ }\bibfield  {title} {\enquote {\bibinfo {title} {Sediment
  movement under unidirectional flows: an assessment of empirical threshold
  curves},}\ }\href
  {http://www.sciencedirect.com/science/article/pii/S0378383901000151}
  {\bibfield  {journal} {\bibinfo  {journal} {Coast. Eng.}\ }\textbf {\bibinfo
  {volume} {43}},\ \bibinfo {pages} {227 -- 245} (\bibinfo {year}
  {2001})}\BibitemShut {NoStop}%
\bibitem [{\citenamefont {Iwagaki}(1956)}]{Iwagaki1956}%
  \BibitemOpen
  \bibfield  {author} {\bibinfo {author} {\bibfnamefont {Y.}~\bibnamefont
  {Iwagaki}},\ }\bibfield  {title} {\enquote {\bibinfo {title} {Hydrodynamical
  study on critical tractive force},}\ }\href@noop {} {\bibfield  {journal}
  {\bibinfo  {journal} {Transaction of the Japanese Society of Civil
  Engineers}\ }\textbf {\bibinfo {volume} {41}},\ \bibinfo {pages} {1--21}
  (\bibinfo {year} {1956})}\BibitemShut {NoStop}%
\bibitem [{\citenamefont {Ling}(1995)}]{Ling1995}%
  \BibitemOpen
  \bibfield  {author} {\bibinfo {author} {\bibfnamefont {C.~H.}\ \bibnamefont
  {Ling}},\ }\bibfield  {title} {\enquote {\bibinfo {title} {Criteria for
  incipient motion of spherical sediment particles},}\ }\href@noop {}
  {\bibfield  {journal} {\bibinfo  {journal} {J. Hydraul. Eng.}\ }\textbf
  {\bibinfo {volume} {121}},\ \bibinfo {pages} {472--478} (\bibinfo {year}
  {1995})}\BibitemShut {NoStop}%
\bibitem [{\citenamefont {Dey}(1999)}]{Dey1999}%
  \BibitemOpen
  \bibfield  {author} {\bibinfo {author} {\bibfnamefont {S.}~\bibnamefont
  {Dey}},\ }\bibfield  {title} {\enquote {\bibinfo {title} {Sediment
  threshold},}\ }\href@noop {} {\bibfield  {journal} {\bibinfo  {journal}
  {Appl. Math. Model.}\ }\textbf {\bibinfo {volume} {23}},\ \bibinfo {pages}
  {399 -- 417} (\bibinfo {year} {1999})}\BibitemShut {NoStop}%
\bibitem [{\citenamefont {Dey}\ and\ \citenamefont {Debnath}(2000)}]{dey2000}%
  \BibitemOpen
  \bibfield  {author} {\bibinfo {author} {\bibfnamefont {S.}~\bibnamefont
  {Dey}}\ and\ \bibinfo {author} {\bibfnamefont {K.}~\bibnamefont {Debnath}},\
  }\bibfield  {title} {\enquote {\bibinfo {title} {Influence of streamwise bed
  slope on sediment threshold under stream flow},}\ }\href@noop {} {\bibfield
  {journal} {\bibinfo  {journal} {J. Irrig. Drain. E.}\ }\textbf {\bibinfo
  {volume} {126}},\ \bibinfo {pages} {255--263} (\bibinfo {year}
  {2000})}\BibitemShut {NoStop}%
\bibitem [{\citenamefont {Kirchner}\ \emph {et~al.}(1990)\citenamefont
  {Kirchner}, \citenamefont {Dietrich}, \citenamefont {Iseya},\ and\
  \citenamefont {Ikeda}}]{kirchner1990}%
  \BibitemOpen
  \bibfield  {author} {\bibinfo {author} {\bibfnamefont {J.~W.}\ \bibnamefont
  {Kirchner}}, \bibinfo {author} {\bibfnamefont {W.~E.}\ \bibnamefont
  {Dietrich}}, \bibinfo {author} {\bibfnamefont {F.}~\bibnamefont {Iseya}}, \
  and\ \bibinfo {author} {\bibfnamefont {H.}~\bibnamefont {Ikeda}},\ }\bibfield
   {title} {\enquote {\bibinfo {title} {The variability of critical shear
  stress, friction angle, and grain protrusion in water-worked sediments},}\
  }\href@noop {} {\bibfield  {journal} {\bibinfo  {journal} {Sedimentology}\
  }\textbf {\bibinfo {volume} {37}},\ \bibinfo {pages} {647--672} (\bibinfo
  {year} {1990})}\BibitemShut {NoStop}%
\bibitem [{\citenamefont {Andrews}(1994)}]{andrews1994}%
  \BibitemOpen
  \bibfield  {author} {\bibinfo {author} {\bibfnamefont {E.~D.}\ \bibnamefont
  {Andrews}},\ }\bibfield  {title} {\enquote {\bibinfo {title} {Marginal bed
  load transport in a gravel bed stream, sagehen creek, california},}\
  }\href@noop {} {\bibfield  {journal} {\bibinfo  {journal} {Water Resour.
  Res.}\ }\textbf {\bibinfo {volume} {30}},\ \bibinfo {pages} {2241--2250}
  (\bibinfo {year} {1994})}\BibitemShut {NoStop}%
\bibitem [{\citenamefont {Buffington}\ \emph {et~al.}(1992)\citenamefont
  {Buffington}, \citenamefont {Dietrich},\ and\ \citenamefont
  {Kirchner}}]{buffington1992}%
  \BibitemOpen
  \bibfield  {author} {\bibinfo {author} {\bibfnamefont {J.~M.}\ \bibnamefont
  {Buffington}}, \bibinfo {author} {\bibfnamefont {W.~E.}\ \bibnamefont
  {Dietrich}}, \ and\ \bibinfo {author} {\bibfnamefont {J.~W.}\ \bibnamefont
  {Kirchner}},\ }\bibfield  {title} {\enquote {\bibinfo {title} {Friction angle
  measurements on a naturally formed gravel streambed: Implications for
  critical boundary shear stress},}\ }\href@noop {} {\bibfield  {journal}
  {\bibinfo  {journal} {Water Resour. Res.}\ }\textbf {\bibinfo {volume}
  {28}},\ \bibinfo {pages} {411--425} (\bibinfo {year} {1992})}\BibitemShut
  {NoStop}%
\bibitem [{\citenamefont {Lamb}\ \emph {et~al.}(2008)\citenamefont {Lamb},
  \citenamefont {Dietrich},\ and\ \citenamefont {Venditti}}]{lamb2008}%
  \BibitemOpen
  \bibfield  {author} {\bibinfo {author} {\bibfnamefont {M.~P.}\ \bibnamefont
  {Lamb}}, \bibinfo {author} {\bibfnamefont {W.~E.}\ \bibnamefont {Dietrich}},
  \ and\ \bibinfo {author} {\bibfnamefont {J.~G.}\ \bibnamefont {Venditti}},\
  }\bibfield  {title} {\enquote {\bibinfo {title} {Is the critical shields
  stress for incipient sediment motion dependent on channel-bed slope?}}\
  }\href@noop {} {\bibfield  {journal} {\bibinfo  {journal} {J. Geophys. Res.
  Earth Surf.}\ }\textbf {\bibinfo {volume} {113}} (\bibinfo {year}
  {2008})}\BibitemShut {NoStop}%
\bibitem [{\citenamefont {Reichardt}(1951)}]{Reichardt1951}%
  \BibitemOpen
  \bibfield  {author} {\bibinfo {author} {\bibfnamefont {H.}~\bibnamefont
  {Reichardt}},\ }\bibfield  {title} {\enquote {\bibinfo {title} {Vollständige
  darstellung der turbulenten geschwindigkeitsverteilung in glatten
  leitungen},}\ }\href@noop {} {\bibfield  {journal} {\bibinfo  {journal} {ZAMM
  - Zeitschrift für Angewandte Mathematik und Mechanik}\ }\textbf {\bibinfo
  {volume} {31}},\ \bibinfo {pages} {208--219} (\bibinfo {year}
  {1951})}\BibitemShut {NoStop}%
\bibitem [{\citenamefont {Schlichting}\ and\ \citenamefont
  {Gersten}(2003)}]{Schlichting2003}%
  \BibitemOpen
  \bibfield  {author} {\bibinfo {author} {\bibfnamefont {H.}~\bibnamefont
  {Schlichting}}\ and\ \bibinfo {author} {\bibfnamefont {K.}~\bibnamefont
  {Gersten}},\ }\href@noop {} {\emph {\bibinfo {title} {Boundary-layer
  theory}}}\ (\bibinfo  {publisher} {Springer Science \& Business Media},\
  \bibinfo {year} {2003})\BibitemShut {NoStop}%
\bibitem [{\citenamefont {Charru}\ \emph {et~al.}(2004)\citenamefont {Charru},
  \citenamefont {Mouilleron},\ and\ \citenamefont {Eiff}}]{charru2004}%
  \BibitemOpen
  \bibfield  {author} {\bibinfo {author} {\bibfnamefont {F.}~\bibnamefont
  {Charru}}, \bibinfo {author} {\bibfnamefont {H.}~\bibnamefont {Mouilleron}},
  \ and\ \bibinfo {author} {\bibfnamefont {O.}~\bibnamefont {Eiff}},\
  }\bibfield  {title} {\enquote {\bibinfo {title} {Erosion and deposition of
  particles on a bed sheared by a viscous flow},}\ }\href@noop {} {\bibfield
  {journal} {\bibinfo  {journal} {J. Fluid Mech.}\ }\textbf {\bibinfo {volume}
  {519}},\ \bibinfo {pages} {55--80} (\bibinfo {year} {2004})}\BibitemShut
  {NoStop}%
\bibitem [{\citenamefont {Lobkovsky}\ \emph {et~al.}(2008)\citenamefont
  {Lobkovsky}, \citenamefont {Orpe}, \citenamefont {Molloy}, \citenamefont
  {Kudrolli},\ and\ \citenamefont {Rothman}}]{lobkovsky2008}%
  \BibitemOpen
  \bibfield  {author} {\bibinfo {author} {\bibfnamefont {A.~E.}\ \bibnamefont
  {Lobkovsky}}, \bibinfo {author} {\bibfnamefont {A.~V.}\ \bibnamefont {Orpe}},
  \bibinfo {author} {\bibfnamefont {R.}~\bibnamefont {Molloy}}, \bibinfo
  {author} {\bibfnamefont {A.}~\bibnamefont {Kudrolli}}, \ and\ \bibinfo
  {author} {\bibfnamefont {D.~H.}\ \bibnamefont {Rothman}},\ }\bibfield
  {title} {\enquote {\bibinfo {title} {Erosion of a granular bed driven by
  laminar fluid flow},}\ }\href@noop {} {\bibfield  {journal} {\bibinfo
  {journal} {J. Fluid Mech.}\ }\textbf {\bibinfo {volume} {605}},\ \bibinfo
  {pages} {47--58} (\bibinfo {year} {2008})}\BibitemShut {NoStop}%
\bibitem [{\citenamefont {Hong}\ \emph {et~al.}(2015)\citenamefont {Hong},
  \citenamefont {Tao},\ and\ \citenamefont {Kudrolli}}]{hong2015}%
  \BibitemOpen
  \bibfield  {author} {\bibinfo {author} {\bibfnamefont {A.}~\bibnamefont
  {Hong}}, \bibinfo {author} {\bibfnamefont {M.}~\bibnamefont {Tao}}, \ and\
  \bibinfo {author} {\bibfnamefont {A.}~\bibnamefont {Kudrolli}},\ }\bibfield
  {title} {\enquote {\bibinfo {title} {Onset of erosion of a granular bed in a
  channel driven by fluid flow},}\ }\href@noop {} {\bibfield  {journal}
  {\bibinfo  {journal} {Phys. Fluids}\ }\textbf {\bibinfo {volume} {27}},\
  \bibinfo {eid} {013301} (\bibinfo {year} {2015})}\BibitemShut {NoStop}%
\bibitem [{\citenamefont {Roseberry}\ \emph {et~al.}(2012)\citenamefont
  {Roseberry}, \citenamefont {Schmeeckle},\ and\ \citenamefont
  {Furbish}}]{roseberry2012}%
  \BibitemOpen
  \bibfield  {author} {\bibinfo {author} {\bibfnamefont {J.~C.}\ \bibnamefont
  {Roseberry}}, \bibinfo {author} {\bibfnamefont {M.~W.}\ \bibnamefont
  {Schmeeckle}}, \ and\ \bibinfo {author} {\bibfnamefont {D.~J.}\ \bibnamefont
  {Furbish}},\ }\bibfield  {title} {\enquote {\bibinfo {title} {A probabilistic
  description of the bed load sediment flux: 2. particle activity and
  motions},}\ }\href@noop {} {\bibfield  {journal} {\bibinfo  {journal} {J.
  Geophys. Res. Earth Surf.}\ }\textbf {\bibinfo {volume} {117}} (\bibinfo
  {year} {2012})}\BibitemShut {NoStop}%
\bibitem [{\citenamefont {Schmeeckle}(2014)}]{schmeeckle2014}%
  \BibitemOpen
  \bibfield  {author} {\bibinfo {author} {\bibfnamefont {M.~W.}\ \bibnamefont
  {Schmeeckle}},\ }\bibfield  {title} {\enquote {\bibinfo {title} {Numerical
  simulation of turbulence and sediment transport of medium sand},}\
  }\href@noop {} {\bibfield  {journal} {\bibinfo  {journal} {J. Geophys. Res.
  Earth Surf.}\ }\textbf {\bibinfo {volume} {119}},\ \bibinfo {pages}
  {1240--1262} (\bibinfo {year} {2014})}\BibitemShut {NoStop}%
\bibitem [{\citenamefont {Joseph}\ \emph {et~al.}(2001)\citenamefont {Joseph},
  \citenamefont {Zenit}, \citenamefont {Hunt},\ and\ \citenamefont
  {Rosenwinkel}}]{joseph2001}%
  \BibitemOpen
  \bibfield  {author} {\bibinfo {author} {\bibfnamefont {G.~G.}\ \bibnamefont
  {Joseph}}, \bibinfo {author} {\bibfnamefont {R.}~\bibnamefont {Zenit}},
  \bibinfo {author} {\bibfnamefont {M.~L.}\ \bibnamefont {Hunt}}, \ and\
  \bibinfo {author} {\bibfnamefont {A.~M.}\ \bibnamefont {Rosenwinkel}},\
  }\bibfield  {title} {\enquote {\bibinfo {title} {Particle-wall collisions in
  a viscous fluid},}\ }\href@noop {} {\bibfield  {journal} {\bibinfo  {journal}
  {J. Fluid Mech.}\ }\textbf {\bibinfo {volume} {433}},\ \bibinfo {pages}
  {329--346} (\bibinfo {year} {2001})}\BibitemShut {NoStop}%
\bibitem [{\citenamefont {Yang}\ and\ \citenamefont {Hunt}(2006)}]{yang2006}%
  \BibitemOpen
  \bibfield  {author} {\bibinfo {author} {\bibfnamefont {F.-L.}\ \bibnamefont
  {Yang}}\ and\ \bibinfo {author} {\bibfnamefont {M.~L.}\ \bibnamefont
  {Hunt}},\ }\bibfield  {title} {\enquote {\bibinfo {title} {Dynamics of
  particle-particle collisions in a viscous liquid},}\ }\href@noop {}
  {\bibfield  {journal} {\bibinfo  {journal} {Phys. Fluids}\ }\textbf {\bibinfo
  {volume} {18}},\ \bibinfo {eid} {121506} (\bibinfo {year}
  {2006})}\BibitemShut {NoStop}%
\bibitem [{\citenamefont {Schmeeckle}\ \emph {et~al.}(2001)\citenamefont
  {Schmeeckle}, \citenamefont {Nelson}, \citenamefont {Pitlick},\ and\
  \citenamefont {Bennett}}]{Schmeeckle2001}%
  \BibitemOpen
  \bibfield  {author} {\bibinfo {author} {\bibfnamefont {M.~W.}\ \bibnamefont
  {Schmeeckle}}, \bibinfo {author} {\bibfnamefont {J.~M.}\ \bibnamefont
  {Nelson}}, \bibinfo {author} {\bibfnamefont {J.}~\bibnamefont {Pitlick}}, \
  and\ \bibinfo {author} {\bibfnamefont {J.~P.}\ \bibnamefont {Bennett}},\
  }\bibfield  {title} {\enquote {\bibinfo {title} {Interparticle collision of
  natural sediment grains in water},}\ }\href@noop {} {\bibfield  {journal}
  {\bibinfo  {journal} {Water Resour. Res.}\ }\textbf {\bibinfo {volume}
  {37}},\ \bibinfo {pages} {2377--2391} (\bibinfo {year} {2001})}\BibitemShut
  {NoStop}%
\bibitem [{\citenamefont {Schmeeckle}\ and\ \citenamefont
  {Nelson}(2003)}]{schmeeckle2003}%
  \BibitemOpen
  \bibfield  {author} {\bibinfo {author} {\bibfnamefont {M.~W.}\ \bibnamefont
  {Schmeeckle}}\ and\ \bibinfo {author} {\bibfnamefont {J.~M.}\ \bibnamefont
  {Nelson}},\ }\bibfield  {title} {\enquote {\bibinfo {title} {Direct numerical
  simulation of bedload transport using a local, dynamic boundary condition},}\
  }\href@noop {} {\bibfield  {journal} {\bibinfo  {journal} {Sedimentology}\
  }\textbf {\bibinfo {volume} {50}},\ \bibinfo {pages} {279--301} (\bibinfo
  {year} {2003})}\BibitemShut {NoStop}%
\bibitem [{\citenamefont {Carneiro}\ \emph {et~al.}(2011)\citenamefont
  {Carneiro}, \citenamefont {P\"ahtz},\ and\ \citenamefont
  {Herrmann}}]{carneiro2011}%
  \BibitemOpen
  \bibfield  {author} {\bibinfo {author} {\bibfnamefont {M.~V.}\ \bibnamefont
  {Carneiro}}, \bibinfo {author} {\bibfnamefont {T.}~\bibnamefont {P\"ahtz}}, \
  and\ \bibinfo {author} {\bibfnamefont {H.~J.}\ \bibnamefont {Herrmann}},\
  }\bibfield  {title} {\enquote {\bibinfo {title} {Jump at the onset of
  saltation},}\ }\href@noop {} {\bibfield  {journal} {\bibinfo  {journal}
  {Phys. Rev. Lett.}\ }\textbf {\bibinfo {volume} {107}},\ \bibinfo {eid}
  {098001} (\bibinfo {year} {2011})}\BibitemShut {NoStop}%
\bibitem [{\citenamefont {Dur\'{a}n}\ \emph {et~al.}(2012)\citenamefont
  {Dur\'{a}n}, \citenamefont {Andreotti},\ and\ \citenamefont
  {Claudin}}]{duran2012}%
  \BibitemOpen
  \bibfield  {author} {\bibinfo {author} {\bibfnamefont {O.}~\bibnamefont
  {Dur\'{a}n}}, \bibinfo {author} {\bibfnamefont {B.}~\bibnamefont
  {Andreotti}}, \ and\ \bibinfo {author} {\bibfnamefont {P.}~\bibnamefont
  {Claudin}},\ }\bibfield  {title} {\enquote {\bibinfo {title} {Numerical
  simulation of turbulent sediment transport, from bed load to saltation},}\
  }\href@noop {} {\bibfield  {journal} {\bibinfo  {journal} {Phys. Fluids}\
  }\textbf {\bibinfo {volume} {24}},\ \bibinfo {eid} {103306} (\bibinfo {year}
  {2012})}\BibitemShut {NoStop}%
\bibitem [{\citenamefont {Capecelatro}\ and\ \citenamefont
  {Desjardins}(2013)}]{capecelatro2013}%
  \BibitemOpen
  \bibfield  {author} {\bibinfo {author} {\bibfnamefont {J.}~\bibnamefont
  {Capecelatro}}\ and\ \bibinfo {author} {\bibfnamefont {O.}~\bibnamefont
  {Desjardins}},\ }\bibfield  {title} {\enquote {\bibinfo {title}
  {Eulerian--{L}agrangian modeling of turbulent liquid-solid slurries in
  horizontal pipes},}\ }\href@noop {} {\bibfield  {journal} {\bibinfo
  {journal} {Intl. J. Multiphase Flow}\ }\textbf {\bibinfo {volume} {55}},\
  \bibinfo {pages} {64--79} (\bibinfo {year} {2013})}\BibitemShut {NoStop}%
\bibitem [{\citenamefont {Nabi}\ \emph {et~al.}(2013)\citenamefont {Nabi},
  \citenamefont {de~Vriend}, \citenamefont {Mosselman}, \citenamefont {Sloff},\
  and\ \citenamefont {Shimizu}}]{nabi2013}%
  \BibitemOpen
  \bibfield  {author} {\bibinfo {author} {\bibfnamefont {M.}~\bibnamefont
  {Nabi}}, \bibinfo {author} {\bibfnamefont {H.~J.}\ \bibnamefont {de~Vriend}},
  \bibinfo {author} {\bibfnamefont {E.}~\bibnamefont {Mosselman}}, \bibinfo
  {author} {\bibfnamefont {C.~J.}\ \bibnamefont {Sloff}}, \ and\ \bibinfo
  {author} {\bibfnamefont {Y.}~\bibnamefont {Shimizu}},\ }\bibfield  {title}
  {\enquote {\bibinfo {title} {Detailed simulation of morphodynamics: 2.
  {S}ediment pickup, transport, and deposition},}\ }\href@noop {} {\bibfield
  {journal} {\bibinfo  {journal} {Water Resour. Res.}\ }\textbf {\bibinfo
  {volume} {49}},\ \bibinfo {pages} {4775--4791} (\bibinfo {year}
  {2013})}\BibitemShut {NoStop}%
\bibitem [{\citenamefont {Mehta}\ \emph {et~al.}(1989)\citenamefont {Mehta},
  \citenamefont {Hayter}, \citenamefont {Parker}, \citenamefont {Krone},\ and\
  \citenamefont {Teeter}}]{mehta1989}%
  \BibitemOpen
  \bibfield  {author} {\bibinfo {author} {\bibfnamefont {A.~J.}\ \bibnamefont
  {Mehta}}, \bibinfo {author} {\bibfnamefont {E.~J.}\ \bibnamefont {Hayter}},
  \bibinfo {author} {\bibfnamefont {W.~R.}\ \bibnamefont {Parker}}, \bibinfo
  {author} {\bibfnamefont {R.~B.}\ \bibnamefont {Krone}}, \ and\ \bibinfo
  {author} {\bibfnamefont {A.~M.}\ \bibnamefont {Teeter}},\ }\bibfield  {title}
  {\enquote {\bibinfo {title} {Cohesive sediment transport. i: Process
  description},}\ }\href@noop {} {\bibfield  {journal} {\bibinfo  {journal} {J.
  Hydraul. Eng.}\ }\textbf {\bibinfo {volume} {115}},\ \bibinfo {pages}
  {1076--1093} (\bibinfo {year} {1989})}\BibitemShut {NoStop}%
\bibitem [{\citenamefont {Diplas}\ \emph {et~al.}(2008)\citenamefont {Diplas},
  \citenamefont {Dancey}, \citenamefont {Celik}, \citenamefont {Valyrakis},
  \citenamefont {Greer},\ and\ \citenamefont {Akar}}]{diplas2008}%
  \BibitemOpen
  \bibfield  {author} {\bibinfo {author} {\bibfnamefont {P.}~\bibnamefont
  {Diplas}}, \bibinfo {author} {\bibfnamefont {C.~L.}\ \bibnamefont {Dancey}},
  \bibinfo {author} {\bibfnamefont {A.~O.}\ \bibnamefont {Celik}}, \bibinfo
  {author} {\bibfnamefont {M.}~\bibnamefont {Valyrakis}}, \bibinfo {author}
  {\bibfnamefont {K.}~\bibnamefont {Greer}}, \ and\ \bibinfo {author}
  {\bibfnamefont {T.}~\bibnamefont {Akar}},\ }\bibfield  {title} {\enquote
  {\bibinfo {title} {The role of impulse on the initiation of particle movement
  under turbulent flow conditions},}\ }\href@noop {} {\bibfield  {journal}
  {\bibinfo  {journal} {Science}\ }\textbf {\bibinfo {volume} {322}},\ \bibinfo
  {pages} {717--720} (\bibinfo {year} {2008})}\BibitemShut {NoStop}%
\bibitem [{\citenamefont {Robinson}(1991)}]{robinson1991}%
  \BibitemOpen
  \bibfield  {author} {\bibinfo {author} {\bibfnamefont {S.~K.}\ \bibnamefont
  {Robinson}},\ }\bibfield  {title} {\enquote {\bibinfo {title} {Coherent
  motions in the turbulent boundary layer},}\ }\href@noop {} {\bibfield
  {journal} {\bibinfo  {journal} {Annu. Rev. Fluid Mech.}\ }\textbf {\bibinfo
  {volume} {23}},\ \bibinfo {pages} {601--639} (\bibinfo {year}
  {1991})}\BibitemShut {NoStop}%
\bibitem [{\citenamefont {Adrian}(2007)}]{adrian2007}%
  \BibitemOpen
  \bibfield  {author} {\bibinfo {author} {\bibfnamefont {R.~J.}\ \bibnamefont
  {Adrian}},\ }\bibfield  {title} {\enquote {\bibinfo {title} {Hairpin vortex
  organization in wall turbulence},}\ }\href@noop {} {\bibfield  {journal}
  {\bibinfo  {journal} {Phys. Fluids}\ }\textbf {\bibinfo {volume} {19}},\
  \bibinfo {pages} {041301} (\bibinfo {year} {2007})}\BibitemShut {NoStop}%
\bibitem [{\citenamefont {Hardy}\ \emph {et~al.}(2009)\citenamefont {Hardy},
  \citenamefont {Best}, \citenamefont {Lane},\ and\ \citenamefont
  {Carbonneau}}]{hardy2009}%
  \BibitemOpen
  \bibfield  {author} {\bibinfo {author} {\bibfnamefont {R.~J.}\ \bibnamefont
  {Hardy}}, \bibinfo {author} {\bibfnamefont {J.~L.}\ \bibnamefont {Best}},
  \bibinfo {author} {\bibfnamefont {S.~N.}\ \bibnamefont {Lane}}, \ and\
  \bibinfo {author} {\bibfnamefont {P.~E.}\ \bibnamefont {Carbonneau}},\
  }\bibfield  {title} {\enquote {\bibinfo {title} {Coherent flow structures in
  a depth-limited flow over a gravel surface: {T}he role of near-bed turbulence
  and influence of {R}eynolds number},}\ }\href@noop {} {\bibfield  {journal}
  {\bibinfo  {journal} {J. Geophys. Res. Earth Surf.}\ }\textbf {\bibinfo
  {volume} {114}},\ \bibinfo {eid} {F01003} (\bibinfo {year}
  {2009})}\BibitemShut {NoStop}%
\bibitem [{\citenamefont {Vowinckel}\ \emph {et~al.}(2016)\citenamefont
  {Vowinckel}, \citenamefont {Jain}, \citenamefont {Kempe},\ and\ \citenamefont
  {Fr{\"o}hlich}}]{vowinckel2016}%
  \BibitemOpen
  \bibfield  {author} {\bibinfo {author} {\bibfnamefont {B.}~\bibnamefont
  {Vowinckel}}, \bibinfo {author} {\bibfnamefont {R.}~\bibnamefont {Jain}},
  \bibinfo {author} {\bibfnamefont {T.}~\bibnamefont {Kempe}}, \ and\ \bibinfo
  {author} {\bibfnamefont {J.}~\bibnamefont {Fr{\"o}hlich}},\ }\bibfield
  {title} {\enquote {\bibinfo {title} {Entrainment of single particles in a
  turbulent open-channel flow: a numerical study},}\ }\href@noop {} {\bibfield
  {journal} {\bibinfo  {journal} {J. Hydraul. Res.}\ }\textbf {\bibinfo
  {volume} {54}},\ \bibinfo {pages} {158--171} (\bibinfo {year}
  {2016})}\BibitemShut {NoStop}%
\bibitem [{\citenamefont {Drake}\ and\ \citenamefont
  {Calantoni}(2001)}]{drake2001}%
  \BibitemOpen
  \bibfield  {author} {\bibinfo {author} {\bibfnamefont {T.~G.}\ \bibnamefont
  {Drake}}\ and\ \bibinfo {author} {\bibfnamefont {J.}~\bibnamefont
  {Calantoni}},\ }\bibfield  {title} {\enquote {\bibinfo {title} {Discrete
  particle model for sheet flow sediment transport in the nearshore},}\
  }\href@noop {} {\bibfield  {journal} {\bibinfo  {journal} {J. Geophys. Res.
  Oceans}\ }\textbf {\bibinfo {volume} {106}},\ \bibinfo {pages} {19859--19868}
  (\bibinfo {year} {2001})}\BibitemShut {NoStop}%
\bibitem [{\citenamefont {Ni{\~n}o}\ and\ \citenamefont
  {Garc{\'\i}a}(1998)}]{nino1998}%
  \BibitemOpen
  \bibfield  {author} {\bibinfo {author} {\bibfnamefont {Y.}~\bibnamefont
  {Ni{\~n}o}}\ and\ \bibinfo {author} {\bibfnamefont {M.}~\bibnamefont
  {Garc{\'\i}a}},\ }\bibfield  {title} {\enquote {\bibinfo {title} {Experiments
  on saltation of sand in water},}\ }\href@noop {} {\bibfield  {journal}
  {\bibinfo  {journal} {J. Hydraul. Eng.}\ }\textbf {\bibinfo {volume} {124}},\
  \bibinfo {pages} {1014--1025} (\bibinfo {year} {1998})}\BibitemShut {NoStop}%
\bibitem [{\citenamefont {Clark}\ \emph {et~al.}(2015)\citenamefont {Clark},
  \citenamefont {Shattuck}, \citenamefont {Ouellette},\ and\ \citenamefont
  {O'Hern}}]{clark2015hydro}%
  \BibitemOpen
  \bibfield  {author} {\bibinfo {author} {\bibfnamefont {A.~H.}\ \bibnamefont
  {Clark}}, \bibinfo {author} {\bibfnamefont {M.~D.}\ \bibnamefont {Shattuck}},
  \bibinfo {author} {\bibfnamefont {N.~T.}\ \bibnamefont {Ouellette}}, \ and\
  \bibinfo {author} {\bibfnamefont {C.~S.}\ \bibnamefont {O'Hern}},\ }\bibfield
   {title} {\enquote {\bibinfo {title} {Onset and cessation of motion in
  hydrodynamically sheared granular beds},}\ }\href@noop {} {\bibfield
  {journal} {\bibinfo  {journal} {Phys. Rev. E}\ }\textbf {\bibinfo {volume}
  {92}},\ \bibinfo {eid} {042202} (\bibinfo {year} {2015})}\BibitemShut
  {NoStop}%
\bibitem [{\citenamefont {Perera}\ and\ \citenamefont
  {Harrowell}(1999)}]{Perera1999}%
  \BibitemOpen
  \bibfield  {author} {\bibinfo {author} {\bibfnamefont {D.~N.}\ \bibnamefont
  {Perera}}\ and\ \bibinfo {author} {\bibfnamefont {P.}~\bibnamefont
  {Harrowell}},\ }\bibfield  {title} {\enquote {\bibinfo {title} {Stability and
  structure of a supercooled liquid mixture in two dimensions},}\ }\href@noop
  {} {\bibfield  {journal} {\bibinfo  {journal} {Phys. Rev. E}\ }\textbf
  {\bibinfo {volume} {59}},\ \bibinfo {pages} {5721--5743} (\bibinfo {year}
  {1999})}\BibitemShut {NoStop}%
\bibitem [{\citenamefont {Speedy}(1999)}]{Speedy1999}%
  \BibitemOpen
  \bibfield  {author} {\bibinfo {author} {\bibfnamefont {R.~J.}\ \bibnamefont
  {Speedy}},\ }\bibfield  {title} {\enquote {\bibinfo {title} {Glass transition
  in hard disc mixtures},}\ }\href@noop {} {\bibfield  {journal} {\bibinfo
  {journal} {J. Chem. Phys.}\ }\textbf {\bibinfo {volume} {110}},\ \bibinfo
  {pages} {4559--4565} (\bibinfo {year} {1999})}\BibitemShut {NoStop}%
\bibitem [{\citenamefont {Zhang}\ \emph {et~al.}(2014)\citenamefont {Zhang},
  \citenamefont {Smith}, \citenamefont {Wang}, \citenamefont {Liu},
  \citenamefont {Schroers}, \citenamefont {Shattuck},\ and\ \citenamefont
  {O'Hern}}]{Zhang2014}%
  \BibitemOpen
  \bibfield  {author} {\bibinfo {author} {\bibfnamefont {K.}~\bibnamefont
  {Zhang}}, \bibinfo {author} {\bibfnamefont {W.~W.}\ \bibnamefont {Smith}},
  \bibinfo {author} {\bibfnamefont {M.}~\bibnamefont {Wang}}, \bibinfo {author}
  {\bibfnamefont {Y.}~\bibnamefont {Liu}}, \bibinfo {author} {\bibfnamefont
  {J.}~\bibnamefont {Schroers}}, \bibinfo {author} {\bibfnamefont {M.~D.}\
  \bibnamefont {Shattuck}}, \ and\ \bibinfo {author} {\bibfnamefont {C.~S.}\
  \bibnamefont {O'Hern}},\ }\bibfield  {title} {\enquote {\bibinfo {title}
  {Connection between the packing efficiency of binary hard spheres and the
  glass-forming ability of bulk metallic glasses},}\ }\href@noop {} {\bibfield
  {journal} {\bibinfo  {journal} {Phys. Rev. E}\ }\textbf {\bibinfo {volume}
  {90}},\ \bibinfo {eid} {032311} (\bibinfo {year} {2014})}\BibitemShut
  {NoStop}%
\bibitem [{\citenamefont {Cundall}\ and\ \citenamefont
  {Strack}(1979)}]{cundall79}%
  \BibitemOpen
  \bibfield  {author} {\bibinfo {author} {\bibfnamefont {P.~A.}\ \bibnamefont
  {Cundall}}\ and\ \bibinfo {author} {\bibfnamefont {O.~D.~L.}\ \bibnamefont
  {Strack}},\ }\bibfield  {title} {\enquote {\bibinfo {title} {A discrete
  numerical model for granular assemblies},}\ }\href@noop {} {\bibfield
  {journal} {\bibinfo  {journal} {G\'eotechnique}\ }\textbf {\bibinfo {volume}
  {29}},\ \bibinfo {pages} {47--65} (\bibinfo {year} {1979})}\BibitemShut
  {NoStop}%
\bibitem [{\citenamefont {Sch\"afer}\ \emph {et~al.}(1996)\citenamefont
  {Sch\"afer}, \citenamefont {Dippel},\ and\ \citenamefont {Wolf}}]{schafer96}%
  \BibitemOpen
  \bibfield  {author} {\bibinfo {author} {\bibfnamefont {J.}~\bibnamefont
  {Sch\"afer}}, \bibinfo {author} {\bibfnamefont {S.}~\bibnamefont {Dippel}}, \
  and\ \bibinfo {author} {\bibfnamefont {D.~E.}\ \bibnamefont {Wolf}},\
  }\bibfield  {title} {\enquote {\bibinfo {title} {Force schemes in simulations
  of granular materials},}\ }\href@noop {} {\bibfield  {journal} {\bibinfo
  {journal} {J. Phys. I France}\ }\textbf {\bibinfo {volume} {6}},\ \bibinfo
  {pages} {5} (\bibinfo {year} {1996})}\BibitemShut {NoStop}%
\bibitem [{\citenamefont {B.}\ and\ \citenamefont {P.}(1994)}]{Buchholtz1994}%
  \BibitemOpen
  \bibfield  {author} {\bibinfo {author} {\bibfnamefont {Volkhard}\
  \bibnamefont {B.}}\ and\ \bibinfo {author} {\bibfnamefont {Thorsten}\
  \bibnamefont {P.}},\ }\bibfield  {title} {\enquote {\bibinfo {title}
  {Numerical investigations of the evolution of sandpiles},}\ }\href@noop {}
  {\bibfield  {journal} {\bibinfo  {journal} {Physica A}\ }\textbf {\bibinfo
  {volume} {202}},\ \bibinfo {pages} {390 -- 401} (\bibinfo {year}
  {1994})}\BibitemShut {NoStop}%
\bibitem [{\citenamefont {Papanikolaou}\ \emph {et~al.}(2013)\citenamefont
  {Papanikolaou}, \citenamefont {O'Hern},\ and\ \citenamefont
  {Shattuck}}]{papanikolaou2013}%
  \BibitemOpen
  \bibfield  {author} {\bibinfo {author} {\bibfnamefont {S.}~\bibnamefont
  {Papanikolaou}}, \bibinfo {author} {\bibfnamefont {C.~S.}\ \bibnamefont
  {O'Hern}}, \ and\ \bibinfo {author} {\bibfnamefont {M.~D.}\ \bibnamefont
  {Shattuck}},\ }\bibfield  {title} {\enquote {\bibinfo {title} {Isostaticity
  at frictional jamming},}\ }\href@noop {} {\bibfield  {journal} {\bibinfo
  {journal} {Phys. Rev. Lett.}\ }\textbf {\bibinfo {volume} {110}},\ \bibinfo
  {eid} {198002} (\bibinfo {year} {2013})}\BibitemShut {NoStop}%
\bibitem [{\citenamefont {Pilotti}\ and\ \citenamefont
  {Menduni}(2001)}]{pilotti2001}%
  \BibitemOpen
  \bibfield  {author} {\bibinfo {author} {\bibfnamefont {M.}~\bibnamefont
  {Pilotti}}\ and\ \bibinfo {author} {\bibfnamefont {G.}~\bibnamefont
  {Menduni}},\ }\bibfield  {title} {\enquote {\bibinfo {title} {Beginning of
  sediment transport of incoherent grains in shallow shear flows},}\
  }\href@noop {} {\bibfield  {journal} {\bibinfo  {journal} {J. Hydraul. Res.}\
  }\textbf {\bibinfo {volume} {39}},\ \bibinfo {pages} {115--124} (\bibinfo
  {year} {2001})}\BibitemShut {NoStop}%
\bibitem [{\citenamefont {Ouriemi}\ \emph {et~al.}(2007)\citenamefont
  {Ouriemi}, \citenamefont {Aussillous}, \citenamefont {Medale}, \citenamefont
  {Peysson},\ and\ \citenamefont {Guazzelli}}]{ouriemi2007}%
  \BibitemOpen
  \bibfield  {author} {\bibinfo {author} {\bibfnamefont {M.}~\bibnamefont
  {Ouriemi}}, \bibinfo {author} {\bibfnamefont {P.}~\bibnamefont {Aussillous}},
  \bibinfo {author} {\bibfnamefont {M.}~\bibnamefont {Medale}}, \bibinfo
  {author} {\bibfnamefont {Y.}~\bibnamefont {Peysson}}, \ and\ \bibinfo
  {author} {\bibfnamefont {E.}~\bibnamefont {Guazzelli}},\ }\bibfield  {title}
  {\enquote {\bibinfo {title} {Determination of the critical shields number for
  particle erosion in laminar flow},}\ }\href@noop {} {\bibfield  {journal}
  {\bibinfo  {journal} {Phys. Fluids}\ }\textbf {\bibinfo {volume} {19}},\
  \bibinfo {eid} {061706} (\bibinfo {year} {2007})}\BibitemShut {NoStop}%
\bibitem [{\citenamefont {Joseph}\ and\ \citenamefont
  {Hunt}(2004)}]{joseph2004}%
  \BibitemOpen
  \bibfield  {author} {\bibinfo {author} {\bibfnamefont {G.~G.}\ \bibnamefont
  {Joseph}}\ and\ \bibinfo {author} {\bibfnamefont {M.~L.}\ \bibnamefont
  {Hunt}},\ }\bibfield  {title} {\enquote {\bibinfo {title} {Oblique
  particle--wall collisions in a liquid},}\ }\href@noop {} {\bibfield
  {journal} {\bibinfo  {journal} {J. Fluid Mech.}\ }\textbf {\bibinfo {volume}
  {510}},\ \bibinfo {pages} {71--93} (\bibinfo {year} {2004})}\BibitemShut
  {NoStop}%
\bibitem [{\citenamefont {Weibull}(1939)}]{weibull1939}%
  \BibitemOpen
  \bibfield  {author} {\bibinfo {author} {\bibfnamefont {W.}~\bibnamefont
  {Weibull}},\ }\href {http://books.google.com/books?id=otVRAQAAIAAJ} {\emph
  {\bibinfo {title} {A Statistical Theory of the Strength of Materials}}},\
  Ingeni{\"o}rsvetenskapsakademiens handlingar\ (\bibinfo  {publisher}
  {Generalstabens litografiska anstalts f{\"o}rlag},\ \bibinfo {year}
  {1939})\BibitemShut {NoStop}%
\bibitem [{\citenamefont {Weibull}(1951)}]{weibull1951}%
  \BibitemOpen
  \bibfield  {author} {\bibinfo {author} {\bibfnamefont {W.}~\bibnamefont
  {Weibull}},\ }\bibfield  {title} {\enquote {\bibinfo {title} {A statistical
  distribution function of wide applicability},}\ }\href@noop {} {\bibfield
  {journal} {\bibinfo  {journal} {J. Appl. Mech.}\ }\textbf {\bibinfo {volume}
  {18}},\ \bibinfo {pages} {293--297} (\bibinfo {year} {1951})}\BibitemShut
  {NoStop}%
\bibitem [{\citenamefont {Franklin}(2014)}]{franklin2014}%
  \BibitemOpen
  \bibfield  {author} {\bibinfo {author} {\bibfnamefont {S.~V.}\ \bibnamefont
  {Franklin}},\ }\bibfield  {title} {\enquote {\bibinfo {title} {Extensional
  rheology of entangled granular materials},}\ }\href
  {http://stacks.iop.org/0295-5075/106/i=5/a=58004} {\bibfield  {journal}
  {\bibinfo  {journal} {EPL}\ }\textbf {\bibinfo {volume} {106}},\ \bibinfo
  {eid} {58004} (\bibinfo {year} {2014})}\BibitemShut {NoStop}%
\bibitem [{\citenamefont {Gonz{\'a}lez}\ \emph {et~al.}(2017)\citenamefont
  {Gonz{\'a}lez}, \citenamefont {Richter}, \citenamefont {Bolster},
  \citenamefont {Bateman}, \citenamefont {Calantoni},\ and\ \citenamefont
  {Escauriaza}}]{gonzalez2016}%
  \BibitemOpen
  \bibfield  {author} {\bibinfo {author} {\bibfnamefont {C.}~\bibnamefont
  {Gonz{\'a}lez}}, \bibinfo {author} {\bibfnamefont {D.~H.}\ \bibnamefont
  {Richter}}, \bibinfo {author} {\bibfnamefont {D.}~\bibnamefont {Bolster}},
  \bibinfo {author} {\bibfnamefont {S.}~\bibnamefont {Bateman}}, \bibinfo
  {author} {\bibfnamefont {J.}~\bibnamefont {Calantoni}}, \ and\ \bibinfo
  {author} {\bibfnamefont {C.}~\bibnamefont {Escauriaza}},\ }\bibfield  {title}
  {\enquote {\bibinfo {title} {Characterization of bedload intermittency near
  the threshold of motion using a lagrangian sediment transport model},}\
  }\href@noop {} {\bibfield  {journal} {\bibinfo  {journal} {Environ. Fluid
  Mech.}\ }\textbf {\bibinfo {volume} {17}},\ \bibinfo {pages} {111--137}
  (\bibinfo {year} {2017})}\BibitemShut {NoStop}%
\bibitem [{\citenamefont {Mitha}\ \emph {et~al.}(1986)\citenamefont {Mitha},
  \citenamefont {Tran}, \citenamefont {Werner},\ and\ \citenamefont
  {Haff}}]{mitha1986}%
  \BibitemOpen
  \bibfield  {author} {\bibinfo {author} {\bibfnamefont {S.}~\bibnamefont
  {Mitha}}, \bibinfo {author} {\bibfnamefont {M.~Q.}\ \bibnamefont {Tran}},
  \bibinfo {author} {\bibfnamefont {B.~T.}\ \bibnamefont {Werner}}, \ and\
  \bibinfo {author} {\bibfnamefont {P.~K.}\ \bibnamefont {Haff}},\ }\bibfield
  {title} {\enquote {\bibinfo {title} {The grain-bed impact process in aeolian
  saltation},}\ }\href@noop {} {\bibfield  {journal} {\bibinfo  {journal} {Acta
  Mech.}\ }\textbf {\bibinfo {volume} {63}},\ \bibinfo {pages} {267--278}
  (\bibinfo {year} {1986})}\BibitemShut {NoStop}%
\bibitem [{\citenamefont {Gao}\ \emph {et~al.}(2006)\citenamefont {Gao},
  \citenamefont {B\l{}awzdziewicz},\ and\ \citenamefont {O'Hern}}]{gao2006}%
  \BibitemOpen
  \bibfield  {author} {\bibinfo {author} {\bibfnamefont {G.~J.}\ \bibnamefont
  {Gao}}, \bibinfo {author} {\bibfnamefont {J.}~\bibnamefont
  {B\l{}awzdziewicz}}, \ and\ \bibinfo {author} {\bibfnamefont {C.~S.}\
  \bibnamefont {O'Hern}},\ }\bibfield  {title} {\enquote {\bibinfo {title}
  {Frequency distribution of mechanically stable disk packings},}\ }\href@noop
  {} {\bibfield  {journal} {\bibinfo  {journal} {Phys. Rev. E}\ }\textbf
  {\bibinfo {volume} {74}},\ \bibinfo {eid} {061304} (\bibinfo {year}
  {2006})}\BibitemShut {NoStop}%
\end{thebibliography}%

\end{document}